\documentclass[11pt]{article}
\usepackage[utf8]{inputenc}
\usepackage{amsfonts,amsmath,amssymb,mathtools}
\usepackage[left=2cm,right=2cm,top=2cm,bottom=2cm,bindingoffset=0cm]{geometry}
\usepackage{graphicx}
\usepackage{cite}
\usepackage{authblk}
\usepackage{slashed}
\usepackage{braket}
\usepackage{xcolor}
\usepackage{hyperref}

\newcommand\bem{\begin{pmatrix}}
\newcommand\eem{\end{pmatrix}}
\newcommand\beq{\begin{equation}}
\newcommand\eeq{\end{equation}}
\newcommand\beqs{\begin{equation*}}
\newcommand\eeqs{\end{equation*}}

\newcommand{\tr}{\text{tr}}
\newcommand{\pd}{\partial}
\newcommand{\sgn}{\text{sgn}}
\newcommand{\Sch}{\text{Sch}}
\newcommand{\TOC}{\text{TOC}}
\newcommand{\OTOC}{\text{OTOC}}

\numberwithin{equation}{section}

\hypersetup{
    colorlinks=true,
    linkcolor=black,
    citecolor=black,   
    urlcolor=blue,
}

\title{\bf Pedagogical introduction to SYK model and 2D Dilaton Gravity}
\author[1,2]{Dmitrii~A.~Trunin\thanks{\href{mailto:dmitriy.trunin@phystech.edu}{dmitriy.trunin@phystech.edu}}}
\affil[1]{Moscow Institute of Physics and Technology, 141700, Institutskii per., 9, Dolgoprudny, Russia}
\affil[2]{Institute for Theoretical and Experimental Physics, 117218, B. Cheremushkinskaya, 25, Moscow, Russia}
\date{\today}

\begin{document}

\maketitle

\begin{abstract}
SYK model and 2D dilaton gravity have recently attracted considerable attention from the high energy and condensed matter physics community. The success of these models is due to their remarkable properties. Following the original papers, we broadly discuss the properties of these models, including large $N$ diagrammatics, emergence of conformal symmetry in the IR limit, effective action, four-point functions and chaos. Also we briefly review some recent results in the topic. On the one hand, we try to be as specific as possible, i.e. reveal every detail and loophole in the discussion. On the other hand, we expect this review to be suitable even for a reader who is not familiar with these models.
\end{abstract}

\newpage
\tableofcontents
\newpage

\section{Introduction}

Sachdev--Ye--Kitaev (SYK) model was proposed by Kitaev~\cite{Kitaev-talks} as a generalization of Sahdev--Ye model~\cite{Sachdev-Ye, Sachdev-1006} and first was extensively studied in~\cite{Polchinski, Maldacena-SYK, Kitaev, Jevicki-1, Jevicki-2}. Ever since it has received a great attention from the high energy and condensed matter physics community.

The success of SYK model is due to its remarkable properties. First, this model is exactly solvable in the large $N$ and IR limit. Second, in this limit the model acquires conformal symmetry and the effective action can be approximated by the Schwarzian one. Third, the leading correction to the out-of-time ordered four-point correlation function exponentially grows with time, with the exponent of this growth saturating the universal bound~\cite{MSS}. This behavior is very unusual; moreover, it coincides with the behavior of similar functions on a black hole background. Finally, SYK model is closely related to 2D dilaton gravity which describes excitations above the near horizon extremal black hole~\cite{Almheiri, Maldacena-JT, Jensen, Engelsoy}. Together these properties make SYK model an excellent toy model for many physical phenomena, including quantum chaos~\cite{Kitaev-talks, MSS}, information scrambling~\cite{Sekino, Susskind, Lashkari}, traversable wormholes~\cite{Maldacena-1704, Maldacena-1804, Maldacena-1807, Maldacena-1912} and strange metals~\cite{Hartnoll, Song, Sachdev}.

In this review we give a pedagogical introduction to SYK model and 2D dilaton gravity. We follow the original papers~\cite{Polchinski, Maldacena-SYK, Kitaev, Jevicki-1, Jevicki-2, MSS, Almheiri, Maldacena-JT, Jensen, Engelsoy} and try to be as specific as possible, i.e. we do our best to reveal every detail and loophole in the discussion. We believe this makes the discussion clear and self-consistent. Due to this reason we also expect the review to be suitable even for a reader who is not familiar with the phenomena under consideration.

The paper is organized as follows. In section~\ref{sec:chaos} we briefly discuss quantum chaos and scrambling, the phenomena that are related to the quantum black hole dynamics and motivate the study of SYK model and 2D dilaton gravity. In particular, we introduce out-of-time ordered correlation functions (OTOCs), which are the main tool for studying these phenomena. This section is relatively sketchy, because for brevity we postpone the discussion of specific examples to the following sections.

In sections~\ref{sec:basics} and~\ref{sec:treatment} we give a comprehensive review of SYK model. We broadly discuss large $N$ diagrammatics, emergence of conformal symmetry in the IR limit, effective and Schwarzian action, exact two-point and four-point functions. Some technical details are discussed in appendices. Also we briefly review recent results in the topic. 

In section~\ref{sec:JT} we attempt to give an equally comprehensive review of 2D dilaton gravity (or Jackiw--Teitelboim gravity). We show that this theory describes excitations above the near horizon extremal black hole, explain that this theory effectively reduces to the one-dimensional theory with Schwarzian action, calculate four-point functions of the matter fields living in the corresponding space.

Finally, instead of conclusion in section~\ref{sec:examples} we briefly review the most notable examples of chaotic behavior. Among them are SYK model and 2D dilaton gravity (we briefly recall the main properties of these models), SYK-like tensor models, BTZ black hole, $CFT_2$ with large central charge and Hermitian matrix quantum field theory with quartic self-interaction.

\section{Motivation}
\label{sec:chaos}

In this section we discuss the main motivation for studying SYK model and 2D dilaton gravity, which is based on the connection to quantum chaos (subsection~\ref{sec:classical}) and scrambling (subsection~\ref{sec:scramblers}). It is believed that these phenomena are related to the black hole information paradox~\cite{Sekino, Susskind}, so they have received a lot of attention.

Here we qualitatively show that both of these phenomena rely on the exponential growth of OTOCs, which were first calculated in~\cite{Larkin} and popularized by~\cite{Almheiri-1304, Shenker-1306, MSS}. Therefore, systems with such a behavior of the correlators are of particular interest. SYK model and 2D dilaton gravity are exactly such type of systems. In section~\ref{sec:examples} we also briefly review other chaotic systems.

Note that this section may seem relatively sketchy, because we do not discuss the limits of applicability of the statements being formulated and do not provide any specific examples. Such examples will be broadly discussed in the following sections. In fact, part of the original motivation to study SYK model was exactly to find a convenient example for which the statements of this section can be verified in a controlled way~\cite{Kitaev-talks}.

\subsection{Quantum chaos}
\label{sec:classical}

In this subsection we discuss a putative connection between some specific correlation functions and classical chaos~\cite{Kitaev-talks,MSS}.

First of all, let us remind what the classical chaos is. Consider a classical system with the following equation of motion:
\beq \label{eq:chaos-1}
\dot{X}^i(t) = F^i\left[X^i(t)\right], \quad i=1 \ldots N, \eeq
where $\mathbf{X}$ is a vector in the $N$-dimensional phase space, $\mathbf{F}$ is a smooth vector function and $\dot X^i \equiv d X^i/dt$. Let us introduce the norm  on the phase space, $\| \cdot \|$, and expand the function $\mathbf{F}$ near a point $\mathbf{X}_0$:
\beq \label{eq:chaos-2}
\delta \dot{X}^i = A^i_j \delta X^j + B^i (\delta X^i), \quad i=1 \ldots N, \eeq
where $\delta X^i \equiv X^i - X^i_0$, $A^i_j \equiv \left( \frac{\pd F^i}{\pd X^j} \right)_{\delta\mathbf{X} = \mathbf{0}}$ and $\mathbf{B}$ is analytical function such that $\left\| \mathbf{B}(\delta \mathbf{X}) \right\| \rightarrow 0$ as $\left\| \delta \mathbf{X} \right\| \rightarrow 0$. The solution of the linearized equation (i.e. equation with ommited $\mathbf{B}$) is straightforward:
\beq \delta \mathbf{X} = \sum_{j = 1}^N c_j \mathbf{h}_j e^{\lambda_j t}, \eeq
where $\lambda_j$ and $\mathbf{h}_j$ are eigenvalues and eigenvectors of the matrix $\mathbf{A}$ (for simplicity we assume that all eigenspaces are one-dimensional), $c_j$ are integration constants that correspond to the initial condition $\delta \mathbf{X}(t = 0) = \delta \mathbf{X}_0$. It is easy to see that for long evolution times but small $\delta \mathbf{X}_0$, such that the condition $\left\| \mathbf{B}(\delta \mathbf{X}) \right\| \ll \| \mathbf{A} \delta \mathbf{X} \|$ is always satisfied, the norm of the final deviation vector grows exponentially:
\beq \| \delta \mathbf{X} (t) \| \le \| \delta \mathbf{X}_0 \| e^{\lambda_{max} t}, \eeq
where $\lambda_{max}$ is the biggest eigenvalue of $\mathbf{A}$. If this eigenvalue is positive, phase space trajectories rapidly diverge, i.e. a small perturbation in the initial conditions leads to a significat change in the future behavior of the system (at least for some set of initial conditions). Such sensitivity to initial conditions is sometimes called the ``butterfly effect'' or ``classical chaos''.

In general, eigenvalues and eigenvectors depend on the point $\mathbf{X}_0$ and the definition of norm $\| \cdot \|$. However, the maximal eigenvalue, which is also referred to as the maximal Lyapunov exponent, can be considered as the general property of the system:
\beq \label{eq:chaos-3}
\lambda_{max} \equiv \lim_{t \rightarrow \infty} \lim_{\| \delta \mathbf{X} \| \rightarrow 0} \sup \left( \frac{1}{t} \log \frac{ \| \delta \mathbf{X}(t) \|}{ \| \delta \mathbf{X}(0) \| } \right). \eeq
This definition can be applied both to linearized~\eqref{eq:chaos-2} and general systems~\eqref{eq:chaos-1}. Since the exponent~\eqref{eq:chaos-3} does not depend on the definition of the norm~\cite{Gur-Ari,Eichhorn}, we can choose it as $\| \mathbf{X} \| = \sum_{i=1}^N |X^i|$. Then the sensitivity to initial conditions can be reformulated as follows:
\beq \label{eq:chaos-4}
\left| \frac{\pd X^i(t)}{\pd X^j(0)} \right| \approx \left| \frac{\delta X^i(t)}{\delta X^j(0)} \right| \sim e^{\lambda t}, \eeq
for some components $X^i$ and $X^j$ of the vector $\mathbf{X}(t)$, which describes the phase trajectory. The first identity is approximately equal for small $\delta \textbf{X}$.

Then let us consider a larger system whose configuration space coincides with the phase space of the initial system: $q^i = X^i$, $i=1 \ldots N$. Here $q^i$ are generalized coordinates, corresponding generalized momenta are denoted as $p^i$. Introducing the Poisson bracket $\{ \cdot, \cdot \}_{PB}$, we can rewrite the property~\eqref{eq:chaos-4} in a form suitable for quantum generalizations:
\beq \left| \{ q^i(t), p^j(0) \}_{PB} \right| = \left| \sum_{k = 1}^N \frac{\pd q^i(t)}{\pd q^k(0)} \frac{\pd p^j(0)}{\pd p^k(0)} -  \frac{\pd q^i(t)}{\pd p^k(0)} \frac{\pd p^j(0)}{\pd q^k(0)} \right| = \left| \frac{\pd q^i(t)}{\pd q^j(0)} \right| \sim e^{\lambda t}. \eeq
So far we have considered classical mechanics. Now let us proceed to the quantum mechanical situation. We remind that in the semiclassical limit the Poisson bracket coincides with the commutator of the corresponding operators:
\beq \label{eq:chaos-5}
\left\{ q^i(t), p^j(0) \right\}_{PB} \sim -\frac{i}{\hbar} \left[ \hat{q}^i(t), \hat{p}^j(0) \right], \quad \text{as} \quad \hbar \rightarrow 0. \eeq
Note that the position and momentum operators act at different moments of time, so the expression~\eqref{eq:chaos-5} is not trivial.

This correspondence allows one to extend the concept of classical chaos and maximal Lyapunov exponent to arbitrary quantum systems~\cite{Wijn, Fine, Kitaev-talks, Aleiner-96}. Roughly speaking, we want to derive a quantity that correctly captures the sensitivity of the quantum system to a change in initial conditions and reproduces the exponential growth~\eqref{eq:chaos-4} in the limit $\hbar \rightarrow 0$ if the system is chaotic. The simplest expression of this kind is the following amplitude:
\beq \label{eq:chaos-6}
A_{in-out} = \langle out | \left[ q^i(t), p^j(0) \right] | in \rangle, \eeq
where $| in \rangle$ and $| out \rangle$ are initial and final wave-functions of the system. Unfortunately, this expression has two drawbacks. First, due to the dependence on the specific states the quantity~\eqref{eq:chaos-6} varies significantly for the same system. Second, in quantum field theory one usually considers the analog of~\eqref{eq:chaos-6} for the vacuum state or thermal ensemble, for which two-point functions exponentially decay rather than grow (in quantum mechanics correlation functions decay or grow algebraically). Thus, we need to eliminate the dependence on $| in \rangle$ and $| out \rangle$.

In order to do this we sum over final states and average over a suitable initial ensemble, e.g. over the thermal one:
\beq \label{eq:chaos-7}
C(t) = \sum_n \sum_{out} \frac{1}{Z} e^{-\beta E_n} \langle n | \left[ q^i(t), p^j(0) \right]^\dagger | out \rangle \langle out | \left[ q^i(t), p^j(0) \right] | n \rangle = -\langle \left[ q^i(t), p^j(0) \right]^2 \rangle_\beta,  \eeq
where $\beta$ is the inverse temperature, $E_n$ is the energy of the $n$-th energy level, $Z = \sum_n e^{-\beta E_n}$ is the partition function, $\langle \cdots \rangle_\beta$ denotes the averaging over the thermal ensemble. Such an average was first considered in the classical paper~\cite{Larkin}.

On the one hand, due to~\eqref{eq:chaos-5} we expect that this quantity exponentially grows: $C(t) \sim \hbar^2 e^{2 \lambda t}$. On the other hand, the semiclassical approximation is applicable only for small enough times, $t < t_* \sim \frac{1}{\lambda} \log \frac{1}{\hbar}$, where $t_*$ is called the ``Ehrenfest time''~\cite{Aleiner-96, Silvestrov, Berman, Zaslavsky}. Note that $t_* \rightarrow \infty$ as $\hbar \rightarrow 0$. One expects that for larger times correlator $C(t)$ approaches some constant value~\cite{MSS,Almheiri-1304}.

The quantity~\eqref{eq:chaos-7} can be easily generalized to an arbitrary quantum system with a large number of degrees of freedom, $N \gg 1$:
\beq \label{eq:chaos-8}
C(t) = -\langle \left[V(t), W(0) \right]^2 \rangle_\beta, \eeq
where $V$ and $W$ are Hermitian operators each of which has vanishing one-point function ($\langle V \rangle_\beta = \langle W \rangle_\beta = 0$) and corresponds to $\mathcal{O}(1)$ degrees of freedom\footnote{E.g. in the case of SYK model such operators are Majorana fermions: $V(t) = \chi_i(t)$, $W(0) = \chi_j(0)$.}. We call the system chaotic if the quantity~\eqref{eq:chaos-8} grows exponentially for \textit{all possible pairs}\footnote{In integrable systems the function $C(t)$ can grow for some, but not all pairs of operators, e.g. see~\cite{Roberts-1412}.} of operators $V$ and $W$ with mentioned properties. The maximal exponent of this growth is referred to as ``quantum Lyapunov exponent''. The time $t_*$ at which $C(t)$ saturates is referred to as ``scrambling time'', which is an analog of the Ehrenfest time. We will discuss the motivation for this terminology in more detail in section~\ref{sec:scramblers}.

Note that in practice the correlator~\eqref{eq:chaos-8} should be regularized because it contains the product of operators at coincident times. A common approach is to uniformly smear the thermal distribution between the two commutators (which is equivalent to smearing of operators in the imaginary time):
\beq \label{eq:chaos-9}
C(t) = -\tr\left( \rho^{\frac{1}{2}} \left[V(t), W(0) \right] \rho^{\frac{1}{2}} \left[V(t), W(0) \right] \right), \eeq
where $\rho = \frac{1}{Z} e^{-\beta H}$ is the density matrix. Of course, one can also consider other types of smearing, but this one has the most natural physical interpretation, see~\cite{Romero-Bermudez} for a more detailed discussion. Therefore, in this paper we are interested in such correlators as~\eqref{eq:chaos-9}. In the main body of this paper we will see how such an expression arises naturally.

Let us expand the commutators in~\eqref{eq:chaos-9} and rewrite $C(t)$ as the sum of four four-point correlation functions:
\beq \begin{aligned}
C(t) &= 2 \tr\left( V(t) \rho^{\frac{1}{2}} V(t) W \rho^{\frac{1}{2}} W \right) - \tr\left( \rho^{\frac{1}{2}} V(t) W \rho^{\frac{1}{2}} V(t) W \right) - \tr\left( \rho^{\frac{1}{2}} W V(t) \rho^{\frac{1}{2}} V(t) W \right) = \\
&= 2 \times  \TOC(t) - \OTOC\left(t - \frac{i\beta}{4}\right) - \OTOC\left(t + \frac{i\beta}{4}\right),
\end{aligned} \eeq
where we denoted $W = W(0)$ for short, introduced time-ordered correlator (TOC) and out-of-time ordered correlator (OTOC):
\beq \TOC(t) \equiv \tr\left( V(t) \rho^{\frac{1}{2}} V(t) W \rho^{\frac{1}{2}} W \right), \quad \OTOC(t) \equiv \tr\left( \rho^{\frac{1}{4}} V(t) \rho^{\frac{1}{4}} W \rho^{\frac{1}{4}} V(t) \rho^{\frac{1}{4}} W \right). \eeq
There are two important time scales for $C(t)$. First one is the dissipation time $t_d$, at which two-point correlation functions exponentially decay: $\langle V(t) V \rangle_\beta \sim \langle W(t) W \rangle_\beta \sim \langle V(t) W \rangle_\beta \sim e^{-t/t_d}$. Typically $t_d \sim \beta$. At this time scale both TOC and OTOC are approximately equal to the product of two disconnected two-point functions, so the commutator $C(t)$ is close to zero~\cite{MSS,Polyakov,Makeenko}:
\beq \TOC(t) \approx \OTOC(t) \approx \langle V V \rangle_\beta \langle W W \rangle_\beta  + \mathcal{O}\left(e^{-t/t_d}\right) + \mathcal{O}\left(\frac{1}{N}\right), \eeq
where we denoted $\langle V V \rangle_\beta = \left\langle V\left(-i\beta/2\right) V \right\rangle_\beta = \tr\left( \rho^{\frac{1}{2}} V \rho^{\frac{1}{2}} V \right)$ for short. We remind that we work in the large $N$ limit, so the number $\frac{1}{N}$ plays the role of Planck's constant $\hbar$.

The second time scale is the scrambling time $t_*$. Typically $t_*$ is parametrically larger than $t_d$, namely $t_* \sim \beta \log N$. If the system is chaotic, well after the dissipation time and well before the scrambling time $C(t)$ exponentially grows and OTOC rapidly decays:
\beq \label{eq:chaos-10}
C(t) \sim \frac{1}{N} e^{\kappa t}, \quad \OTOC(t) \sim \langle V V \rangle_\beta \langle W W \rangle_\beta - \frac{A}{N} e^{\kappa t}, \eeq
where $A$ is some numerical coefficient. At greater times $C(t)$ is saturated and OTOC approaches zero. Since TOC at such times is approximately constant, growth of $C(t)$ and decay of OTOC are qualitatively identical.

Thus, such a behavior of OTOC and of the function $C(t)$ can be considered as an indicator of quantum chaos. In particular, it allows one to extract the quantum Lyapunov exponent $\kappa$, which is expected to coincide with the classical exponent~\eqref{eq:chaos-3} in the semiclassical limit.

However, we would like to emphasize two important points regarding OTOCs and quantum chaos. First, one should keep in mind that the argumentation of this subsection is quite naive and in fact the connection between the exponential growth of $C(t)$ and classical chaos is questionable. There is an evidence both in favor of this interpretation~\cite{Cotler-1704} and against it~\cite{Hashimoto, Xu}. For this reason notions of ``scrambling'' (exponential growth of OTOC) and ``chaos'' (exponential growth of the average distance between phase trajectories) should be distinguished, although they are often considered as the same.

Second, OTOCs are not the only possible measure of chaos; in fact, there were several attempts to extend the concept of classical chaos to quantum systems. The most notable alternative approach\footnote{In fact, this idea is old and well developed enough to be included in textbooks on chaos, e.g. see~\cite{Haake,Ott, Stockmann}.} to quantum chaos is based on the level statistics at small energy separation: if this statistics agrees with Random Matrix Theory, one can consider the system as chaotic~\cite{Gharibyan, Haake, Ott, Stockmann}. This approach is also closely related to the Eigenstate Thermalization Hypothesis~\cite{Deutsch, Srednicki, DAlessio}, which states that under some assumptions any local operator in an isolated quantum system eventually approaches its thermal form:
\beq V_{ij} = \langle E_i | V | E_j \rangle = \overline{V}(E) \delta_{ij} + e^{-\frac{1}{2} S(E)} f(E, \omega) R_{ij}, \eeq
where $| E_i \rangle$ is the state with the energy $E_i$, $S(E) = -\tr\left( \rho \log \rho\right)$, $\overline{V}(E) = \tr \left(\rho V\right)$, thermal density matrix $\rho$ is fixed by $E = \tr\left(\rho H\right)$, $f(E, \omega) = f(E, -\omega)$ is a smooth real function and $R_{ij}$ is a Hermitian random matrix with zero mean and unit variance. It is still not known whether this old approach is related to OTOCs or not, although there is some evidence in favor of this~\cite{Foini, Murthy, Parker, Avdoshkin, Huang-2}. In particular, it was shown that SYK model and 2D $CFT$ with large central charge under some assumptions behave like a random-matrix theory~\cite{Sonner, Vielma, Anous}, whereas correlation functions in these models have the form~\eqref{eq:chaos-10}.

\subsection{Fast scramblers}
\label{sec:scramblers}

The original motivation for studying of OTOCs was based on the fast scrambling conjecture, which was proposed in~\cite{Sekino,Susskind}, proved in~\cite{Lashkari} and adapted for correlators in~\cite{MSS}. In this subsection we briefly review this conjecture. Please note that this subsection may seem relatively vague if the reader does not have a specific example in mind. Such examples are discussed in the following sections.

First of all, consider a complex quantum system with a large number of degrees of freedom $N$, prepare a pure state $| \Psi \rangle$ and let this state freely evolve under the action of unitary operator $U$. Due to the Eigenstate Thermalization Hypothesis one expects that after a long enough time the system thermalizes although its state remains pure. By this we mean that density matrix of every small subsystem (with number of degrees of freedom $m < N/2$) is close to thermal density matrix, or, equivalently, the entanglement entropy\footnote{\label{foot:S} We remind that the entanglement entropy of subsystem $L$ is defined as $S_L = -\tr_L \big(\rho_L \log \rho_L\big)$, where the trace is taken over the Hilbert space of $L$, $\rho_L = \tr_R| \Psi \rangle \langle \Psi |$ and $R$ is the complement of $L$.} of every small subsystem is close to the maximal value~\cite{Page,Nishioka}. Roughly speaking, by this time the information about the initial state has been smeared throughout the system, so one needs to measure $\mathcal{O}(N)$ degrees of freedom to restore it. For this reason such a system was proposed to be called ``scrambled''~\cite{Sekino}.

Then let us perturb a small amount of degrees of freedom in a scrambled system and again let the system evolve freely. We expect that after some time the information about the perturbation is also smeared across all degrees of freedom, and system returns to a scrambled state. This time is referred to as ``scrambling time''.

The fast scrambling conjecture~\cite{Sekino,Susskind,Lashkari} states that scrambling time of any system cannot be less than $t_*^{min} \sim \beta \log N$. Moreover, the bound is saturated for black holes (if they satisfy all the explicit and implicit assumptions of the conjecture), which makes them ``the fastest scramblers in nature by a wide margin''\footnote{For finite-dimensional systems the bound can be tightened: $t_*^{min} \sim \beta N^{\frac{2}{d}}$, where $d$ is dimensionality of the system~\cite{Sekino}. For instance, in 3D $t_*^{min} \sim \beta N^{\frac{2}{3}}$. Thus, from this point of view black holes seem to be infinite-dimensional systems.}. Later it was argued that Rindler and de Sitter spaces also saturate this bound~\cite{Susskind}, but subsequent direct calculations did not confirm\footnote{The original argumentation of~\cite{Susskind} was based on the fact that the clock close to the event horizon goes as $e^{2 \pi t/\beta}$, where $t$ is the asymptotic observer's time. However, later it was shown that this is not enough. This is a good reminder that it is important to clarify all the assumptions in which the hypothesis is formulated.} this conjecture~\cite{Anninos, Aalsma}. This conjecture has important implications for information cloning and black hole information paradox~\cite{Hayden,Almheiri-1212,Mathur}.

To estimate scrambling time, one needs to find how quickly a small perturbation spreads over the entire system. In some special cases this process can be studied directly~\cite{Roberts-1802,Qi}, but much more often one needs to rely on implicit signs of scrambling. In essense, there are two such indicators.

One way to capture the rate of scrambling is to prepare a thermofield double (TFD) state, which describes two identical thermal subsystems:
\beq \label{eq:TFD}
| TFD \rangle = \frac{1}{\sqrt{Z}} \sum_n e^{-\frac{1}{2} \beta E_n} | n \rangle_L \otimes | n \rangle_R, \quad \text{so that} \quad \rho_L = \rho_R = \frac{1}{Z} \sum_n e^{-\beta E_n} | n \rangle \langle n |, \eeq
perturb one subsystem by a local operator and check how the mutual information, $I_{LR} = S_L + S_R - S_{L \cup R}$, evolves in time (see footnote~\ref{foot:S} for the definition of $S$). Usually subsystems are called ``left'' (L) and ``right'' (R) which explains the subscripts of $S$. Before the perturbation both subsystems are highly correlated, so the mutual information is non-zero. However, gradually the perturbation grows and affects more and more degrees of freedom. For instance, for a local operator $V$ and a generic Hamiltonian $H$ with local interactions, the $k$-th term in the expansion of the evolved operator $V(t) = e^{i H t} V e^{-i H t}$ can lead to a product of $k$ local operators:
\beq V(t) = V + i t [H, V] + \frac{(i t)^2}{2!} \left[ H, [H, V] \right] + \cdots + \frac{(i t)^k}{k!} \left[ H, \left[H, \cdots [H, V] \right] \right] + \cdots. \eeq 
Thus, one expects that eventually the perturbation spreads throughout the entire system, the entanglement between L and R subsystems disappears and mutual information becomes close to zero. Therefore, the moment $t_*$ at which $I_{LR}(t_*) \approx 0$ can be considered as an estimate of the scrambling time. An example of such calculation can be found e.g. in~\cite{Shenker-1306,Roberts-1412,Hartman,Asplund,Arefeva}. In particular, this calculation reproduces the conjectured bound $t_* \sim \beta \log N$ for black holes~\cite{Shenker-1306,Roberts-1412}.

Another way to evaluate $t_*$ is based on calculation of out-of-time-ordered correlators introduced in the previous subsection. Let us qualitatively explain why such correlators are sensitive to scrambling. As was noticed in~\cite{Almheiri-1304,Shenker-1306,Roberts-1412,MSS}, OTOC can be rewritten as a two-sided correlation function in a perturbed thermofield double state:
\beq \label{eq:scramblers-1}
\OTOC(t) = \left\langle V\Big(t - \frac{i\beta}{4}\Big) W\Big(0\Big) V\Big(t + \frac{i\beta}{4}\Big) W\Big(\frac{i\beta}{2}\Big) \right\rangle_\beta = \langle \psi | W_L W_R | \psi \rangle, \eeq
where $V$ and $W$ are local Hermitian operators, $W_L = W^\dagger \otimes 1$ acts on the left subsystem, $W_R = 1 \otimes W$ acts on the right subsystem and the perturbed state is as follows:
\beq \label{eq:scramblers-2}
| \psi \rangle = V_L\Big(t + \frac{i \beta}{4}\Big) | TFD \rangle = \frac{1}{\sqrt{Z}} \sum_{mn} e^{-\frac{\beta}{4} (E_m + E_n)} V(t)_{nm} | m \rangle_L \otimes | n \rangle_R. \eeq
At small times the operator $V$ affects only $\mathcal{O}(1)$ degrees of freedom and cannot significantly change the global pattern of correlations, so the perturbed state is close to pure $| TFD \rangle$. Thus, left and right subsystems are highly entangled and the correlator is big, i.e. $\OTOC(t) \approx \langle V V \rangle_\beta \langle W W \rangle_\beta$. However, over time the perturbation involves other degrees of freedom and destroys the fragile pattern of correlations, so eventually OTOC decays to zero. In this setting, scrambling time is the time at which OTOC saturates: $\OTOC(t_*) \approx 0$ or $C(t_*) \approx 2 \langle V V \rangle_\beta \langle W W \rangle_\beta$.

What is interesting here is the rate at which OTOC approaches zero. On general grounds, one expects that in the large $N$ limit and for small evolution times the first correction to OTOC is of the order of $\mathcal{O}\left(\frac{1}{N}\right)$:
\beq \frac{\OTOC(t)}{\langle V V \rangle_\beta \langle W W \rangle_\beta} = 1 - \frac{A}{N} f(t) + \mathcal{O}\left(\frac{1}{N^2}\right), \eeq
where $A$ is some positive $\mathcal{O}(1)$ numerical factor and $f(t)$ is some monotonously growing function. Extending this approximation to large times, one can qualitatively estimate the scrambling time as $t_* \sim f^{-1}\left(N/A\right)$, where $f^{-1}$ is the inverse of $f$, $f \circ f^{-1} = f^{-1} \circ f = 1$. At the same time, the fast scrambling conjecture states that $t_* \gtrsim \beta \log N$. Therefore, the function $f$ cannot grow faster than exponentially in time, $f(t) \lesssim e^{\kappa t}$. The exponent of this growth is also bounded, $\kappa \le \frac{B}{\beta}$, where $B$ is a universal positive $\mathcal{O}(1)$ numerical constant. This analog of the fast scrambling conjecture for OTOCs was proven in~\cite{MSS} and called a ``bound on chaos''\footnote{In fact, for gravitational scattering of massive particles with spin $J > 2$ one expects that $\kappa \sim \frac{2 \pi}{\beta} (J - 1)$. However, it was argued that such processes violate causality and unitarity~\cite{MSS, Zhiboedov}.}:
\beq \frac{d}{dt} \Big[ \langle V V \rangle_\beta \langle W W \rangle_\beta - \OTOC(t) \Big] \le \frac{2 \pi}{\beta} \Big[ \langle V V \rangle_\beta \langle W W \rangle_\beta - \OTOC(t) \Big], \quad \text{i.e.} \quad \kappa \le \frac{2 \pi}{\beta}. \eeq
Note that for systems that saturate the bound on $f$, the number $\kappa$ can be considered as an analog of classical Lyapunov exponent from subsection~\ref{sec:classical}.

Furthermore, OTOC is a very convenient measure of the spatial growth of operators. In $(d>1)$-dimensional chaotic systems (i.e. systems with $f(t) \sim e^{\kappa t}$) the exponential growth in time is typically supplemented~\cite{Roberts-1603} by a coordinate-dependent factor: $f(t) \sim e^{\kappa (t - |x|/v_B)}$, where $|x|$ is the distance to the initial perturbation caused by the operator $V$ and $v_B$ is some positive constant. It is easy to see that OTOC significantly deviates from the initial value only inside the ball of radius $r < v_B t$. This ball can be interpreted as an area affected by the perturbation, i.e. the ``size'' of the operator $V$. For this reason constant $v_B$ is called the ``butterfly velocity''. The discussion and examples of spatial operator growth can be found e.g. in~\cite{Roberts-1603,Roberts-1409,Shenker-1306,Hosur,Nahum,Mezei}.

Of course, compared to mutual information, OTOCs are a very crude measure of scrambling. In particular, $I_{LR}$ drops to zero almost immediately after $t_*$, whereas OTOCs at such times merely start to decay~\cite{Shenker-1306}. However, in practice it is much easier to calculate correlation functions than mutual information, which makes OTOCs a very popular tool. To the present moment OTOCs were calculated in a large variety of models, including BTZ black hole~\cite{Shenker-1306, Shenker-1312, Roberts-1409, Shenker-1412}, 2D $CFT$~\cite{Roberts-1412, Turiaci, Fitzpatrick}, de Sitter space~\cite{Anninos, Aalsma}, SYK model~\cite{Maldacena-SYK, Kitaev, Jevicki-1, Jevicki-2, Kitaev-talks, Polchinski} and its analogs~\cite{SUSY-SYK, Fu, Gross-1610, Gu}, 2D dilaton gravity~\cite{Maldacena-JT, Jensen}, matrix models~\cite{Stanford-1512, Gur-Ari}, and of course in plenty of quantum many-body systems~\cite{Aleiner, Wijn, Fine, Yao, Huang, Swingle, Shen, Dora, Lin, Bohrdt, Patel-1, Patel-2, vonKeyserlingk, Hosur, Nahum, Mezei}. In the following sections we will take a closer look at the two most notable examples: SYK model (sections~\ref{sec:basics} and~\ref{sec:treatment}) and 2D dilaton gravity (section~\ref{sec:JT}).

Finally, let us emphasize that arguments of~\cite{Sekino, Susskind, Lashkari, MSS} work only for nearly equilibrium situations (e.g. large, semiclassical black hole or eternal black hole in $AdS$ space), assuming that a small perturbation induced by operator $V$ cannot significantly change the initial state. Usually OTOCs are also calculated for such situations. Due to this assumption one can use equilibrium (Matsubara) diagrammatic technique and apply the analytic continuation procedure to correlation functions. However, this intuition does not work if the perturbation is big or the system is far from equilibrium (e.g. for small black holes). In this case one needs to use non-equilibrium (Schwinger--Keldysh) diagrammatic technique and take into account that the state of the system can evolve in time~\cite{Arseev, Kamenev}. An example of such calculation for black holes and de Sitter space can be found in~\cite{Maldacena-1912, Krotov, Akhmedov, Akhmedov-H, Akhmedov-1701, Akhmedov-1901}, a generalization of non-equilibrium technique for OTOCs can be found in~\cite{Aleiner, Haehl}. However, it is still unknown whether arguments of~\cite{Sekino, Susskind, Lashkari, MSS} can be extended to non-equilibrium systems or not.

\section{SYK basics}
\label{sec:basics}

SYK model is one of the most notable models for quantum chaos and holography. Due to its remarkable properties it is an excellent toy model for many physical phenomena, including traversable wormholes~\cite{Maldacena-1704, Maldacena-1804, Maldacena-1807, Maldacena-1912} and strange metals~\cite{Hartnoll, Song, Sachdev}. Due to this reason we review this model in great detail.

This section is mostly based on the pioneer papers~\cite{Polchinski, Maldacena-SYK, Kitaev} and talks of Kitaev~\cite{Kitaev-talks}. Reviews~\cite{Sarosi, Rosenhaus-1807} also come in handy. For simplicity we consider the model with four-fermion interaction vertex ($q = 4$), which is the simplest non-trivial and non-degenerate case. The generalization to other cases ($q\ge2$) is straightforward and can be found in the mentioned papers.

In this section we discuss the basic properties of SYK model: large $N$ diagrammatics, emergence of conformal symmetry in the IR limit, effective and Schwarzian action. The calculation of the four-point function is placed in the separate section (section~\ref{sec:treatment}) because of its bulkiness.

\subsection{Main definitions}
\label{sec:SYK-defs}

SYK model is quantum mechanics of $N \gg 1$ Majorana fermions with all-to-all random couplings:
\beq
\label{eq:SYK-action}
I_{SYK} = \int d\tau \left[ \frac{1}{2} \sum_{i=1}^N \chi_i(\tau) \dot{\chi}_i(\tau) - \frac{1}{4!} \sum_{i,j,k,l=1}^N J_{ijkl} \, \chi_i(\tau) \chi_j(\tau) \chi_k(\tau) \chi_l(\tau) \right],
\eeq
where $\dot{\chi}_i = d \chi_i/d\tau$. Let us clarify the notations. Letter~$\tau$ denotes Euclidean time, which is related to Lorentzian time~$t$ by the Wick rotation: $\tau = i t$. In this section we work in Euclidean time if not stated otherwise. Operators~$\chi_i$ are Hermitian: $\chi_i = \chi_i^\dagger$, and obey the usual anticommutation relations:
\beq
\label{eq:anticommutator}
\left\{ \chi_i, \chi_j \right\} = \delta_{ij}, \quad i, j = 1 \ldots N.
\eeq
One can find more information about representations of one-dimensional Clifford algebra in appendix~\ref{sec:majorana}. Note that in one-dimensional case Majorana fermions are dimensionless. The couplings $J_{ijkl}$ are distributed randomly and independently, i.e. accordingly to the gaussian distribution\footnote{A generalization to non-gaussian distributions can be found in~\cite{Krajewski}.} with the following probability density function:
\beq
\label{eq:PDF}
P(J_{ijkl}) = \exp\left( -\frac{N^3 J_{ijkl}^2}{12 J^2} \right) \quad \text{for every} \quad J_{ijkl}.
\eeq
We emphasize that the summation over $i$, $j$, $k$ and $l$ is not assumed. This distribution leads to several important properties. First, it fixes the average and average square of couplings:
\beq
\label{eq:J1}
\overline{J_{ijkl}} = 0, \quad \overline{J_{ijkl}^2} = \frac{3! J^2}{N^3},
\eeq
where $J$ is a constant with dimension of mass. Second, the even moments of couplings split into the sum of all possible products of the second moments (average squares), i.e. there is a Wick-type decomposition for average of even number of couplings. For instance,
\beq
\label{eq:J2}
\begin{aligned}
\overline{J_{i_1 i_2 i_3 i_4} J_{j_1 j_2 j_3 j_4} J_{k_1 k_2 k_3 k_4} J_{l_1 l_2 l_3 l_4}} &= \overline{J_{i_1 i_2 i_3 i_4} J_{j_1 j_2 j_3 j_4}} \; \overline{J_{k_1 k_2 k_3 k_4} J_{l_1 l_2 l_3 l_4}} + \overline{J_{i_1 i_2 i_3 i_4} J_{k_1 k_2 k_3 k_4}} \; \overline{J_{j_1 j_2 j_3 j_4} J_{l_1 l_2 l_3 l_4}} + \\ &+ \overline{J_{i_1 i_2 i_3 i_4} J_{l_1 l_2 l_3 l_4}} \overline{J_{j_1 j_2 j_3 j_4} J_{k_1 k_2 k_3 k_4}}.
\end{aligned} \eeq
Roughly speaking, to perform such an averaging one should create many copies of the system with randomly choosen couplings\footnote{In fact, if one is interested only in extensive quantities, such as energy or entropy, for large $N$ it is sufficient to consider only one specific realization with randomly distributed couplings. Roughly speaking, the large $N$ system can be divided into a large number of large subsystems that average themselves in such quantities.}, calculate the expressions in question and average the final results\footnote{One can also consider a generalization of the model with dynamical couplings. In particular, large $N$ fermionic tensor models reproduce all main properties of SYK model without the trick with disorder average. For review see subsubsection~\ref{sec:tensor} and papers~\cite{Klebanov-1, Klebanov-2, Klebanov-3, Witten-1610, Gurau-1611, Nishinaka}.}. The reasons why one requires properties~\eqref{eq:J1} and~\eqref{eq:J2} will become clear in the next subsection.

Note that anticommutation relations~\eqref{eq:anticommutator} imply the antisimmetry of the couplings:
\beq J_{ijkl} = \sgn \sigma J_{\sigma(i) \sigma(j) \sigma(k) \sigma(l)}, \quad \text{where} \quad \sigma: i \rightarrow \sigma(i), \quad i =1 \ldots N. \eeq
First, this reduces the number of independent non-zero components of $J_{ijkl}$ to $\frac{N!}{4!(N-4)!}$. Second, this allows one to define the disorder average of two arbitrary couplings:
\beq
\label{eq:J3}
\overline{J_{i_1 i_2 i_3 i_4} J_{j_1 j_2 j_3 j_4}} = \frac{3! J^2}{N^3} \sum_{\sigma} \sgn \sigma \delta_{i_1 \sigma(j_1)} \delta_{i_2 \sigma(j_2)} \delta_{i_3 \sigma(j_3)} \delta_{i_4 \sigma(j_4)},
\eeq
where the sum is performed over all possible permutations of indices. Essentially, this sum just checks whether indices of $J_{i_1 i_2 i_3 i_4}$ and $J_{j_1 j_2 j_3 j_4}$ coincide or not.

The important particular case in applications below is the case of three coincident indices:
\beq
\label{eq:J4}
\sum_{k,l,m = 1}^N \overline{J_{iklm} J_{jklm}} = \frac{3! J^2}{N^3} \sum_{k,l,m=1}^N \delta_{ij} \delta_{kk} \delta_{ll} \delta_{mm} + \cdots = \frac{3! J^2}{N^3} \left( N^3 \delta_{ij} + \mathcal{O}(N^2) \right) = 3! J^2 \delta_{ij} + \mathcal{O}\left(\frac{1}{N}\right).
\eeq
Let us also specify the interval where the Euclidean time $\tau$ runs. In this paper we consider two closely related cases: Euclidean line $\tau_{line} \in (-\infty, \infty)$ and Euclidean circle: $\tau_{circle} \in \left[-\frac{\beta}{2}, \frac{\beta}{2}\right)$, $\tau + \beta \sim \tau$. The first case describes the zero-temperature quantum mechanics, whereas the second case corresponds to the thermal state with the inverse temperature $\beta = \frac{1}{T}$. Below we will use the following map to change between the Euclidean line and circle:
\beq
\label{eq:circle-line}
\tau_{line} = \tan \frac{\pi \tau_{circle}}{\beta}. \eeq
Note that this mapping function is real and monotonic, i.e. it preserves the order of times: $d\tau_{line}/d\tau_{circle} > 0$.

Finally, note that in the free theory the Hamiltonian is zero, $H_0(\tau) = 0$. Hence, operators are constant even in Heisenberg picture: $\chi_i(\tau) = e^{\tau H_0} \chi_i(0) e^{-\tau H_0} = \chi_i(0)$. Therefore, one can use anticommutation relations~\eqref{eq:anticommutator} to find the two-point correlation functions in the zero-temperature free theory:
\beq
\label{eq:2p-free-0}
\langle 0 | \mathcal{T} \chi_i(\tau) \chi_j(0) | 0 \rangle \equiv  \theta(\tau) \langle 0 | \chi_i \chi_j | 0 \rangle - \theta(-\tau) \langle 0 | \chi_j \chi_i | 0 \rangle = \frac{1}{2} \sgn \tau \delta_{ij},
\eeq
and finite-temperature free theory:
\beq \langle \mathcal{T} \chi_i(\tau) \chi_j(0) \rangle_\beta = \frac{1}{2} \sgn \left( \sin \frac{\pi \tau}{\beta} \right) \delta_{ij}. \eeq
Here $| 0 \rangle$ denotes the vacuum state in the free theory and  $\langle \cdots \rangle_\beta$ denotes the averaging over the thermal distribution together with the quantum averaging:
\beq \langle \cdots \rangle_\beta \equiv \frac{\tr \left[ e^{-\beta H} \cdots \right]}{\tr \left[ e^{-\beta H} \right]}. \eeq
A more accurate derivation of the propagators can be found in appendix~\ref{sec:majorana}.

Note that thermal fermion propagator is antiperiodic due to anticommutation rule~\eqref{eq:anticommutator}. For instance, for $\tau > 0$:
\beq \tr \left[ e^{-\beta H} \chi(\tau + \beta) \chi(0) \right] = \tr \left[ \chi(\tau) e^{-\beta H} \chi(0) \right] = \tr \left[ e^{-\beta H} \chi(0) \chi(\tau) \right] = - \tr \left[ e^{-\beta H} \chi(\tau) \chi(0) \right], \eeq
Finally, it is convenient to define the averaged correlation functions:
\begin{align}
G_0(\tau) &\equiv \frac{1}{N} \sum_{i=1}^N \langle \mathcal{T} \chi_i(\tau) \chi_i(0) \rangle = \frac{1}{2} \sgn \tau, \label{eq:bare-0} \\
G_0^\beta(\tau) &\equiv \frac{1}{N} \sum_{i=1}^N \langle \mathcal{T} \chi_i(\tau) \chi_i(0) \rangle_\beta = \frac{1}{2} \sgn \left( \sin \frac{\pi \tau}{\beta} \right). \label{eq:bare-t}
\end{align}
Note that for $\tau \in \left[ -\frac{\beta}{2}, \frac{\beta}{2} \right)$ the finite-temperature propagator~\eqref{eq:bare-t} coincides with the zero-temperature propagator~\eqref{eq:bare-0}. Also note that any fermion Green function is antisymmetric: $G(\tau) = - G(-\tau)$.

\subsection{Two-point function and diagrammatics}
\label{sec:SYK-diagrams}

Let us turn on the interaction term:
\beq H(\tau) = \frac{1}{4!} \sum_{i,j,k,l} J_{ijkl} \chi_i(\tau) \chi_j(\tau) \chi_k(\tau) \chi_l(\tau), \eeq
and calculate averaged over disorder loop corrections to the free propagators. For greater clarity, we turn back to the Lorentzian time for a while, expand the evolution operators and calculate few first orders in $J$. The evolution operator is given by the following expression:
\beq U(t_1, t_2) \equiv \mathcal{T}\text{exp} \left[-i \int_{t_2}^{t_1} dt H(t)\right] = 1 - i \int_{t_2}^{t_1} dt H(t) - \int_{t_2}^{t_1} dt \int_{t_2}^t dt' H(t) H(t') + \cdots. \eeq
The exact propagator $G(t)$ can be transformed to the following form:
\beq G(t) \delta_{ij} = \left\langle \mathcal{T} U^\dagger(t, -\infty) \chi_i(t) U(t, 0) \chi_j(0) U(0, -\infty) \right\rangle = \frac{\langle \mathcal{T} \chi_i(t) \chi_j(0) U(+\infty, -\infty) \rangle}{\langle U(+\infty, -
\infty) \rangle}, \eeq 
Here we use the unitarity of $U(t_1, t_2)$ and suppose that the vacuum state is not disturbed under adiabatic turning on and switching off the interaction term~\cite{Peskin,Akhmedov}. Note that we do not need to use the interaction picture since $H_0 = 0$. Now let us expand this expression and average it over the disorder:
\beq 
\label{eq:2p-1}
\begin{aligned}
G(t) \delta_{ab} &= \Big\langle \mathcal{T} \Big[ \chi_a(t) \chi_b(0) - \frac{i}{4!} \sum_{i,j,k,l} \overline{J_{ijkl}} \int_{-\infty}^{+\infty} dt' \chi_a(t) \chi_b(0) \chi'_i \chi'_j \chi'_k \chi'_l - \\ &- \frac{1}{2} \frac{1}{(4!)^2} \sum_{i,j,k,l,p,q,r,s}\overline{J_{ijkl} J_{pqrs}} \int_{-\infty}^{+\infty} dt' \int_{-\infty}^{+\infty} dt'' \chi_a(t) \chi_b(0) \chi'_i \chi'_j \chi'_k \chi'_l \chi''_p \chi''_q \chi''_r \chi''_s + \mathcal{O}(J^3) \Big] \Big\rangle,
\end{aligned} \eeq
where we denoted $\chi_i(t') \equiv \chi'_i$ and $\chi_i(t'') \equiv \chi''_i$ for short. We also used that in the large $N$ limit the averaging over the disorder in the numerator and denonimator in~\eqref{eq:2p-1} can be done independently.

\begin{figure}[t]
\begin{center}
\begin{minipage}[t]{0.45\linewidth}
\includegraphics[width=1\linewidth]{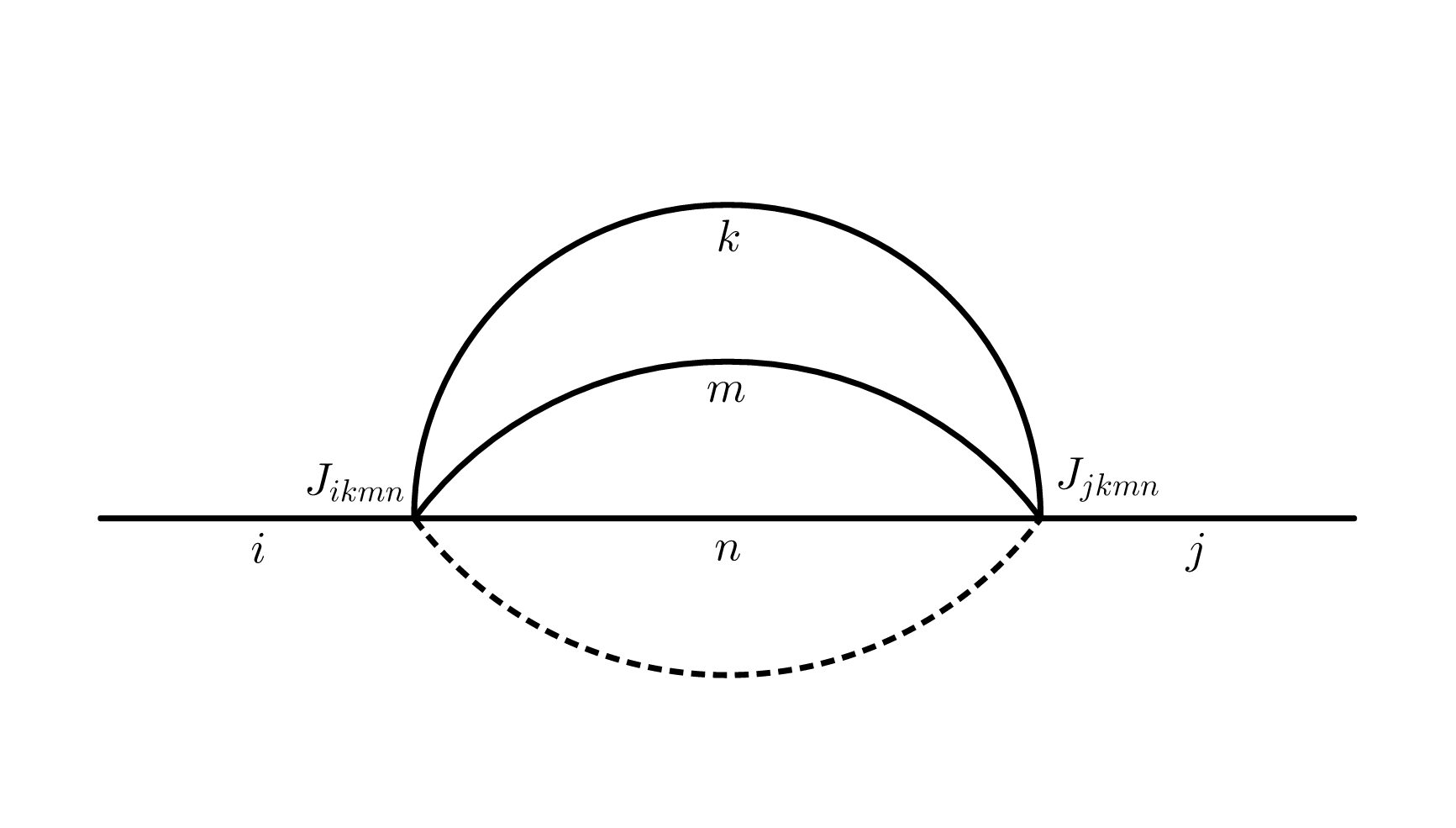}
\caption{Melonic diagram} 
\label{fig:melonic-1}
\end{minipage}
\hfill 
\begin{minipage}[t]{0.45\linewidth}
\includegraphics[width=1\linewidth]{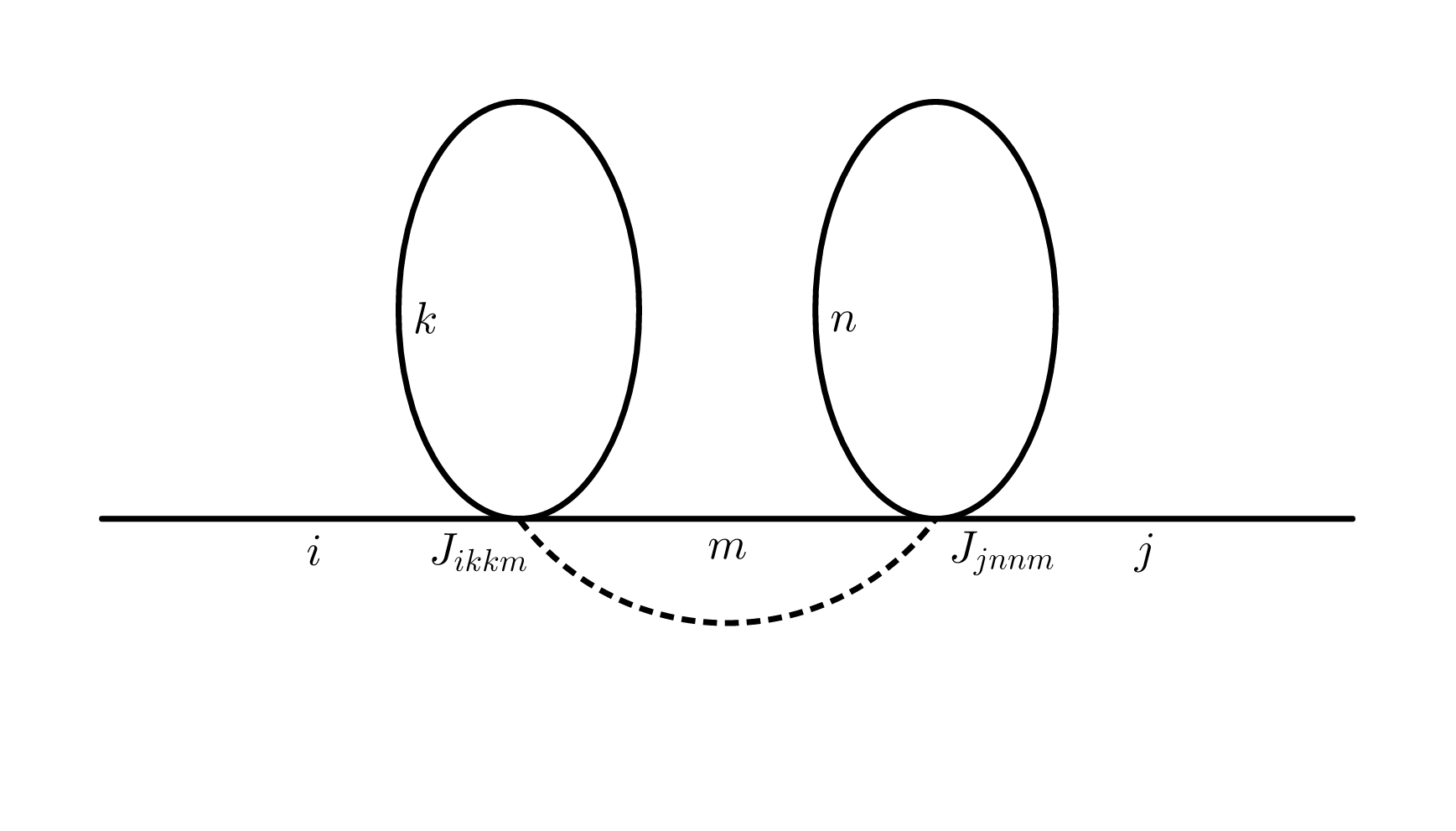}
\caption{Double tadpole diagram that is identically zero}
\label{fig:melonic-2}
\end{minipage}
\end{center}
\end{figure}

Now one sees that the rules~\eqref{eq:J1},~\eqref{eq:J2} and~\eqref{eq:J3} single out the very special type of vacuum expectation values. First, the disconnected part of the averages factorizes as usual. Second, odd orders in $J_{ijkl}$ die out after the disorder averaging. Third, the connected part of the expression~\eqref{eq:2p-1} reduces to the following expression:
\beq \label{eq:2p-3} \begin{aligned}
G(t) - G_0(t) &= \frac{2 \cdot 4 \cdot 4!}{2 (4!)^2} \frac{1}{N} \sum_{i,j,k,m,n}\overline{J_{ikmn} J_{jkmn}} \delta_{ij} \int dt' dt'' G_0(t-t') G_0(t'-t'')^3 G_0(t'') + \mathcal{O}(J^4) = \\ &= J^2 \int dt' dt'' G_0(t-t') G(t'-t'')^3 G(t'') + \mathcal{O}\left(\frac{J^2}{N}\right) + \mathcal{O}(J^4).
\end{aligned} \eeq
Here we have applied the Wick's theorem for the vacuum expectation values, contracted couplings with Kronecker deltas which come from the free propagators~\eqref{eq:2p-free-0}, used antisimmetry of $J_{ijkl}$ to find the numerical coefficient\footnote{All possible contractions give~$4\cdot4\cdot3\cdot2$ and the symmetry under the change~$t'\leftrightarrow t''$ gives~$2$.} and used the relation~\eqref{eq:J4} to single out the leading order in~$N$. The expression~\eqref{eq:2p-3} can be schematically represented by the so-called melonic diagram (Fig.~\ref{fig:melonic-1}). The other second-order diagram (Fig.~\ref{fig:melonic-2}) identically equals zero, because it contains couplings with coincident indices.

\begin{figure}[t]
\center{\includegraphics[scale=0.4]{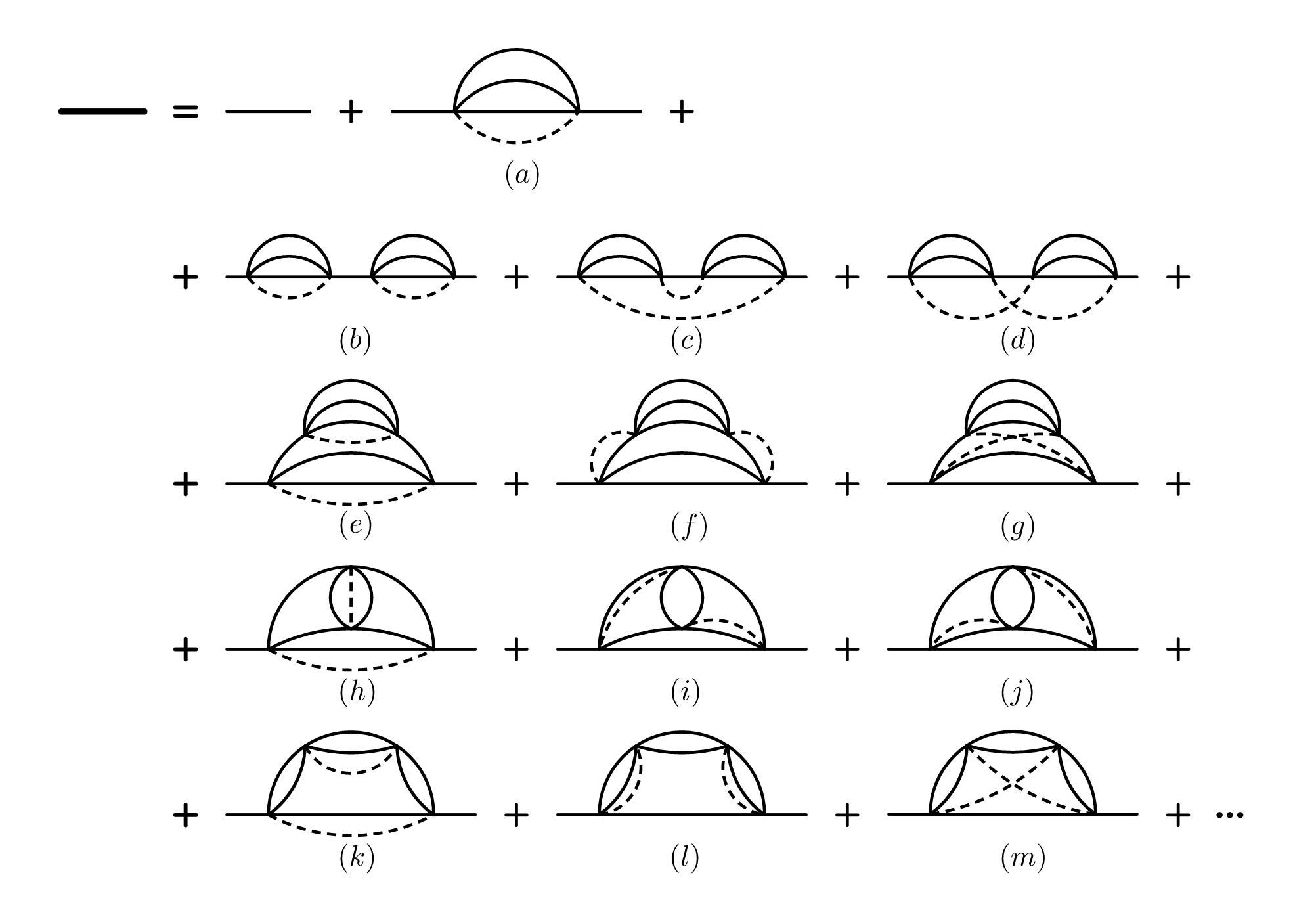}}\caption{Second-order (a) and fourth-order (b--m) corrections to the propagator. The only diagrams that survive in the limit $N \rightarrow \infty$ are (a), (b) and (e).} \label{fig:melonic-3}
\end{figure}

Using Wick's theorem and the relation~\eqref{eq:J3} one can write down higher order corrections, which correspond to higher-order diagrams (Fig.~\ref{fig:melonic-3}). Each diagram is proportional to the certain power of~$J$ and~$N$. The power of~$J$ is simply equals to the number of vertices of the diagram (each vertex gives~$J$). The power of~$N$ has no simple connection with the shape of the diagram. However, it is easy to see that the only diagrams which survive in the limit $N \rightarrow \infty$ are melonic diagrams, because the expression~\eqref{eq:J4} is the only one of the order~$J^2 N^0$. Roughly speaking, Kronecker deltas in~\eqref{eq:J4} are contracted directly, whereas Kronecker deltas in other averages are contracted through the other deltas. The longer the ``path'' of contraction of the indexes via Kronecker symbols, the lower is the power of~$N$.

For instance, compare the double melon (Fig.~\ref{fig:melonic-3}b or Fig.~\ref{fig:melonic-3}e) with non-melonic diagram (e.g. Fig.~\ref{fig:melonic-3}h). Double melonic diagram contains the following disorder average:
\beq \frac{J^4}{N^6} \sum \overline{J_{iklm} J_{nklm}} \; \overline{J_{npqr} J_{jpqr}} \propto \frac{J^4}{N^6} \sum \delta_{in} \delta_{kk} \delta_{ll} \delta_{mm} \delta_{jn} \delta_{pp} \delta_{qq} \delta_{rr} + \cdots = J^4 + \mathcal{O} \left(\frac{J^4}{N}\right). \eeq
Obviously, the contraction of six Kronecker deltas of the form $\delta_{nn}$ gives~$N^6$, so that the overall order of the diagram is~$J^4 N^0$. At the same time, the diagram depicted on the Fig.~\ref{fig:melonic-3}h contains a slightly modified average:
\beq \frac{J^4}{N^6} \sum \overline{J_{iklm} J_{jqrm}} \; \overline{J_{krnp} J_{qlnp}} \propto \frac{J^4}{N^6} \sum \delta_{ij} \delta_{kq} \delta_{lr} \delta_{mm} \delta_{kq} \delta_{lr} \delta_{nn} \delta_{pp} + \cdots = \mathcal{O} \left(\frac{J^4}{N}\right). \eeq
Here the power~$N^5$ comes from the contraction of $\delta_{mm}$, $\delta_{nn}$, $\delta_{pp}$, $\delta_{kq} \delta_{kq}$ and $\delta_{lr} \delta_{lr}$. One can see that two ``paths'' of the contraction lengthened and one ``path'' shortened, which reduced the power of~$N$ by one. The other possible products of the Kronecker deltas, which follow from~\eqref{eq:J3}, give even longer ``paths'' of the contraction.

Thus, the only type of diagrams which survive in the limit $N \rightarrow \infty$ are melonic diagrams (Fig.~\ref{fig:melonic-3}a,~\ref{fig:melonic-3}b and~\ref{fig:melonic-3}e). Moreover, one needs not to care about the signs and numerical coefficients in front of such diagrams, because all melons come with the same numerical coefficient. In fact, the correction (Fig.~\ref{fig:melonic-1}) can be thought of as a single block that can be inserted into any tree-level line of itself.

Recently the dominance of melonic diagrams also was rigorously proved, e.g. based on combinatorial analysis~\cite{Bonzom} and generalizations of the model~\cite{Gurau-1702}. We will not discuss such proof.

Note that in this subsection we worked in the zero-temperature limit, $\beta = \infty$, i.e. calculated the vacuum expectation values. However, the obtained results can be easily generalized to the finite-temperature case, because the averaging over the disorder does not depend on the temperature and always singles out melonic diagrams. It does not matter whether Feynman or Matsubara technique is used, the prefactors of diagrams are the same.

\subsection{Dyson--Schwinger equation and IR limit}
\label{sec:DS}

Using the results of the previous section one can straightforwardly write down the Dyson--Schwinger (DS) equation in the limit~$N \rightarrow \infty$:
\beq
\label{eq:DS-1}
\begin{aligned}
G(\tau_1, \tau_2) &= G_0(\tau_1, \tau_2) + \int d\tau_3 d\tau_4 G_0(\tau_1, \tau_3) \Sigma(\tau_3, \tau_4) G(\tau_4, \tau_2), \\
\Sigma(\tau_1, \tau_2) &\equiv J^2 G(\tau_1, \tau_2)^3.
\end{aligned} \eeq
This equation sums up only the melonic diagrams, which dominate in the limit in question. Here we turned back to the Euclidean time and took into account that corrections to each propagator of the melon endlessly grow upwards (as in Fig.~\ref{fig:melonic-3}e) and to the right (as in Fig.~\ref{fig:melonic-3}b), i.e. corresponding tree-level propagators are replaced with the exact ones (Fig.~\ref{fig:Dyson}). This equation (with the appropriate limits of the integration over $\tau$) holds both for zero- and finite-temperature propagators. Due to the translational invariance the exact propagator depends on the time difference: $G(\tau_1, \tau_2) = G(\tau_1 - \tau_2)$, $\Sigma(\tau_1, \tau_2) = \Sigma(\tau_1 - \tau_2)$. Hence, we can make the Fourier transformation of the equation~\eqref{eq:DS-1}:
\begin{figure}[t]
\center{\includegraphics[scale=0.3]{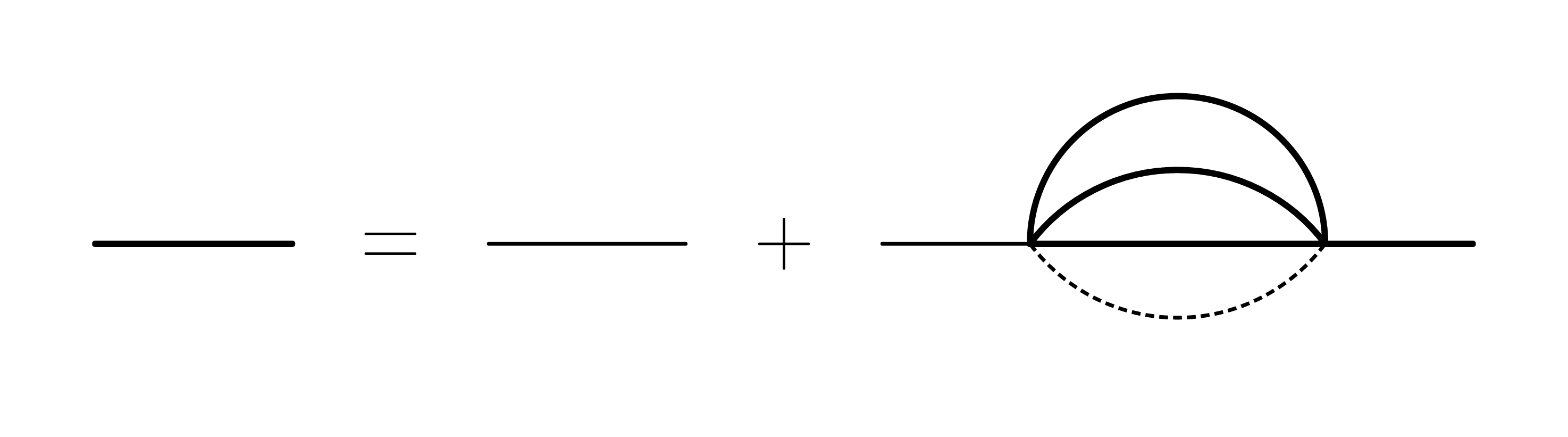}}\caption{Dyson--Schwinger equation which sums up melonic diagrams. Thin lines correspond to tree-level propagators, thick lines correspond to exact ones.} \label{fig:Dyson}
\end{figure}
\beq
\label{eq:DS-2}
G^{-1}(\omega) = -i \omega - \Sigma(\omega),
\eeq
where we used the explicit form of the tree-level propagator:
\beq G_0(\omega) \equiv \int_{-\infty}^\infty d\tau e^{i \omega \tau} \frac{1}{2} \sgn \tau = \frac{i}{\omega + i 0}, \quad \text{i.e.} \quad G_0^{-1}(\omega) = - i \omega. \eeq
The equation~\eqref{eq:DS-1} can be solved numerically. However, in the low frequency limit, $\omega \ll J$ (i.e. $J \tau \gg 1$), and strong coupling, $\beta J \gg 1$, one can also find its analitical solution. Let us first consider the zero-temperature case $\beta = \infty$. On dimensional grounds, we expect that in the limit under consideration the exact propagator decays as $G(\tau) \sim \tau^{-\frac{1}{2}}$. Hence, the left-hand side of the equation~\eqref{eq:DS-1} is negligible and the equation reduces to the following form (the result below shows that this assumption is correct):
\beq \label{eq:DS-3}
0 = G_0(\tau_1, \tau_2) + \int_{-\infty}^\infty d\tau_3 \int_{-\infty}^{\infty} d\tau_4 G_0(\tau_1, \tau_3) \Sigma(\tau_3, \tau_4) G(\tau_4, \tau_2), \eeq
hence,
\beq \label{eq:DS-9}
\int d\tau \Sigma(\tau_1, \tau) G(\tau, \tau_2) = -\delta(\tau_1 - \tau_2). \eeq
To obtain the second identity we have differentiated~\eqref{eq:DS-3} over $\tau_1$, used the relation $\pd_{\tau_1} G_0(\tau_1, \tau_2) = \delta(\tau_1 - \tau_2)$ and then took the integral over $\tau_3$. Obviously, the same equation arises when one dropes off the inverse tree-level propagagtor in~\eqref{eq:DS-2}:
\beq
\label{eq:DS-8}
G^{-1}(\omega) \approx -\Sigma(\omega), \quad \text{or} \quad \Sigma(\omega) G(\omega) \approx -1.
\eeq
This is just a Fourier transformation of the equation~\eqref{eq:DS-9}. Note that in the limit in question the DS equation~\eqref{eq:DS-3} is invariant under reparametrizations of time, $\tau \rightarrow f(\tau)$, $ f'(\tau) > 0$:
\beq
\label{eq:DS-4}
\begin{aligned}
G(\tau_1, \tau_2) &\rightarrow G\left[f(\tau_1), f(\tau_2)\right] f'(\tau_1)^\Delta f'(\tau_2)^\Delta, \\
\Sigma(\tau_1, \tau_2) &\rightarrow \Sigma\left[f(\tau_1), f(\tau_2)\right] f'(\tau_1)^{3\Delta} f'(\tau_2)^{3\Delta}, 
\end{aligned} \eeq
where $\Delta = \frac{1}{4}$. In fact, 
\beq
\label{eq:DS-5}
\int d f(\tau) \Sigma\left[f(\tau'), f(\tau)\right] G\left[f(\tau),f(\tau'')\right] = \frac{\int d\tau \Sigma(\tau', \tau) G(\tau, \tau'')}{f'(\tau')^{\frac{1}{4}} f'(\tau'')^{\frac{3}{4}}} = \frac{-\delta(\tau' - \tau'')}{f'(\tau')} = -\delta\left[f(\tau') - f(\tau'')\right].
\eeq
We emphasize that these reparametrizations should respect the orientation of the Euclidean circle: otherwise, the last equality in~\eqref{eq:DS-5} does not hold. 

Thus, we obtain that in the IR limit fermions acquire an anomalous conformal dimension\footnote{In general, in the model with $q$-fermion interaction term fermions acquire a conformal dimension $\Delta = \frac{1}{q}$.} $\Delta = \frac{1}{4}$, which hints at the following ansatz to solve the DS equation:
\beq G_c(\tau_1, \tau_2) = B \frac{\sgn \tau_{12}}{|J \tau_{12}|^{2\Delta}}, \eeq
where $\tau_{12} \equiv \tau_1 - \tau_2$ and $B$ is some numerical constant to be determined. The letter ``c'' stands for ``conformal''. Keeping in mind the following integral, which reduces to the gamma-function after the $\frac{\pi}{2}$ rotation in the complex plane:
\beq \int_{-\infty}^\infty d\tau e^{i \omega \tau} \frac{\sgn\tau}{|\tau|^{2D}} = 2 i \Gamma(1 - 2D) \cos(\pi D) |\omega|^{2D - 1} \sgn\omega, \eeq
we confirm that our ansatz does solve the equation~\eqref{eq:DS-3}, and find the numerical factor~$B$:
\beq
\label{eq:DS-6}
G_c(\tau) = \frac{1}{(4\pi)^{\frac{1}{4}}} \frac{\sgn \tau}{|J \tau|^{2\Delta}}. \eeq
Note that this solution decays as $J (\tau_1 - \tau_2) \rightarrow \infty$, which confirms the self-consistency of the approximation in which the equation~\eqref{eq:DS-3} was obtained. This solution was originally found by Sachdev and Ye in the system of randomly coupled spins~\cite{Sachdev-Ye}.

Finally, reparametrization invariance~\eqref{eq:DS-4} allows one to find the finite-temperature exact propagator without solving the corresponding DS equation~\cite{Parcollet}. In fact, zero- and finite-temperature propagators are connected by the map~\eqref{eq:circle-line}, which does satisfy the condition $f'(\tau) > 0$. Therefore, we can simply use this map in the expression~\eqref{eq:DS-6}:
\beq 
\label{eq:DS-7}
G_c^\beta(\tau) = \frac{\pi^{\frac{1}{4}}}{\sqrt{2 \beta J}} \frac{\sgn \left(\sin \frac{\pi \tau}{\beta} \right)}{|\sin \frac{\pi \tau}{\beta}|^{2\Delta}}, \quad \tau \in \left[-\frac{\beta}{2}, \frac{\beta}{2} \right). \eeq
Here we substituted the correct $\sgn$ function from subsection~\ref{sec:SYK-defs}. However, note that $\sgn \left( \sin\frac{\pi \tau}{\beta} \right) = \sgn \left( \tan\frac{\pi \tau}{\beta} \right) = \sgn(\tau)$ for $\tau \in \left[-\frac{\beta}{2}, \frac{\beta}{2}\right)$. Also note that in the limit $\tau \ll \beta$ expressions~\eqref{eq:DS-6} and~\eqref{eq:DS-7} coincide.

\begin{figure}[t]
\center{\includegraphics[scale=0.25]{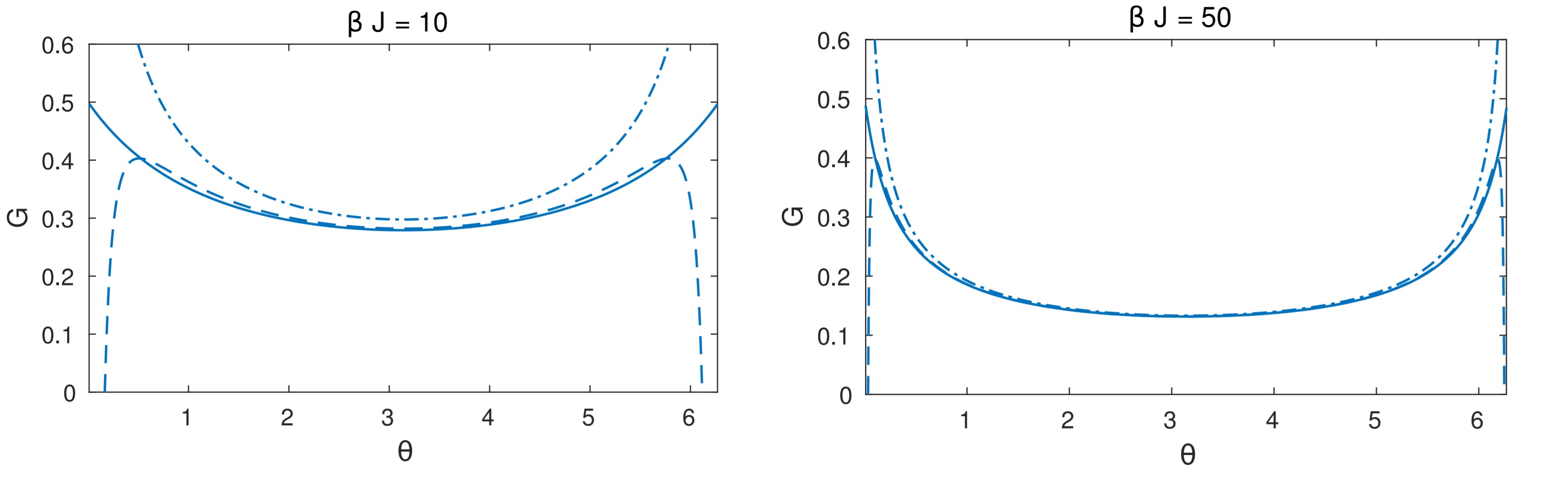}}\caption{A reprint of numerical solutions to the large $N$ Dyson--Schwinger equation~\eqref{eq:DS-1} obtained in~\cite{Maldacena-SYK} for $\beta J = 10$ and $\beta J = 50$. The exact solution is shown in solid lines, conformal approximation in dash-dotted lines and conformal approximation plus the first correction (which breaks the reparametrization invariance) in dashed lines. For convenience the variable $\theta = \frac{2 \pi \tau}{\beta}$ is introduced.} \label{fig:G-numerical}
\end{figure}

We remind that $G_c(\tau)$ and $G_c^\beta(\tau)$ are approximately equal to the exact propagators $G(\tau)$ and $G^\beta(\tau)$ only for relatively large times $\tau \gg 1/J$. At the same time, in the UV limit ($\tau \ll 1/J$) exact propagators are approximately equal to the bare ones, $G_0(\tau)$ and $G_0^\beta(\tau)$ correspondingly. In the intermediate region $G(\tau)$ and $G^\beta(\tau)$ interpolate between these functions (e.g. see Fig.~\ref{fig:G-numerical}).

After the analytic continuation of~\eqref{eq:DS-7} to the Lorentzian time $t = -i \tau$ one obtains the following two-point function\footnote{Note that in the Lorentzian signature one should specify the propagator (i.e. ordering of the operators into the correlation function) using $i \epsilon$ prescription~\cite{Maldacena-SYK, Peskin}. The analytical behavior of different propagators is different, but the overall exponential factor is unique.}:
\beq G_c^\beta(t) = \frac{(\pi)^{\frac{1}{4}}}{\sqrt{2 \beta J}} \frac{1}{|\sinh \frac{\pi t}{\beta}|^{2\Delta}} \propto e^{-\frac{2 \pi \Delta}{\beta} t}, \quad \text{as} \quad t \gg \frac{1}{J}. \eeq
This function becomes exponentially small after the time $t_d = \frac{\beta}{2 \pi \Delta} \sim \beta$, which is usualy called as dissipation time. This is quite an unusual behavior for a 1D system, but note that we consider large $N$ limit. In fact, it was shown in~\cite{Lunkin} that the exponential decay is replaced by the correct power-like one: $G_c^\beta(t) \sim (t/t_M)^{-3/2}$, for times larger than $t_M \sim N/J$. We will return to this expression when we discuss four-point funcitons (Sec.~\ref{sec:treatment}).

\subsection{Effective action}
\label{sec:effective}

In this subsection we derive the effective action and DS equations~\eqref{eq:DS-1} directly from the path integral. Here we assume the Gaussian distribution for coupling constants $J_{ijkl}$, which gives the following averaging rule:
\beq \overline{f(J_{ijkl})} \equiv \int \mathcal{D} J_{ijkl} f\left( J_{ijkl} \right), \quad \text{where} \quad \mathcal{D} J_{ijkl} \equiv \exp\left[ - \frac{N^3}{12 J^2} \sum_{i<j<k<l} J_{ijkl}^2 \right] \prod_{i<j<k<l} \sqrt{\frac{N^3}{3! J^2}} \frac{dJ_{ijkl}}{\sqrt{2\pi}}. \eeq
There are two physically distinct ways to realize the disorder average. First, one can average the partition functions itself, i.e. find $\overline{Z}$. Second, one can average the free energy using so-called replica trick:
\beq
\label{eq:effective-1}
\beta \overline{F} \equiv -\overline{\log Z} = -\lim_{M \rightarrow 0} \pd_M \overline{Z^M}.
\eeq
In this approach one introduces $M$ copies of the system ($\chi_i \rightarrow \chi_i^\alpha$, $i=1 \ldots N$, $\alpha = 1 \ldots M$), calculates the extended partition function $Z^M$, averages over the disorder, analitically continues to non-integer $M$ and take the formal limit~\eqref{eq:effective-1}. If one wants to find the free energy, entropy and other thermodinamic functions that are in some sense directly observable quantities, one should consider the second average.

However, in SYK model both methods of averaging give the same effective action~\cite{Sarosi,Jevicki-1,Sachdev}, because the replica-nondiagonal contributions to the replica action are suppressed by higher powers of $\frac{1}{N}$, and replica partition function simply splits into the product of $M$ naively-averaged partition functions: $\overline{Z^M} = \left(\overline{Z}\right)^M + \mathcal{O}\left(\frac{1}{N}\right)$. One can find the details on replica calculation in~\cite{Kitaev,Jevicki-1,Bagrets-1607, Arefeva-1811}. Thus, for simplicity we consider the disorder average of the partition function itself:
\beq \label{eq:effective-0}
\begin{aligned}
\overline{Z} &= \int \mathcal{D} J_{ijkl} \mathcal{D} \chi_i \exp \left[ \int d\tau \left( \frac{1}{2} \sum_{i=1}^N \chi_i \pd \chi_i - \frac{1}{4!} \sum_{i,j,k,l=1}^N J_{ijkl} \, \chi_i \chi_j \chi_k \chi_l \right) \right] = \\
& = \int \mathcal{D} \chi_i \exp\left[ \frac{1}{2} \sum_i \int d\tau \chi_i \pd \chi_i  + \frac{3! J^2}{2 N^3} \frac{1}{4!} \sum_{i,j,k,l} \left( \int d\tau \chi_i \chi_j \chi_k \chi_l \right)^2 \right] = \\ &= \int \mathcal{D} \chi_i \exp\Bigg[ \frac{1}{2} \sum_i \int d\tau \chi_i \pd \chi_i  + \frac{N J^2}{8} \int d\tau d\tau'\Bigg(\frac{1}{N} \sum_i \chi_i(\tau) \chi_i(\tau')\Bigg)^4 \Bigg] = \\ &= \int \mathcal{D} \chi_i \exp\left[ \int d\tau d\tau' \left( \frac{N}{2} G_0^{-1}(\tau, \tau') \Xi(\tau, \tau') + \frac{N J^2}{8} \Xi(\tau, \tau')^4 \right) \right].
\end{aligned} \eeq
Here we performed the gaussian integration over $J_{ijkl}$, reorganized the integrals over $d\tau$ and the sum over fermion indexes. For convenience we also introduced the inverse tree-level propagator $G_0^{-1}(\tau, \tau')$ and mean field variable $\Xi(\tau, \tau')$:
\beq \label{eq:effective-4}
G_0^{-1}(\tau, \tau') = \delta(\tau - \tau') \pd_{\tau}, \qquad \Xi(\tau, \tau') = \frac{1}{N} \sum_{i=1}^N \chi_i(\tau) \chi_i(\tau'). \eeq
Then we formally apply the following identity:
\beq f(\Xi) = \int dx f(x) \delta(x-\Xi) = \frac{N}{2\pi} \int dx dy f(x) e^{i N (x - \Xi) y}, \eeq
for the functional variables
\beq \label{eq:effective-v}
x = G(\tau, \tau'), \quad y = i \Sigma(\tau, \tau'),  \eeq
with the following normalization condition:
\beq \int \mathcal{D} G \mathcal{D} \Sigma \exp \left[ -\frac{N}{2} \int d\tau d\tau' \Sigma(\tau, \tau') G(\tau, \tau') \right] = 1, \eeq
to the function
\beq \begin{aligned}
\exp&\left( \frac{N J^2}{8} \int d\tau d\tau' \Xi(\tau, \tau')^4 \right) = \\ &= \int \mathcal{D} G \mathcal{D} \Sigma \exp \left\{ \frac{N}{2} \int d\tau d\tau' \left[ \frac{J^2}{4} G(\tau, \tau')^4 - \Sigma(\tau, \tau') \Big( G(\tau, \tau') - \Xi(\tau, \tau') \Big) \right] \right\}.
\end{aligned} \eeq
In this way we reorganize the nonlinear term $\Xi(\tau, \tau')^4$ in~\eqref{eq:effective-0}:
\beq \begin{aligned}
\overline{Z} &= \int \mathcal{D} G \mathcal{D} \Sigma \int \mathcal{D} \chi_i \exp \left\{ \frac{N}{2} \int d\tau d\tau' \left[ \left( G_0^{-1}(\tau, \tau') + \Sigma(\tau, \tau') \right) \Xi(\tau, \tau') + \frac{J^2}{4} G(\tau, \tau')^4 - \Sigma(\tau, \tau') G(\tau, \tau') \right] \right\} \\ &= \int \mathcal{D} G \mathcal{D} \Sigma \int \mathcal{D} \chi_i \exp\Bigg[ \frac{1}{2} \sum_i \int d\tau d\tau' \, \chi_i(\tau) \Big( \delta(\tau - \tau') \pd_\tau + \Sigma(\tau, \tau') \Big) \chi_i(\tau') + \\ &\phantom{= \int \mathcal{D} G \mathcal{D} \Sigma \int \mathcal{D} \chi_i \exp\Bigg[}+ \frac{N}{2} \int d\tau d\tau' \left( \frac{J^2}{4} G(\tau, \tau')^4 - \Sigma(\tau, \tau') G(\tau, \tau') \right) \Bigg].
\end{aligned} \eeq
In the last line we substituted the explicit form of the inverse tree-level propagator and mean field variable~\eqref{eq:effective-4}. Finally, after the integration over $\chi_i(\tau)$ we obtain the effective action:
\begin{align} 
\label{eq:effective-2}
\overline{Z} &= \int \mathcal{D} G \mathcal{D} \Sigma \, e^{-I_{eff}[G, \Sigma]}, \\
\label{eq:effective-3}
\frac{I_{eff}}{N} &= -\frac{1}{2} \log \det \Big( -\delta(\tau-\tau')\pd_\tau - \Sigma(\tau,\tau')\Big) + \frac{1}{2} \int d\tau d\tau' \left( \Sigma(\tau, \tau') G(\tau, \tau') - \frac{J^2}{4} G(\tau, \tau')^4 \right).
\end{align}
This effective action clearly reproduces the DS equation~\eqref{eq:DS-1} after the variations over $G$ and $\Sigma$. Indeed, variation wrt $G$ gives the expression for the self-energy, whereas variation wrt $\Sigma$ gives the equation itself\footnote{In the second line we used that $G_0^{-1}(\tau', \tau) = G_0^{-1}(\tau,\tau')$ and $\Sigma(\tau', \tau) = - \Sigma(\tau,\tau')$.}:
\beq \begin{aligned}
\delta_{\Sigma} I_{eff} &= -\frac{1}{2} \tr \log \left( 1 - \left(-\pd_\tau - \Sigma\right)^{-1} \delta \Sigma \right) + \frac{1}{2} \int d\tau d\tau' G(\tau, \tau') \delta \Sigma(\tau, \tau') = \\ &= \frac{1}{2} \int d\tau d\tau' \left[ G(\tau, \tau') - \left[G_0^{-1}(\tau, \tau') - \Sigma(\tau, \tau')\right]^{-1} \right] \delta \Sigma(\tau, \tau'), \quad \text{hence}, \quad G^{-1} = G_0^{-1} - \Sigma.
\end{aligned} \eeq
Practically, this means that we need not to rigorously explain the calculations that have been performed above, because the only important property which we require from the effective action is the correct DS equation. As soon as we find such an action, we entirely define the theory in the limit $N \rightarrow \infty$. In principle, we could just guess the action~\eqref{eq:effective-3} from the equation~\eqref{eq:DS-1}.

We emphasize that the solution of the DS equation~\eqref{eq:DS-1} is a true saddle point of the effective action~\eqref{eq:effective-3}, i.e. it is maximum on $G$ and minimum on $\Sigma$. This is due to the specific choice of the integration variable $y$, which is pure imaginary~\eqref{eq:effective-v}. Such a saddle point should be treated with caution. However, the numerical calculation shows that the solution of the DS equation does converge to this point~\cite{Maldacena-SYK, Jevicki-1, Jevicki-2, Arefeva-1811, Wang}.

Note that the functional integration over one-dimensional Majorana fermions is defined badly, because such fermions cannot be described by neither normal nor Grassmann numbers. In practice one should redefine Majorana fermions in terms of the ordinary Dirac fermions and reduce the integral~\eqref{eq:effective-0} to the integral over Grassmann variables. For the details on this calculation see appendix~\ref{sec:majorana-integral}.

Also note that the number $\frac{1}{N}$ plays the role of Planck's constant $\hbar$ in the functional integral~\eqref{eq:effective-2}, i.e. the limit $N \rightarrow \infty$ is equivalent to the classical limit $\hbar \rightarrow 0$. 

\begin{figure}[t]
\center{\includegraphics[scale=0.2]{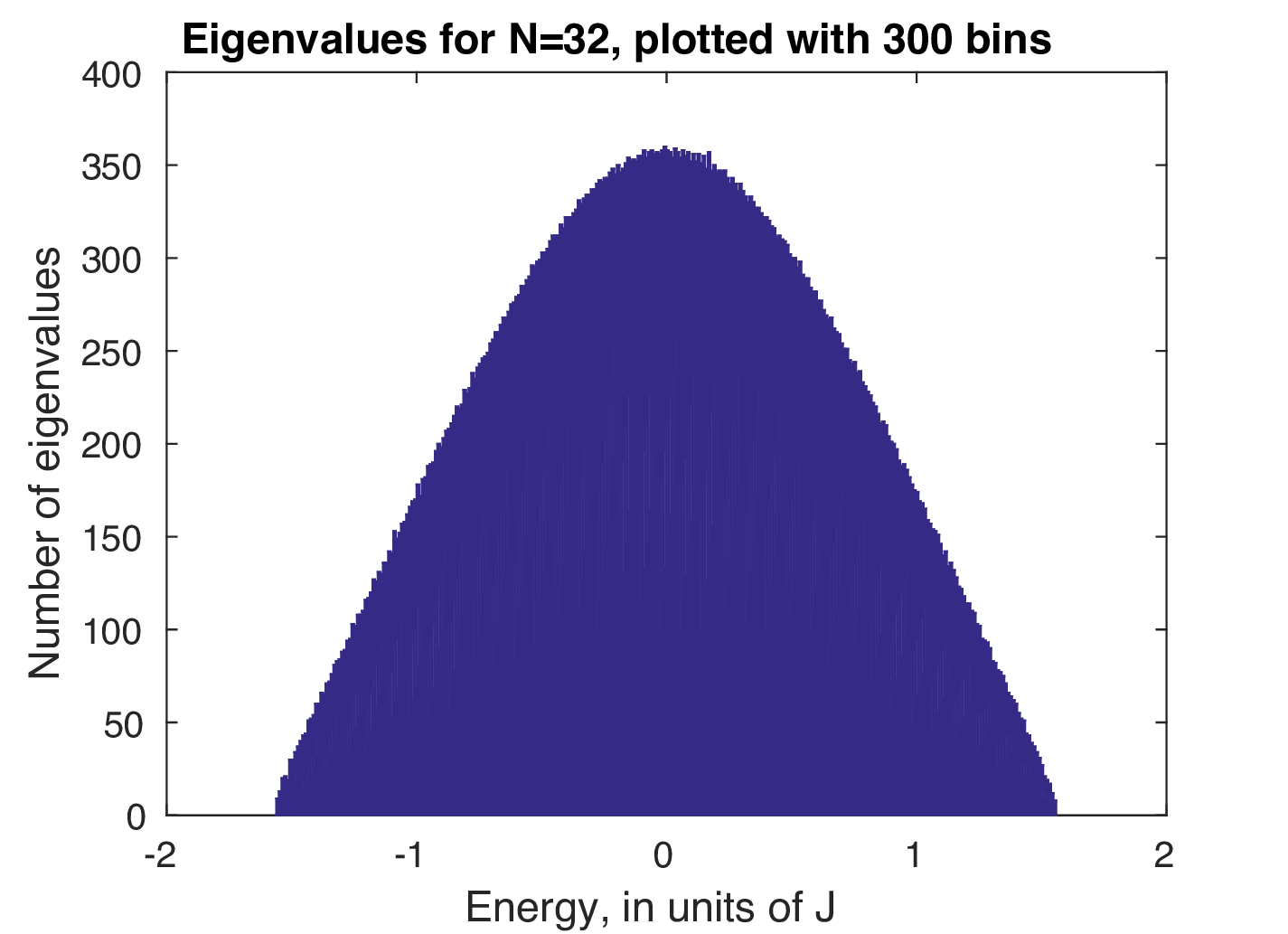}}\caption{A reprint of the energy spectrum numerically calculated in~\cite{Maldacena-SYK} for a single realization of the couplings in the model~\eqref{eq:SYK-action} with $N=32$ fermions.} \label{fig:DOS}
\end{figure}

Finally, the effective action~\eqref{eq:effective-3} allows one to calculate the entropy and free energy of the system, which determine its thermodynamic properties~\cite{Maldacena-SYK,Jevicki-2,Cotler}:
\beq \label{eq:effective-5}
\beta F = \beta E_0 + N \left[ -S_0 - \frac{2 \pi^2 C}{\beta J} + \mathcal{O}\left(\frac{1}{(\beta J)^2}\right) \right] + \frac{3}{2} \log(\beta J) + \text{const} + \mathcal{O}\left(\frac{1}{N}\right), \eeq
where $E_0$ is the ground state energy, $S_0 \approx 0.232$ is the low temperature entropy per site and $C$ is a numerical coefficient, the origin of which will be explained below. Note that the entropy of the system is large ($S \sim N$) even at low temperatures, which is not a common property. This is due to a specific form of the density of states, which resembles the random matrix semicircle and smoothly goes to zero at low energies (Fig.~\ref{fig:DOS}). In other words, even near the ground state the density of states is large ($\rho \sim e^{S_0 N}$) and energy gaps are small ($\sim e^{-S_0 N}$).

\subsection{Schwarzian action}
\label{sec:Schwarzian}

As we have seen in the subsection~\ref{sec:DS}, the presence of the inverse tree-level propagator in~\eqref{eq:DS-1} breaks the reparametrization invariance of DS equation~\eqref{eq:DS-4}. In this subsection we study this breaking more carefully. First let us make the change $\Sigma \rightarrow \Sigma - G_0^{-1}$ in the effective action~\eqref{eq:effective-3} and separate the conformally-invariant and non-invariant parts $I_{eff} = I_{CFT} + I_S$:
\begin{align}
\label{eq:Sch-CFT} \frac{I_{CFT}}{N} &= -\frac{1}{2} \log \det \Big(- \Sigma(\tau,\tau')\Big) + \frac{1}{2} \int d\tau d\tau' \left( \Sigma(\tau, \tau') G(\tau, \tau') - \frac{J^2}{4} G(\tau, \tau')^4 \right), \\
\label{eq:Sch-S} \frac{I_S}{N} &= -\frac{1}{2} \int d\tau d\tau'G_0^{-1}(\tau, \tau') G(\tau, \tau').
\end{align}
Now it is easy to see that conformal part~$I_{CFT}$ reproduces DS equation~\eqref{eq:DS-3} or~\eqref{eq:DS-8}, which is invariant wrt reparametrizations $\tau \rightarrow f(\tau)$, $f'(\tau) > 0$. Furthermore, delta-function in $G_0^{-1}(\tau, \tau')$ picks up small time differences $|\tau - \tau'| \ll J^{-1}$, therefore it can be neglected in the IR limit. Hence, conformal invariance emerges in the deep IR limit and disappears when one moves avay from it.

However, one cannot simply throw away the non-invariant part of the effective action, because it contains the essential information about the theory. In order to see this, let us consider fluctuations of the effective action~\eqref{eq:effective-3} near the saddle point $(\tilde{G}, \tilde{\Sigma})$. We emphasize that $\tilde{G} \ne G_c$; $G_c$ is only IR limit of $\tilde{G}$. It is convenient to parametrize the fluctuations\footnote{Note that the measure of the functional integration does not change if we choose fluctuations in this form. } in the form $G = \tilde{G} + \frac{\delta G}{|\tilde{G}|}$, $\Sigma = \tilde{\Sigma} + |\tilde{G}| \delta \Sigma$:
\beq  \begin{aligned}
\frac{I_{eff}}{N} &\approx \frac{1}{4} \int d\tau_1 d\tau_2 d\tau_3 d\tau_4 \, \delta\Sigma(\tau_1, \tau_2) \left(\big|\tilde{G}(\tau_1, \tau_2)\big| \tilde{G}(\tau_1, \tau_3) \tilde{G}(\tau_2, \tau_4) \big|\tilde{G}(\tau_3, \tau_4)\big| \right) \delta\Sigma(\tau_3, \tau_4) + \\ &+ \frac{1}{2} \int d\tau_1 d\tau_2 \left( \delta G(\tau_1, \tau_2) \delta \Sigma(\tau_1, \tau_2) - \frac{3 J^2}{2} \delta G(\tau_1, \tau_2)^2 \right) \equiv \\ &\equiv -\frac{1}{12 J^2} \langle \delta \Sigma | K | \delta \Sigma \rangle + \frac{1}{2} \langle \delta G | \delta \Sigma \rangle - \frac{3 J^2}{4} \langle \delta G | \delta G \rangle. 
\end{aligned} \eeq
Here $K$ is the operator that acts on the space of antisymmetric two-point functions (and generates ladder diagrams, as we will see in Sec.~\ref{sec:4p-CFT}). The integral kernel of this operator looks as follows:
\beq \label{eq:Sch-K} \begin{gathered}
K(\tau_1, \tau_2, \tau_3, \tau_4) \equiv -3 J^2 \big|\tilde{G}(\tau_1, \tau_2)\big| \tilde{G}(\tau_1, \tau_3) \tilde{G}(\tau_2, \tau_4) \big|\tilde{G}(\tau_3, \tau_4)\big|, \\
K | A \rangle = \int d\tau_3 d\tau_4 \, K(\tau_1, \tau_2, \tau_3, \tau_4) A(\tau_3, \tau_4).
\end{gathered} \eeq
It is straightforward to see that this kernel is antisymmetric under the changes $\tau_1 \leftrightarrow \tau_2$ and $\tau_3 \leftrightarrow \tau_4$ but symmetric under the change $(\tau_1, \tau_2) \leftrightarrow (\tau_3, \tau_4)$ (recall that $G(\tau_2, \tau_1) = - G(\tau_1, \tau_2)$). Also we introduce the identity operator~\cite{Kitaev, Kitaev-reps}:
\beq \label{eq:Sch-I} \begin{aligned}
I(\tau_1, \tau_2, \tau_3, \tau_4) &\equiv \frac{1}{2} \left[ \delta(\tau_1 - \tau_3) \delta(\tau_2 - \tau_4) - \delta(\tau_1 - \tau_4) \delta(\tau_2 - \tau_3) \right], \\  I | A \rangle &= | A \rangle,
\end{aligned} \eeq
and the inner product of two-point functions:
\beq \label{eq:inner} \langle A | B \rangle \equiv \int d\tau_1 d\tau_2 A^*(\tau_1, \tau_2) B(\tau_1, \tau_2). \eeq
We remind that $\Sigma$ is a Lagrange multiplier, i.e. it does not appear in physical quantities. Hence, we can just integrate out its fluctuations from the functional integral with the action~\eqref{eq:effective-3} to obtain in the semiclassical approximation:
\beq \label{eq:Sch-2} 
\frac{I_{eff}[\delta G]}{N} = -\log \int \mathcal{D} \delta \Sigma \, e^{- I_{eff}[\delta G,\delta \Sigma]} \simeq \frac{3 J^2}{4} \big\langle \delta G \big| (K^{-1} - I) \big| \delta G \big\rangle. \eeq
Let us check what happens with the action~\eqref{eq:Sch-2} in the conformal (IR) limit. Naively one thinks that non-invariant part of the action is negligible in this limit, i.e. action~\eqref{eq:effective-3} approximately equals~\eqref{eq:Sch-CFT}. This means that conformally invariant propagator replaces the exact saddle point, $\tilde{G} \approx G_c$. The fluctuations of the effective action in this limit are as follows:
\beq \label{eq:Sch-7}
\frac{I_{eff}[\delta G]}{N} \approx \frac{I_{CFT}[\delta G]}{N} \approx \frac{3 J^2}{4} \langle \delta G | K_c^{-1} - I | \delta G \rangle, \eeq
where the operator $K_c$ has the form~\eqref{eq:Sch-K} with the functions $G_c$ instead of $\tilde{G}$. Unfortunately, such a naively trancated effective action does not appropriately treat all fluctuations around the saddle point. Indeed, let us consider such fluctuations $\delta G$ that conserve the conformal symmetry~\eqref{eq:DS-4}. In this case $G = G_c + \frac{\delta G}{|G_c|}$ and $\Sigma = J^2 G_c^3 + 3 J^2 |G_c| \delta G$ solve the conformal Dyson--Schwinger equation~\eqref{eq:DS-9}:
\beq \int d\tau_4 \Big(\Sigma_c(\tau_3, \tau_4) + 3 J^2 |G_c(\tau_3, \tau_4)| \delta G(\tau_3, \tau_4) \Big) \left(G_c(\tau_4, \tau_2) + \frac{\delta G(\tau_4, \tau_2)}{|G_c(\tau_4, \tau_2)|} \right) = -\delta(\tau_3 - \tau_2), \eeq
Substracting the DS equation on the conformal functions $G_c$ and $\Sigma_c$, multiplying by $G_c(\tau_3, \tau_1)$ and integrating over $\tau_3$ we obtain the following identity:
\beq \int d\tau_3 d\tau_4 \left( \frac{\delta G(\tau_4, \tau_2)}{|G_c(\tau_4, \tau_2)|} \Sigma_c(\tau_3, \tau_4) G_c(\tau_3, \tau_1) + 3 J^2 G_c(\tau_1, \tau_3) G_c(\tau_2, \tau_4) |G_c(\tau_3, \tau_4)| \delta G(\tau_3, \tau_4) \right) = 0, \eeq
which straightforwardly reduces to
\beq \left( I - K_c \right) \delta G = 0. \eeq
Thus, on such fluctuations the conformally-invariant action~\eqref{eq:Sch-7} or~\eqref{eq:Sch-CFT} is zero, i.e. non-invariant part~\eqref{eq:Sch-S} cannot be omitted. Therefore, we have to move away from IR limit and estimate how the action~\eqref{eq:Sch-S} changes under the conformal transformations~\eqref{eq:DS-4}.

Let us first consider zero temperature case ($\beta = \infty$). As the first approximation, we expand the conformal propagator:
\beq
G_c(\tau_1, \tau_2) \rightarrow G_c\left[f(\tau_1), f(\tau_2)\right] \approx \frac{\sgn (\tau_1 - \tau_2)}{(4\pi)^{\frac{1}{4}} J^{2\Delta}} \frac{f'(\tau_1)^\Delta f'(\tau_2)^\Delta}{\left| f(\tau_1) - f(\tau_2)\right|^{2\Delta}}, \eeq
near $\tau = \frac{\tau_1 + \tau_2}{2}$ into the powers of $\tau_{12} = \tau_1 - \tau_2$:
\beq 
\label{eq:Sch-3}
G(\tau_1, \tau_2) = G_c(\tau_1, \tau_2) \left( 1 + \frac{\Delta}{6} \tau_{12}^2 \Sch\left[f(\tau), \tau\right] + \mathcal{O} (\tau_{12}^3) \right), \quad \text{where} \quad \Sch\left(f(\tau),\tau\right) \equiv \frac{f'''}{f'} - \frac{3}{2} \left(\frac{f''}{f'}\right)^2. \eeq
We do this expansion, because delta-function from $G_0^{-1}(\tau_1, \tau_2)$ in~\eqref{eq:Sch-S} picks up values around $\tau_{12} \approx 0$. We will use this property below. Then we substract the untransformed part from~\eqref{eq:Sch-3} and substitute the final result into the action~\eqref{eq:Sch-S} to obtain that:
\beq \label{eq:Sch-6} \begin{aligned}
\frac{I_S}{N} = -\frac{1}{2} \langle G_0^{-1} | \delta G \rangle &= -\frac{1}{2} \int d\tau d\tau_{12} \, G_0^{-1}(\tau_{12}) \tilde{G}(\tau_{12}) \left[ \frac{\Delta}{6} \tau_{12}^2 \Sch\left[f(\tau), \tau\right] + \mathcal{O} (\tau_{12}^3) \right] \approx \\ &\approx -\frac{\Delta}{12} \int d\tau_{12} \delta(\tau_{12}) \pd_{\tau_{12}} \left( \tau_{12}^2 \tilde{G}(\tau_{12}) \right) \int d\tau \, \Sch\left[ f(\tau), \tau) \right] = \\ &= -\frac{1}{J} \underbrace{\frac{\Delta}{12} \int d\eta \delta(\eta) \pd_\eta \left( \eta^2 \tilde{G}(\eta) \right)}_C \int d\tau \, \Sch\left[ f(\tau), \tau) \right],
\end{aligned} \eeq
where we have changed to the dimensionless variable $\eta = J \tau_{12}$. Now it is easy to see that the integral over $d\eta$ is undefined:
\beq \label{eq:Sch-C} C = \frac{\Delta}{12} \int d\eta \delta(\eta) \left[ (\eta^2 g(\eta))' \sgn\eta + \eta^2 g(\eta) \delta(\eta) \right] = \frac{\Delta}{12} \int d\eta g(\eta) \eta^2 \delta(\eta)^2 = \frac{\Delta}{12} \delta(0) \cdot g(0) \cdot 0^2 = 0 \cdot \infty, \eeq
where we singled out the relevant part of the saddle point value, $\tilde{G}(\eta) = g(\eta) \sgn\eta$.

There is no simple way to resolve this uncertainty, because we cannot analytically find the function $g(\eta)$ for all times. However, this problem can be solved by smearing the delta-function (i.e. by replacing the term $G_0^{-1}$ with the other suitable source which is big at small times, $\eta \ll 1$) and introducing gentle UV and IR cut-offs for the integral~\eqref{eq:Sch-C}. This was done in~\cite{Kitaev}.

The other way is to calculate the leading non-conformal corrections to the eigenfunctions and eigenvalues of the operator $K$, substitute them into the action~\eqref{eq:Sch-2} and directly evaluate $I_S = I_{eff} - I_{CFT} \approx \delta I_{CFT}$. This calculation was performed in~\cite{Maldacena-SYK,Jevicki-2}. Both these methods lead to the action of the form~\eqref{eq:Sch-6} with the coefficient $C \approx 0.48 \times \frac{\Delta}{12} > 0$. In summary, for the zero-temperature theory we obtain:
\beq
\label{eq:Sch-4}
\frac{I_S}{N} \approx -\frac{C}{J} \int_{-\infty}^\infty \Sch\left[f(\tau), \tau\right] d\tau. \eeq
As usual, one can change to the finite-temperature version of~\eqref{eq:Sch-4} using the map~\eqref{eq:circle-line}:
\beq
\label{eq:Sch-5}
 \frac{I_S}{N} = -\frac{C}{J} \int_{-\frac{\beta}{2}}^{\frac{\beta}{2}}\Sch\left[\tan\frac{\pi \varphi(\tau)}{\beta}, \tau \right] d\tau. \eeq
In this case the saddle point values of the effective action are parametrized by the function $\varphi(\tau)$, which maps the time circle to itself and preserves its orientation. Note that the coefficient $C$ is exactly the coefficient in the thermodynamic identity~\eqref{eq:effective-5}. This is because the low energy dynamics of SYK model is determined by the Schwarzian action.

Note that conformal invariance does not completely disappear when one moves away from IR limit. Indeed, exact propagators and the effective action must be invariant under the transformations from the $SL(2,\mathbb{R})$ group: these transformations are the rotations of the time circle (or time line in the limit $\beta \rightarrow \infty$) and do not correspond to any physical degrees of freedom. Both the action~\eqref{eq:Sch-2} and the Schwarzian action~\eqref{eq:Sch-5} are zero on the reparametrizations from $SL(2,\mathbb{R})$ group.

Thus, the apparent conformal symmetry of the IR theory is actually broken down to the symmetry wrt the transformations from the $SL(2,\mathbb{R})$ group. The dynamics of the pseudo-Goldstone boson which is associated to this broken symmetry (so-called ``soft mode'') is approximately described by the Schwarzian action~\eqref{eq:Sch-5}.

\section{SYK spectrum and four-point functions}
\label{sec:treatment}

This section has two main purposes. First, on a simple example we show how to calculate quantum corrections (which are suppressed by the powers of $\frac{1}{N}$) to many-point correlation functions. For this reason we keep as many details of the calculation as possible. Second, we show that OTOC exponentially saturates with time, with the main growing contribution being provided by the Schwarzian action. This is one of the most striking properties of SYK, as soon as this growth saturates the ``bound on chaos'' and coincides with the behavior of similar correlators calculated on black hole background (see subsection~\ref{sec:scramblers} and paper~\cite{MSS}). This section is mostly based on the pioneer papers~\cite{Maldacena-SYK,Kitaev,Polchinski}. A generalization to $n$-point functions with arbitrary $n$ can be found in~\cite{Gross-1710}. 

Let us consider the following four-point correlation function:
\beq \label{eq:4p-1} \begin{aligned}
&\frac{1}{N^2} \sum_{i,j=1}^N \left\langle \mathcal{T} \chi_i(\tau_1) \chi_i(\tau_2) \chi_j(\tau_3) \chi_j(\tau_4) \right\rangle = \\ &=\frac{1}{Z} \int \mathcal{D} G \mathcal{D} \Sigma \left[ G(\tau_1, \tau_2) G(\tau_3, \tau_4) + \frac{1}{N} \left( G(\tau_1, \tau_4) G(\tau_2, \tau_3) - G(\tau_1, \tau_3) G(\tau_2, \tau_4) \right) \right] e^{- I_{eff}[G, \Sigma]}, 
\end{aligned} \eeq
where we have used the approach of subsection~\ref{sec:effective} to transform from the functional integrals over $\mathcal{D} \chi_i$ on the LHS to those over $\mathcal{D} G$ and $\mathcal{D} \Sigma$ on the RHS. Letter $Z$ denotes the partition function~\eqref{eq:effective-2}. As usual, we work in the limit $J\tau \gg 1$, $N \gg 1$ and keep the leading quantum correction ($\sim \frac{1}{N}$) to the classical expression:
\beq \mathcal{F}(\tau_1, \tau_2, \tau_3, \tau_4) \equiv \frac{1}{N^2} \sum_{i,j=1}^N \left\langle \mathcal{T} \chi_i(\tau_1) \chi_i(\tau_2) \chi_j(\tau_3) \chi_j(\tau_4) \right\rangle - \tilde{G}(\tau_1, \tau_2) \tilde{G}(\tau_3, \tau_4), \eeq
where $\tilde{G}$ denotes the saddle point value of the effective action~\eqref{eq:effective-3}, which in the IR limit approximately equals the conformal propagator~\eqref{eq:DS-7}. For clarity we consider the theory at finite temperature, i.e. $\tau_{1,2,3,4} \in \left[-\frac{\beta}{2}, \frac{\beta}{2}\right)$.

Without loss of generality we restrict ourselves to the region $\tau_1 > \tau_2$, $\tau_3 > \tau_4$ and $\tau_1 > \tau_3$. First, function $\mathcal{F}(\tau_1, \tau_2, \tau_3, \tau_4)$ does not depend on the choise of the coordinates on the time circle, i.e. does not change under the cyclic permutation of its arguments. Second, this function is antisymmetric under the changes $\tau_1 \leftrightarrow \tau_2$ and $\tau_3 \leftrightarrow \tau_4$ and symmetric under the simultaneous change $(\tau_1, \tau_2) \leftrightarrow (\tau_3, \tau_4)$, which follows from the anticommutation relations of $\chi_i$'s. Together these two symmetries allow one to recover the behavior of this function in the regions with the other order of $\tau_{1,2,3,4}$.

As we have shown in the subsection~\ref{sec:Schwarzian}, it is convenient to separate conformally-invarinant and non-invariant fluctuations near the saddle point value $\tilde{G}$. We denote these fluctuations as $\delta G^\parallel$ and $\delta G^\perp$ correspondingly. Unlike the subsection~\ref{sec:Schwarzian} in this section we do not divide the fluctuations by $\tilde{G}$. I.e. the fluctuations $\delta G^\parallel$ are defined in such way that the function $G_c + \delta G^\parallel$ solves the conformal DS equation~\eqref{eq:DS-9}, and the subspace of non-invariant fluctuations $\delta G^\perp$ is the orthogonal complement to the subspace of conformally-invariant fluctuations. Note that due to the symmetry~\eqref{eq:DS-4} all conformal fluctuations can be parametrized by the function $\varphi(\tau)$, which maps the time circle into itself:
\beq \delta G_\varphi^\parallel(\tau_1, \tau_2) = G_c^\beta\left[\varphi(\tau_1), \varphi(\tau_2)\right] - G_c^\beta(\tau_1, \tau_2), \quad \text{for some reparametrisation} \quad \tau \rightarrow \varphi(\tau). \eeq
In these notations the functional integral for the four-point function looks as follows:
\beq \begin{aligned}
\mathcal{F} &\approx \mathcal{F}_0 + \frac{1}{Z} \int \mathcal{D} \delta G^\parallel \mathcal{D} \delta G^\perp \mathcal{D} \Sigma \left( \delta G^\parallel(\tau_1, \tau_2) + \delta G^\perp(\tau_1, \tau_2) \right) \left( \delta G^\parallel(\tau_3, \tau_4) + \delta G^\perp(\tau_3, \tau_4) \right) e^{-I_{CFT} - I_S} = \\
&= \mathcal{F}_0 + \mathcal{F}_S + \mathcal{F}_{CFT} + \mathcal{O}\left(\frac{1}{N^2}\right),
\end{aligned} \eeq
where we expanded the integrand near the saddle point and introduced the following expectation values:
\begin{align}
\label{eq:4p-4} \mathcal{F}_0 &\equiv  \frac{1}{N} \left( \tilde{G}(\tau_1, \tau_4) \tilde{G}(\tau_2, \tau_3) - \tilde{G}(\tau_1, \tau_3) \tilde{G}(\tau_2, \tau_4) \right), \\
\label{eq:4p-2} \mathcal{F}_S &\equiv \left\langle \delta G^\parallel(\tau_1, \tau_2) \delta G^\parallel(\tau_3, \tau_4) \right\rangle_S = \frac{\int \mathcal{D} \varphi  \, \delta G_\varphi^\parallel(\tau_1, \tau_2) \delta G_\varphi^\parallel(\tau_3, \tau_4) e^{-I_S[\varphi]}}{\int \mathcal{D} \varphi \, e^{-I_S[\varphi]}}, \\
\label{eq:4p-3} \mathcal{F}_{CFT} &\equiv \left\langle \delta G^\perp(\tau_1, \tau_2) \delta G^\perp(\tau_3, \tau_4) \right\rangle_{CFT} = \frac{\int \mathcal{D} \delta G^\perp \, \delta G^\perp(\tau_1, \tau_2) \delta G^\perp(\tau_3, \tau_4) e^{-I_{eff}[\delta G^\perp]}}{\int \mathcal{D} \delta G^\perp \, e^{-I_{eff}[\delta G^\perp]}}.
\end{align}
We will clarify the meaning of the notations in subsections~\ref{sec:4p-S} and~\ref{sec:4p-CFT}. To obtain the average~\eqref{eq:4p-2}, we use that the Jacobian
\beq J = \left[\frac{\mathcal{D} G_\varphi^\parallel}{\mathcal{D} \varphi}\right]_{\varphi(\tau) = \frac{2 \pi \tau}{\beta}} \eeq
is constant and non-zero, because for reparametrisations which are infinitesimally close to the identity, $\varphi(\tau) = \frac{2 \pi \tau}{\beta} + \delta \varphi(\tau)$, fluctuations $\delta G_\varphi^\parallel$ depend only on $\delta \varphi$ (see eq.~\eqref{eq:4p-S-1}). The integral $\int \mathcal{D} \delta G^\perp \mathcal{D} \Sigma \, e^{-I_{CFT}}$ in the numerator and denominator of~\eqref{eq:4p-2} is also constant and non-zero. For the average~\eqref{eq:4p-3} we repeated the argumentation around the formula~\eqref{eq:Sch-2} and used the action $I_{eff}$ evaluated on the conformal functions $\tilde{G} = G_c^\beta$. We remind that for the conformally-invariant fluctuations $\left(I - K \right) \delta G^\parallel = 0$, hence, $I_{eff}[\delta G^\perp + \delta G^\parallel] = I_{eff}[\delta G^\perp]$.

For convenience in this section we rescale the fields and map the finite-temperature time circle into the unit circle by the following transformation:
\beq \label{eq:4p-5}
\begin{gathered}
\tau \rightarrow \frac{2 \pi \tau}{\beta}, \quad \chi_i \rightarrow \left( \frac{\beta J}{2 \pi}\right)^\Delta \chi_i, \\
G(\tau, \tau') \rightarrow \left( \frac{\beta J}{2 \pi}\right)^{2 \Delta} G(\tau, \tau'), \quad \Sigma(\tau, \tau') \rightarrow  \frac{1}{J^2} \left( \frac{\beta J}{2 \pi}\right)^{6 \Delta} \Sigma(\tau, \tau').
\end{gathered} \eeq
In this case the Schwarzian and conformally-invariant actions acquire the following form:
\beq \begin{aligned}
\frac{I_{CFT}}{N} &= -\frac{1}{2} \log\det\Big(-\Sigma(\tau, \tau')\Big) + \frac{1}{2} \int_{-\pi}^\pi d\tau \int_{-\pi}^\pi d\tau' \left( \Sigma(\tau, \tau') G(\tau, \tau') - \frac{1}{4} G(\tau, \tau')^4 \right), \\
\frac{I_S}{N} &= -\frac{2 \pi C}{\beta J} \int_{-\pi}^\pi \Sch\left(\tan\frac{\varphi(\tau)}{2},\tau\right) d\tau.
\end{aligned} \eeq
Thus, all the prefactors and their dependence on $N$, $J$ and $\beta$ become explicit. Both contributions from the conformally-invariant and non-invariant parts are of the order $\mathcal{O}\left(\frac{1}{N}\right)$, because both actions $I_S$ and $I_{CFT}$ are proportional to $N$. However, in the case of strong coupling $\beta J \gg 1$ the leading contribution to the correlation function comes from the Schwarzian action due to the additional small factor. Roughly speaking, due to this small factor soft mode fluctuations are easiest to excite. We calculate this contribution in subsection~\ref{sec:4p-S} and compare it with the contribution from the conformal part in subsection~\ref{sec:4p-CFT}.

\subsection{Soft mode contribution}
\label{sec:4p-S}

In this subsection we review the argumentation of~\cite{Kitaev} to estimate the correlator~\eqref{eq:4p-2} in the limit $1 \ll \tau J < \beta J \ll N$. In this limit the fluctuations are small, so we use gaussian approximation for the functional integrals. Note that this limit does not hold in zero temperature case. In fact, we have to work in the limit of small but non-zero temperatures: $\frac{J}{N} \ll T \ll J$.

Consider conformally-invariant fluctuations of the saddle point value $\tilde{G} \approx G_c^\beta$. For the infinitesimal transformations $\delta \varphi(\tau) \equiv \varphi(\tau) - \tau$ the fluctuations look as follows:
\beq 
\label{eq:4p-S-1}
\begin{aligned} 
\delta G_\varphi^\parallel(\tau_1, \tau_2) &= G_c^\beta\left[\varphi(\tau_1), \varphi(\tau_2)\right] - G_c^\beta(\tau_1, \tau_2) = \\
&= \left[ \delta \varphi(\tau_1) \pd_{\tau_1} + \frac{1}{4} \delta \varphi'(\tau_1) + \delta \varphi(\tau_2) \pd_{\tau_2} + \frac{1}{4} \delta \varphi'(\tau_2) \right] G_c^\beta(\tau_1, \tau_2) = \\
&= \frac{1}{4} \left[ \delta \varphi'(\tau_1) + \delta \varphi'(\tau_2) - \frac{\delta \varphi(\tau_1) - \delta \varphi(\tau_2)}{\tan \frac{\tau_1 - \tau_2}{2}}\right] G_c^\beta(\tau_1, \tau_2).
\end{aligned} \eeq
To obtain the last line we have used the expression~\eqref{eq:DS-7}.

Let us expand the function $\delta \varphi$ in Fourier modes:
\beq \delta \varphi(\tau) = \sum_{m \in \mathbb{Z}} (\delta \varphi)_m e^{i m \tau}, \eeq
and rewrite the expression~\eqref{eq:4p-S-1} as:
\beq \frac{\delta G_\varphi^\parallel(\tau_1, \tau_2)}{G_c^\beta(\tau_1, \tau_2)} = -\frac{i}{2} \sum_{m \in \mathbb{Z}} e^{i m \frac{\tau_1 + \tau_2}{2}} \left[ \frac{\sin\left(\frac{m \tau_{12}}{2}\right)}{\tan \frac{\tau_{12}}{2}} - m \cos\left(\frac{m \tau_{12}}{2}\right) \right] (\delta \varphi)_m, \eeq
where $\tau_{12} \equiv \tau_1 - \tau_2$. Then we use the following integral:
\beq \int_{\tau_2}^{\tau_1} \frac{s_{10} s_{02}}{s_{12}} e^{i m \tau_0} \frac{d \tau_0}{2\pi} = \frac{2}{\pi} \frac{1}{m (m^2 - 1)} e^{i m \frac{\tau_1 + \tau_2}{2}} \left[ \frac{\sin\left(\frac{m \tau_{12}}{2}\right)}{\tan \frac{\tau_{12}}{2}} - m \cos\left(\frac{m \tau_{12}}{2}\right) \right], \eeq
which allows one to write:
\beq \label{eq:4p-S-3} 
\frac{\delta G_\varphi^\parallel(\tau_1, \tau_2)}{G_c^\beta(\tau_1, \tau_2)} =
\frac{- i \pi}{2} \sum_{m \in \mathbb{Z}} \int_{\tau_2}^{\tau_1}  \frac{s_{10} s_{02}}{s_{12}} m (m^2 - 1) (\delta \varphi)_m e^{i m \tau_0} \frac{d \tau_0}{2\pi}, \eeq
where we have denoted $s_{12} \equiv 2 \sin \frac{\tau_1 - \tau_2}{2}$ and assumed that $2\pi > \tau_1 - \tau_2 > 0$. Finally, we introduce the $SL(2, \mathbb{R})$-invariante observable:
\beq \label{eq:4p-S-2}
O(\tau) = \Sch\left(\tan\frac{\varphi(\tau)}{2},\tau\right) = \frac{1}{2} + \delta \varphi' + \delta \varphi''' + \frac{1}{2} (\delta \varphi')^2 - (\delta \varphi') (\delta \varphi''') - \frac{3}{2} (\delta \varphi'')^2 + \mathcal{O}(\delta \varphi^3), \eeq
do the Fourier transformation of the non-invariant part:
\beq \delta O(\tau) \equiv O(\tau) - \frac{1}{2} = -i \sum_{m \in \mathbb{Z}} m (m^2 - 1) (\delta \varphi)_m e^{i m \tau} + \mathcal{O}\left(\delta\varphi^2\right), \eeq
and compare this expression to the expression~\eqref{eq:4p-S-3}. As a result, we obtain the following integral representation for the variation of the variable $G$:
\beq \frac{\delta G_\varphi^\parallel(\tau_1, \tau_2)}{G_c^\beta(\tau_1, \tau_2)} = \frac{\pi}{2} \int_{\tau_2}^{\tau_1}  \frac{s_{10} s_{02}}{s_{12}} \delta O(\tau_0) \frac{d \tau_0}{2\pi}. \eeq
Using this representation we can rewrite the correlator~\eqref{eq:4p-2} as:
\beq \label{eq:4p-S-5} \begin{aligned}
\frac{\mathcal{F}_S(\tau_1, \tau_2, \tau_3, \tau_4)}{G_c^\beta(\tau_1, \tau_2) G_c^\beta(\tau_3, \tau_4)} &= \frac{\left\langle \delta G_\varphi^\parallel(\tau_1, \tau_2) \delta G_\varphi^\parallel(\tau_3, \tau_4) \right\rangle_S}{G_c^\beta(\tau_1, \tau_2) G_c^\beta(\tau_3, \tau_4)} = \\ &= \frac{\pi^2}{4} \int_{\tau_2}^{\tau_1} \frac{d\tau_5}{2\pi} \int_{\tau_4}^{\tau_3} \frac{d\tau_6}{2\pi} \langle \delta O(\tau_5) \delta O(\tau_6) \rangle_S \frac{s_{15} s_{52}}{s_{12}} \frac{s_{36} s_{64}}{s_{34}}.
\end{aligned} \eeq
Recall that we have restricted ourselves to the region $\tau_1 > \tau_2$, $\tau_3 > \tau_4$ and $\tau_1 > \tau_3$.

Let us estimate the correlation function of two $\delta O$'s in the gaussian approximation. Using the expansion~\eqref{eq:4p-S-2} we find the Schwarzian action~\eqref{eq:Sch-5} up to the boundary and $\mathcal{O}(\delta \varphi^3)$ terms:
\beq \frac{I_S}{N} = -\frac{2 \pi C}{\beta J} \int_{-\pi}^\pi \left[ \frac{1}{2} + \frac{(\delta \varphi')^2 - (\delta \varphi'')^2 }{2} \right] d\tau = -\frac{\pi C}{\beta J} + \frac{\pi C}{\beta J} \sum_{m \in \mathbb{Z}} m^2 (m^2 - 1) (\delta \varphi)_m (\delta \varphi)_{-m}. \eeq
Therefore, in the gaussian approximation the correlation function of two $\delta \varphi$'s looks as follows:
\beq \label{eq:4p-S-6}
\langle (\delta \varphi)_m (\delta \varphi)_n \rangle_S = \frac{1}{2 \pi C} \frac{\beta J}{N} \frac{\delta_{m, -n}}{m^2 (m^2 - 1)}, \quad \text{where} \quad m, n \ne -1, 0, 1. \eeq
Note that modes with $m = -1, 0, 1$ are $SL(2, \mathbb{R})$ generators, i.e. they correspond to the non-physical degrees of freedom and cancel out from all physical observables. These are zero modes of the Schwarzian action, which we have mentioned at the end of subsection~\ref{sec:Schwarzian}.

Using the identity~\eqref{eq:4p-S-6} we find the correlation function of two $\delta O$'s:
\beq \label{eq:4p-S-7} \begin{aligned}
\langle \delta O(\tau_5) \delta O(\tau_6) \rangle_S &= -\sum_{m,n \in \mathbb{Z}} m (m^2 - 1) n (n^2 - 1) \langle (\delta \varphi)_m (\delta \varphi)_n \rangle_S e^{i m \tau_5 + i n \tau_6} = \\ &= \frac{1}{2 \pi C} \frac{\beta J}{N} \sum_{m \ne 0} (m^2 - 1) e^{i m (\tau_5 - \tau_6)} = \frac{1}{2 \pi C} \frac{\beta J}{N} \Big[ 1 - 2 \pi \delta(\tau_{56}) - 2 \pi \delta''(\tau_{56}) \Big],
\end{aligned} \eeq
where we have used that $\delta(\tau) = \frac{1}{2\pi} \sum e^{i m \tau}$. Please note that delta-functions in~\eqref{eq:4p-S-7} are zero if the integration intervals over $d\tau_5$ and $d\tau_6$ do not overlap. Therefore, it is convenient to separately consider two different orderings:
\beq \begin{aligned}
\text{OPE:} \quad 2 \pi > \tau_1 > \tau_2 > \tau_3 > \tau_4 > 0, \\
\text{OTO:} \quad 2 \pi > \tau_1 > \tau_3 > \tau_2 > \tau_4 > 0.
\end{aligned} \eeq
The abbreviation OPE stands for ``operator product expansion'', which is applicable for the corresponding time ordering (see~\cite{Maldacena-SYK,Sarosi,Gross} and subsubsection~\eqref{sec:OPE} for the details). The abbreviation OTO stands for ``out of time ordered'' for obvious reasons.

For the OPE ordering the integrals over $d\tau_5$ and $d\tau_6$ decouple, and the result of the integration in~\eqref{eq:4p-S-5} reduces to:
\beq \label{eq:4p-S-8}
\frac{\mathcal{F}_S(\tau_1, \tau_2, \tau_3, \tau_4)}{G_c^\beta(\tau_1, \tau_2) G_c^\beta(\tau_3, \tau_4)} = \frac{1}{8 \pi C} \frac{\beta J}{N} \left( \frac{\tau_{12}}{2 \tan \frac{\tau_{12}}{2}} - 1 \right) \left( \frac{\tau_{34}}{2 \tan \frac{\tau_{34}}{2}} - 1 \right), \eeq
In fact, this correlator describes the fluctuations of the total energy in the thermal ensemble, so it could be expected to factorize. More detailed explanation can be found in appendix~\ref{sec:energy-fluctuations} and paper~\cite{Maldacena-SYK}. 

For the OTO ordering we obtain\footnote{A useful relation is $\pd_{\tau_{56}}^2 = \frac{1}{4} \pd_{\tau_5}^2 + \frac{1}{4} \pd_{\tau_6}^2 - \frac{1}{2} \pd_{\tau_5} \pd_{\tau_6}$} the contribution~\eqref{eq:4p-S-8} plus the additional term due to the delta-functions in~\eqref{eq:4p-S-7}:
\beq \label{eq:4p-S-9} \begin{aligned}
\frac{\mathcal{F}_S(\tau_1, \tau_2, \tau_3, \tau_4)}{G_c^\beta(\tau_1, \tau_2) G_c^\beta(\tau_3, \tau_4)} = \frac{1}{8 \pi C} \frac{\beta J}{N} \Bigg[ &-\frac{3 \pi}{8} \frac{\sin\left(\Delta \tau\right)}{\sin\left(\frac{\tau_{12}}{2}\right) \sin\left(\frac{\tau_{34}}{2}\right)} + \frac{\pi}{16} \frac{\sin\left(\Delta \tau - \tau_{12} \right)}{\sin\left(\frac{\tau_{12}}{2}\right) \sin\left(\frac{\tau_{34}}{2}\right)} + \frac{\pi}{16} \frac{\sin\left(\Delta \tau - \tau_{34} \right)}{\sin\left(\frac{\tau_{12}}{2}\right) \sin\left(\frac{\tau_{34}}{2}\right)} - \\ &-\frac{\pi}{8} \frac{ 2 \Delta \tau - \tau_{12} - \tau_{34} }{\tan\left(\frac{\tau_{12}}{2}\right) \tan\left(\frac{\tau_{34}}{2}\right)} + \frac{3 \pi}{8} \frac{1}{\tan\left(\frac{\tau_{12}}{2}\right)} + \frac{3 \pi}{8} \frac{1}{\tan\left(\frac{\tau_{34}}{2}\right)} + \\ &+ \left( \frac{\tau_{12}}{2 \tan \frac{\tau_{12}}{2}} - 1 \right) \left( \frac{\tau_{34}}{2 \tan \frac{\tau_{34}}{2}} - 1 \right) \Bigg],
\end{aligned} \eeq
where we have introduced the time $\Delta \tau \equiv \frac{\tau_1 + \tau_2}{2} - \frac{\tau_3 + \tau_4}{2}$. It is convenient to take $\tau_1 - \tau_2 = \pi$ and $\tau_3 - \tau_4 = \pi$, because in this case the expression for the correlator~\eqref{eq:4p-S-9} significantly simplifies to:
\beq \label{eq:4p-S-10}
\frac{\mathcal{F}_S\left(\tau_1, \tau_2, \tau_3, \tau_4 \right)}{G_c^\beta\left(\frac{\beta}{2}\right) G_c^\beta\left(\frac{\beta}{2}\right)} = \frac{1}{8 \pi C} \frac{\beta J}{N} \left[1 - \frac{\pi}{2} \sin\left(\frac{2 \pi \Delta \tau}{\beta}\right) \right]. \eeq
Here we have restored $\beta$ in the exponent, i.e. mapped unit circle back to the $\beta$-circle~\eqref{eq:4p-5}.

To understand the physical relevance of the obtained result, let us analitycally continue the four-point function to the Lorentzian time and check the behavior of the correlator at large values of $t = -i \Delta \tau \gg J^{-1}$. A particularly important case is when $\tau_1 = \frac{\beta}{4} + i t$, $\tau_2 = - \frac{\beta}{4} + i t$, $\tau_3 = 0$, $\tau_4 = -\frac{\beta}{2}$, which describes the regularized out-of-time-ordered correlation function (OTOC):
\beq \label{eq:OTOC} \begin{aligned}
\OTOC(t) &\equiv \frac{1}{N^2} \sum_{i,j=1}^N \tr \left[ \rho^{\frac{1}{4}} \chi_i(t) \rho^{\frac{1}{4}} \chi_j(0) \rho^{\frac{1}{4}} \chi_i(t) \rho^{\frac{1}{4}} \chi_j(0) \right] = \\ &= \tilde{G}\left(\frac{\beta}{2}\right) \tilde{G}\left(\frac{\beta}{2}\right) + \mathcal{F}\left( \frac{\beta}{4} + i t, -\frac{\beta}{4} + i t, 0, - \frac{\beta}{2} \right) = \\ &= \tilde{G} \tilde{G} + \mathcal{F}_S + \mathcal{F}_{CFT} + \mathcal{F}_0 + \mathcal{O}\left(\frac{1}{N^2}\right),
\end{aligned} \eeq
where we have defined the density matrix as $\rho \equiv \frac{1}{Z} e^{-\beta H}$. For brevity we omitted arguments of four-point functions in the last line. At $t=0$ this choice corresponds to the OTO region, so the correlator is given by the analytical continuation of~\eqref{eq:4p-S-10} to the non-zero real $t$. Now it is straightforward to see that in the leading order the corrected OTOC rapidly decays:
\beq \label{eq:OTOC-S} \begin{aligned}
\OTOC(t) &\approx \tilde{G}\left(\frac{\beta}{2}\right) \tilde{G}\left(\frac{\beta}{2}\right) + \mathcal{F}_S\left( \frac{\beta}{4} + i t, -\frac{\beta}{4} + i t, 0, - \frac{\beta}{2} \right) \approx \\ &\approx \frac{\sqrt{\pi}}{2 \beta J} \left[ 1 + \frac{1}{8 \pi C} \frac{\beta J}{N} \left( 1 - \frac{\pi}{2} \cos\left(\frac{2 \pi i t}{\beta}\right)\right) \right] \approx \\
&\approx \frac{\sqrt{\pi}}{2 \beta J}\left[1  - \frac{\Delta^2}{2 C} \frac{\beta J}{N} e^{\frac{2 \pi}{\beta} t} \right], \quad \text{for} \quad \beta \ll t \ll \beta \log \frac{N}{\beta J}.
\end{aligned} \eeq
Here we restored the conformal dimension $\Delta = \frac{1}{4}$ and substituted the approximate saddle value, $\tilde{G} \approx G_c^\beta$. However, for bigger times gaussian approximation breaks down and one has to take into account corrections to this result. In general, one expects that the decay is saturated due to the contribution of multiple parallel ladders (see subsection~\ref{sec:4p-CFT}), but we will not discuss this point here.

Note that the contribution of the soft mode to the regularized time-ordered correlation function (TOC) does not change with $t$:
\beq \label{eq:TOC} \begin{aligned}
\TOC(t) &\equiv \frac{1}{N^2} \sum_{i,j=1}^N \tr \left[ \chi_i(t) \rho^{\frac{1}{2}} \chi_i(t) \chi_j(0) \rho^{\frac{1}{2}} \chi_j(0) \right] = \\ &= \tilde{G}\left(\frac{\beta}{2}\right) \tilde{G}\left(\frac{\beta}{2}\right) + \mathcal{F}\left( \frac{\beta}{2} + i t, i t, 0, - \frac{\beta}{2} \right) \approx \\ &\approx \tilde{G}\left(\frac{\beta}{2}\right) \tilde{G}\left(\frac{\beta}{2}\right) + \mathcal{F}_S\left( \frac{\beta}{2} + i t, i t, 0, - \frac{\beta}{2} \right) \approx \frac{\sqrt{\pi}}{2 \beta J} + \frac{\text{const}}{N}.
\end{aligned} \eeq
Finally, one also should take into account the $\mathcal{F}_0$ and $\mathcal{F}_{CFT}$ corrections to the connected four-point function, which are also of the order $\mathcal{O}\left(\frac{1}{N}\right)$. However, at the end of subsection~\ref{sec:DS} we have shown that two-point correlation functions exponentially decay for big Lorentzian times, $t \gg \beta$. Thus, for such times the contribution of $\mathcal{F}_0$ to OTOC and TOC also exponentially decays and therefore can be neglected. The contribution of $\mathcal{F}_{CFT}$ will be discussed in the next subsection.

\subsection{Conformal action contribution}
\label{sec:4p-CFT}

In this subsection we estimate the conformal contribution to the four-point correlation function, which is given by~\eqref{eq:4p-3}. As usual, we work in the IR and large $N$ limit. We remind that in this limit the theory is conformally invariant in the sense~\eqref{eq:DS-4}, so we can freely change between the zero temperature and finite temperature cases using the map~\eqref{eq:circle-line}. Due to this reason in the most of this subsection we work with zero-temperature functions.

At the same time, integrands in both numerator and denominator of~\eqref{eq:4p-3} are invariant wrt arbitrary reparametrizations and $I_S$ is non-zero for all but $SL(2,\mathbb{R})$ reparametrizations. One can integrate such reparametrizations out and obtain a non-zero constant that cancels when one calculates correlation functions. Therefore, the full reparametrization symmetry of the four-point function is effectively broken down to $SL(2,\mathbb{R})$.

Taking the integral over the fluctuations\footnote{We remind that in this section we parametrize fluctuations as $G = \tilde{G} + \delta G$ while in the subsection~\eqref{sec:Schwarzian} we used the notation $G=\tilde{G} + \frac{\delta G}{|\tilde{G}|}$.} of the variable $G$ in the functional integral~\eqref{eq:4p-3} with the effective action~\eqref{eq:Sch-2}, we obtain:
\beq \label{eq:4p-CFT-1}
\mathcal{F}_{CFT} = \frac{2}{3 J^2 N} \frac{(K_c^{-1} - I)^{-1} I}{| G_c(\tau_1, \tau_2) G_c(\tau_3, \tau_4) |} = \frac{2}{3 J^2 N} \frac{(I - K_c)^{-1} K_c I}{| G_c(\tau_1, \tau_2) G_c(\tau_3, \tau_4) |}. 
\eeq
Here $K_c$ denotes the conformal kernel that is defined by~\eqref{eq:Sch-K} with conformal two-point functions $\tilde{G} = G_c$. From~\eqref{eq:Sch-K},~\eqref{eq:Sch-I} and~\eqref{eq:4p-4} it follows that:
\beq K_c I = \frac{3 J^2 N}{2} \mathcal{F}_0(\tau_1, \tau_2, \tau_3, \tau_4) |G_c(\tau_1, \tau_2) G_c(\tau_3, \tau_4) |. \eeq
Now it is easy to see that $\mathcal{F}_{CFT}$ is simply the sum of all possible ladder diagrams from the Fig.~\ref{fig:ladder}:
\begin{figure}[t]
\center{\includegraphics[scale=0.3]{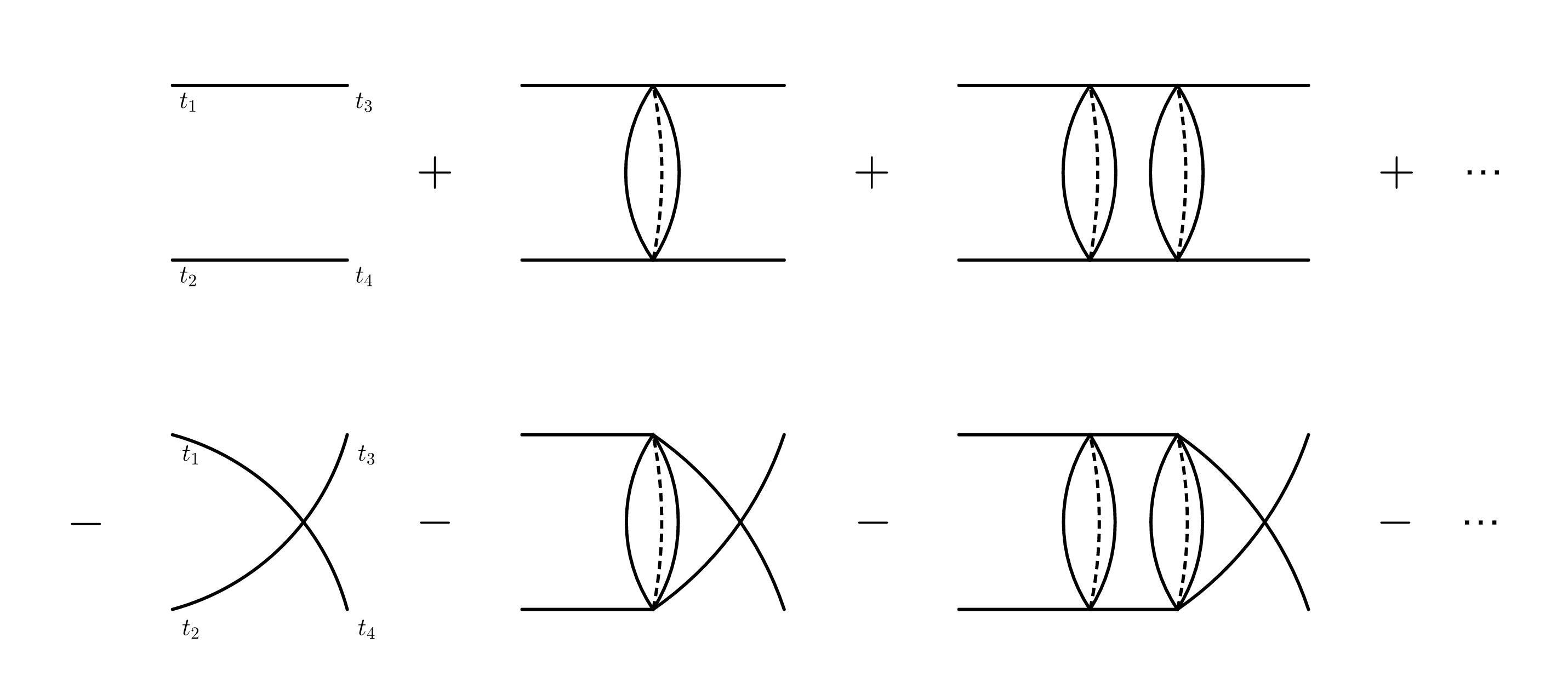}}\caption{Sum of the ladder diagrams which contribute to $\mathcal{F}_{CFT}$} \label{fig:ladder}
\end{figure}\beq  \mathcal{F}_{CFT} = \sum_{n=0}^\infty \mathcal{F}_n = (I - K_c)^{-1} \mathcal{F}_0, \eeq
where $\mathcal{F}_n \equiv K_c^n \mathcal{F}_0$ corresponds to the $n$-ladder diagrams. Indeed, one can check that in the diagrammatic technique introduced in Sec.~\ref{sec:SYK-diagrams} ladder diagrams as in Fig.~\ref{fig:ladder} are the only contributions to the 4-point correlation functions of the order $\frac{1}{N}$.

Note that the kernel $K_c$, which we use, is conjugated to the natural kernel, which follows from the diagrams Fig.~\ref{fig:ladder}, by the power of the propagator:
\beq  K_c(\tau_1, \tau_2, \tau_3, \tau_4) = | G_c(\tau_1, \tau_2) | K_{diagram}(\tau_1, \tau_2, \tau_3, \tau_4) | G_c(\tau_3, \tau_4) |^{-1}. \eeq
We did such a conjugation to make the symmetry $(\tau_1, \tau_2) \leftrightarrow (\tau_3, \tau_4)$ explicit. It is straightforward to check that under reparametrizations $\tau \rightarrow f(\tau)$, $f'(\tau) > 0$, operator $K_c$ transforms as a four-point function of the fields with conformal fimension $\Delta = \frac{1}{2}$.

Note that the diagrammatics with conformal two-point functions naively leads to the divergent expression, because in the conformal limit operator $K$ has unit eigenvalue: $(I - K_c) \delta G = 0$ (see Sec.~\ref{sec:Schwarzian}). In the section~\ref{sec:4p-S} we treated this divergence directly, moving away from the IR limit and considering non-conformal corrections to the effective action. The alternative approach is to calculate the leading correction to the unit eigenvalue~\cite{Maldacena-SYK}.

To calculate the expression~\eqref{eq:4p-CFT-1} we need to determine a complete set\footnote{I.e. such set, in which $I = \sum_h \frac{1}{\langle \Psi_h | \Psi_h \rangle} | \Psi_h \rangle \langle \Psi_h |$.} of antisymmetric eigenfunctions $\Psi_h(\tau_1, \tau_2)$, find eigenvalues $K_c \Psi_h = k(h) \Psi_h$ and calculate the following sum:
\beq  \label{eq:4p-CFT-2}
(I - K_c)^{-1} K_c I = \sum_{k(h) \ne 1} \frac{k(h)}{1 - k(h)} \frac{1}{\langle \Psi_h | \Psi_h \rangle} |\Psi_h \rangle \langle \Psi_h |, \eeq
where $h$ in this expression is an abstract label that numerates eigenvalues and eigenfunctions (this label will be specified below). In other words, we need to find the spectrum of the conformal kernel $K_c$. We remind that we have to exclude the unit eigenvalue subspace, because during the integration over this subspace the effective action~\eqref{eq:Sch-2} is zero, i.e. the dominant contribution to the full four-point function is given by~\eqref{eq:4p-2}.

\subsubsection{$SL(2,\mathbb{R})$ generators and casimir}

It is difficult to directly solve the integral equation $K_c \Psi_h = k(h) \Psi_h$. Fortunately, the $SL(2,\mathbb{R})$ invariance significantly simplifies this task. This invariance implies that $K_c$ commutes with the casimir $C$ of the $SL(2,\mathbb{R})$ group --- therefore, eigenfunctions of $K_c$ and $C$ coincide. This allows one to find eigenfunctions and eigenvalues separately. First, one solves the simpler equation\footnote{It is convenient but not necessary to choose the eigenvalue of the casimir as $h(h-1)$} $C \Psi_h = h(h-1) \Psi_h$, and then determines the eigenvalues $k(h)$ for the known functions $\Psi_h$.

The $SL(2,\mathbb{R})$ algebra can be presented using the following generators:
\beq \label{eq:casimir-1}
L_0^\tau = - \tau \pd_\tau - \Delta, \quad L_{-1}^\tau = \pd_\tau, \quad L_1^\tau = \tau^2 \pd_\tau + 2 \Delta \tau. \eeq
It is straightforward to check that these operators obey the proper commutation relations:
\beq \left[ L_m^\tau, L_n^\tau \right] = (m-n) L_{m+n}^\tau \quad \text{for} \quad m,n = -1, 0, 1. \eeq
Note that in this definition an operator with conformal dimension $\Delta$ is annihilated by the generator $L_0^\tau$.

Please note that in the case $\Delta = \frac{1}{2}$ these generators should commute with the kernel $K_c$:
\beq \label{eq:casimir-2}\begin{aligned}
\left \langle \left(L_m^{\tau_1} + L_m^{\tau_2}\right) K_c(\tau_1, \tau_2, \tau_3, \tau_4) | \Psi_h(\tau_3, \tau_4) \right \rangle &= \left \langle K_c(\tau_1, \tau_2, \tau_3, \tau_4) \left(L_m^{\tau_3} + L_m^{\tau_4} \right) | \Psi_h(\tau_3, \tau_4) \right\rangle + \\ &+ 2 \int_{-\infty}^\infty d\tau_4 \left[ \tau_3^{m+1} K_c(\tau_1, \tau_2, \tau_3, \tau_4) \Psi_h(\tau_3, \tau_4) \right]_{\tau_3 = -\infty}^{\tau_3 = \infty},
\end{aligned} \eeq
where $\langle \cdot | \cdot \rangle$ denotes the inner product~\eqref{eq:inner}. This condition implies that $SL(2,\mathbb{R})$ generators are zero modes of the operator $K_c$. To ensure this commutation relation, the term in the second line must vanish for all basis functions $\Psi_h$ and all generators. Below we will see that this condition imposes an important restriction on the functions $\Psi_h$. 

Finally, using the generators~\eqref{eq:casimir-1} we build the casimir operator:
\beq  \begin{aligned}
C &= \left( L_0^{\tau_1} + L_0^{\tau_2} \right)^2 - \frac{1}{2} \left( L_{-1}^{\tau_1} + L_{-1}^{\tau_2} \right) \left( L_1^{\tau_1} + L_1^{\tau_2} \right) - \frac{1}{2} \left( L_1^{\tau_1} + L_1^{\tau_2} \right) \left( L_{-1}^{\tau_1} + L_{-1}^{\tau_2} \right) = \\ &= 2 \left(\Delta^2 - \Delta\right) + 2 L_0^{\tau_1} L_0^{\tau_2} - L_{-1}^{\tau_1} L_1^{\tau_2} - L_1^{\tau_1} L_{-1}^{\tau_2}.
\end{aligned} \eeq

\subsubsection{Eigenfunctions and eigenvalues}

Let us solve the equation $C \Psi_h = h(h-1) \Psi_h$. Substituting generators~\eqref{eq:casimir-1} and $\Delta = \frac{1}{2}$ we obtain the following differential equation:
\beq \label{eq:casimir-3}
\left[ - \left(\tau_1 - \tau_2\right)^2 \pd_{\tau_1} \pd_{\tau_2} + \left(\tau_1 - \tau_2\right) \left(\pd_{\tau_1} - \pd_{\tau_2}\right) \right] \Psi_h(\tau_1, \tau_2) =  h(h-1) \Psi_h(\tau_1, \tau_2). \eeq
We propose the following ansatz to solve this equation:
\beq \label{eq:eigen-0}
\Psi_{h\omega}(\tau_1, \tau_2) = \frac{\sgn (\tau_1 - \tau_2)}{\sqrt{|\tau_1 - \tau_2|}} \psi_{h}\left(\frac{\left|\omega (\tau_1 - \tau_2)\right|}{2}\right) e^{- i \omega \frac{\tau_1 + \tau_2}{2}} . \eeq
This ansatz is inspired by the following properties of the casimir operator and function $\Psi_h$. First, $\Psi_h$ is an antisimmetric function with the conformal weight $\Delta = \frac{1}{2}$, so we expect the factor $\frac{\sgn(\tau_1 - \tau_2)}{\sqrt{|\tau_1 - \tau_2|}}$. Second, the structure of the equation~\eqref{eq:casimir-3} points that it is convenient to use variables $\tau \equiv \tau_1 - \tau_2$ and $T \equiv \frac{1}{2}(\tau_1 + \tau_2)$ rather than $\tau_1$ and $\tau_2$. Third, the result of the action of the casimir operator~\eqref{eq:casimir-3} on~\eqref{eq:eigen-0} does not depend on $\omega$, and, finally, $\psi_h$ solves the Bessel equation:
\beq \label{eq:eigen-1}
\left[ x^2 \pd_x^2 + x \pd_x + \left( x^2 - h (h-1) - \frac{1}{4} \right) \right] \psi_{h}(x) = 0, \quad \text{where} \quad x \equiv \frac{|\omega \tau|}{2}. \eeq
This means that for each $h$ one has an infinite set of eigenfunctions parametrized by the frequency $\omega$. In the zero temperature case frequency is continious ($\omega \in \mathbb{R}$), in the finite temperature case it is descrete ($\omega = \frac{\pi}{\beta}(2n+1)$, $n\in\mathbb{Z}$). This also implies that in the decomposition~\eqref{eq:4p-CFT-2} one has to sum over the set $\Psi_{h\omega}$ instead of the set $\Psi_h$:
\beq \label{eq:4p-CFT-4}
(I-K_c)^{-1}K_c I = \sum_{k(h) \ne 1} \sum_\omega \frac{k(h,\omega)}{1 - k(h,\omega)} \frac{1}{\langle \Psi_{h\omega} | \Psi_{h\omega} \rangle} |\Psi_{h\omega} \rangle \langle \Psi_{h\omega} |. \eeq
The general solution of the equation~\eqref{eq:eigen-1} is the sum of Bessel functions:
\beq \psi_{h}(x) = -A_h J_{h - \frac{1}{2}} \left(x\right) - B_h Y_{h - \frac{1}{2}} \left(x\right) = \frac{B_{1-h}}{\cos(\pi h)} J_{h-\frac{1}{2}} \left(x\right) - \frac{B_h}{\cos(\pi h)} J_{\frac{1}{2}-h} \left(x\right). \eeq
Here $B_h$ is some function of $h$. To obtain the second equality we required $\Psi_{1-h} = \Psi_h$, because the equation~\eqref{eq:eigen-1} is invariant under the change $h \rightarrow 1 - h$. Also we have used the following relation between Bessel functions of the first and second kinds:
\beq Y_\alpha(x) = \frac{J_\alpha(x) \cos(\pi \alpha) - J_{-\alpha}(x)}{\sin(\pi \alpha)}. \eeq
Then we recall that the kernel $K_c$ must commute with $SL(2,\mathbb{R)}$ generators. This implies that the term in the second line of~\eqref{eq:casimir-2} should be identically zero for $m=-1,0,1$ and all $h$. For $m=-1,0$ this condition is always satisfied, so it does not restrict anything. Indeed, the expression under the square brackets is proportional to $|\tau_3|^{m-2}$ as $\tau_3 \rightarrow \pm \infty$, i.e. in the case $m=-1,0$ the integrand is identically zero. However, in the case $m=1$ this condition imposes an additional restriction on the coefficients $B_h$:
\beq \begin{aligned}
&\int_{-\infty}^\infty d\tau_4 \left[ \tau_3^2 K(\tau_1, \tau_2, \tau_3, \tau_4) \psi_{h \omega}\left(\frac{\left|\omega ( \tau_3 - \tau_4)\right|}{2}\right) \cos\left(\frac{\omega(\tau_3 + \tau_4)}{2}\right) \right]_{\tau_3 = -\infty}^{\tau_3 = \infty} = \\
&= - 3 \sqrt{\pi} J^2 \int_{-\infty}^\infty d\tau_4 \frac{\sgn(\tau_2 - \tau_4)}{\sqrt{|\tau_1 - \tau_2| |\tau_2 - \tau_4|}} \sin\frac{\omega \tau_4}{2} \left[ \frac{B_h}{\cos(\pi h)} \cos\frac{\pi h}{2} - \frac{B_{1-h}}{\cos(\pi h)} \sin\frac{\pi h}{2} \right] = 0,
\end{aligned} \eeq
hence,
\beq \frac{B_h}{B_{1-h}} = \tan\frac{\pi h}{2}. \eeq
Thus, the eigenfunctions have the following form (up to the numerical factor to be fixed below):
\beq \label{eq:eigen-2}
\Psi_{h\omega}(\tau_1, \tau_2) = \frac{\sgn\tau}{\sqrt{|\tau|}} e^{-i \omega T} \left[ \frac{\cos\frac{\pi h}{2}}{\cos(\pi h)} J_{h-\frac{1}{2}} \left(\frac{|\omega \tau|}{2}\right) - \frac{\sin\frac{\pi h}{2}}{\cos(\pi h)} J_{\frac{1}{2}-h} \left(\frac{|\omega \tau|}{2}\right)\right]. \eeq
Integrating the function~\eqref{eq:eigen-2} with the kernel $K_c$ as in~\eqref{eq:Sch-K} one finds the corresponding eigenvalue, $K_c \Psi_h = k(h,\omega) \Psi_h$:
\beq \label{eq:eigen-4}
k(h,\omega) = -\frac{3}{2} \frac{\tan\left[\frac{\pi}{2} \left(h - \frac{1}{2}\right)\right]}{h - \frac{1}{2}}. \eeq
This calculation is cumbersome but straightforward, so we do not reproduce it here. A detailed calculation\footnote{The authors of~\cite{Polchinski} use a different kernel and obtain slightly different eigenfunctions, but the integral for the eigenvalue coincides with our case.} can be found in appendices C and D of~\cite{Polchinski}.

Note that the eigenvalue~\eqref{eq:eigen-4} does not depend on the frequency $\omega$ due to the conformal invariance of the kernel. However, it does depend on the frequency when one moves away from the IR limit. In the paper~\cite{Maldacena-SYK} this dependence was established and used to calculate the leading non-conformal correction to the four-point correlation functions. The result of this calculation coincides with the result of subsection~\ref{sec:4p-S}.

\subsubsection{The complete set of eigenfunctions}

Eigenfunctions of the Hermitian operator form the complete set (e.g. see~\cite{Reed}). Keeping this fact in mind, we demand the  hermiticity of the Casimir operator wrt the inner product~\eqref{eq:inner}:
\beq \label{eq:set-1}
\langle C \Psi_{h\omega}(\tau_1, \tau_2) | \Psi_{h'\omega'}(\tau_1, \tau_2) \rangle = \langle \Psi_{h\omega}(\tau_1, \tau_2) | C \Psi_{h'\omega'}(\tau_1, \tau_2) \rangle. \eeq
On the one hand, the hermiticity mens that the eigenvalue of the casimir is real:
\beq \text{Im}\left[h(h-1)\right]=0, \quad \text{i.e.} \quad \text{Im}[h] \left(2 \text{Re}[h] - 1\right) = 0. \eeq
In other words, variable $h$ either pure real or has the fixed real part: $h = \frac{1}{2} + i s$, $s \in \mathbb{R}$, $s > 0$ (without the last inequality the eigenfunctions are ambiguous: $\Psi_{\frac{1}{2} +is,\omega} = \Psi_{\frac{1}{2}-is,\omega}$). On the other hand, identity~\eqref{eq:set-1} implies that the corresponding boundary term vanishes for arbitrary $\omega$, $\omega'$ and $h$, $h'$ from the spectrum:
\beq \begin{aligned}
\langle C \Psi_{h\omega}(\tau_1, \tau_2) | \Psi_{h'\omega'}(\tau_1, \tau_2) \rangle &- \langle \Psi_{h\omega}(\tau_1, \tau_2) | C \Psi_{h'\omega'}(\tau_1, \tau_2) \rangle = \\ &=\frac{8 \pi \delta(\omega-\omega')}{\omega} \Big[ x \Big(\psi_{h'}(x) \pd_x \psi_{h}^*(x) - \psi_{h}^*(x) \pd_x \psi_{h'}(x) \Big) \Big]_{x = 0}^{x = \infty} = 0.
\end{aligned} \eeq
Here we substituted the ansatz~\eqref{eq:eigen-0} and denoted $x=\frac{|\omega \tau|}{2}$. Substituting the asymptotics of the bessel function~\cite{Gradshteyn} we find that:
\beq \lim_{x \rightarrow \infty} \Big[ x \Big(\psi_{h'}(x) \pd_x \psi_{h}^*(x) - \psi_{h}^*(x) \pd_x \psi_{h'}(x) \Big) \Big] = 0 \eeq
for arbitrary $h$ and $h'$, and
\small \beq \begin{aligned}
\lim_{x \rightarrow 0} &\Big[ x \Big(\psi_{h'}(x) \pd_x \psi_{h}^*(x) - \psi_{h}^*(x) \pd_x \psi_{h'}(x) \Big) \Big] = \lim_{x \rightarrow 0}  \frac{\cos\frac{\pi h^*}{2} \cos\frac{\pi h'}{2}}{\cos(\pi h^*) \cos(\pi h')} \times \\
\times &\Bigg[ \frac{h^* - h'}{\Gamma\left(h^*+\frac{1}{2}\right) \Gamma\left(h'+\frac{1}{2}\right)} \left( \frac{x}{2} \right)^{h^* + h' - 1} - \tan\frac{\pi h^*}{2} \tan\frac{\pi h'}{2} \frac{h^* - h'}{\Gamma\left(\frac{3}{2} - h^*\right) \Gamma\left(\frac{3}{2} - h'\right)}\left( \frac{x}{2} \right)^{1 - h^* - h'} + \\ &+ \tan\frac{\pi h^*}{2} \frac{h^* + h' - 1}{\Gamma\left(\frac{3}{2} - h^*\right) \Gamma\left(h'+\frac{1}{2}\right)} \left( \frac{x}{2} \right)^{h' - h^*} - \tan\frac{\pi h'}{2} \frac{h^* + h' - 1}{\Gamma\left(h^*+\frac{1}{2}\right) \Gamma\left(\frac{3}{2} - h'\right)} \left( \frac{x}{2} \right)^{h^* + h' - 1}\Bigg] = 0
\end{aligned} \eeq \normalsize
for values of the form $h = \frac{1}{2} + is$, $s \in \mathbb{R}$, $s>0$ (in this case one obtains an oscillating expression) or $h = 2 n$, $n = 1,2,3,\cdots$ (in this case the divergent terms are multiplied by zeroes). We conclude that together these two sets form the complete set in the space of antisimmetric two-point functions.

Let us find the normalization in the decomposition~\eqref{eq:4p-CFT-4}, i.e. calculate the inner product of two eigenfunctions:
\beq \begin{aligned}
\langle \Psi_{h\omega}(\tau_1,\tau_2) | \Psi_{h'\omega'}(\tau_1,\tau_2)\rangle = 2 \int_{-\infty}^\infty dT \int_0^\infty \frac{d\tau}{\tau} \psi_h^*\left(\frac{|\omega| \tau}{2}\right) \psi_{h'}\left(\frac{|\omega'| \tau}{2}\right) e^{i (\omega - \omega')T} = \\ = 4 \pi \delta(\omega - \omega') \int_0^\infty \frac{d\tau}{\tau} \left[ \frac{\sin\frac{\pi h}{2}}{\cos(\pi h)} J_{\frac{1}{2}-h} \left(\frac{\omega \tau}{2}\right) - \frac{\cos\frac{\pi h}{2}}{\cos(\pi h)} J_{h-\frac{1}{2}} \left(\frac{\omega \tau}{2}\right) \right]^* \times \\ \times \left[ \frac{\sin\frac{\pi h'}{2}}{\cos(\pi h')} J_{\frac{1}{2}-h'} \left(\frac{\omega \tau}{2}\right) - \frac{\cos\frac{\pi h'}{2}}{\cos(\pi h')} J_{h'-\frac{1}{2}} \left(\frac{\omega \tau}{2}\right) \right].
\end{aligned} \eeq
For the discrete set this integral gives the Kronecker delta:
\beq \langle \Psi_{h\omega}(\tau_1,\tau_2) | \Psi_{h'\omega'}(\tau_1,\tau_2)\rangle = \frac{2 \pi^2}{2h - 1} \delta_{h h'} \cdot 2\pi\delta(\omega - \omega'), \eeq
and for the continuum set it gives the Dirac delta\footnote{We introduce UV cut-off $\epsilon \rightarrow 0$ and use that $\lim_{\epsilon \rightarrow 0} \frac{2}{p-s} \sin\left(\frac{s-p}{2}\log\frac{\epsilon}{2}\right) = \pi \delta(s-p)$. More details can be found in~\cite{Polchinski}.}:
\beq \langle \Psi_{h\omega}(\tau_1,\tau_2) | \Psi_{h'\omega'}(\tau_1,\tau_2)\rangle = \frac{2 \pi \tan(\pi h)}{2h - 1} 2 \pi \delta(h - h') \cdot 2\pi\delta(\omega - \omega'). \eeq
Furthermore, the identity operator~\eqref{eq:Sch-I} on the space of antisimmetric two-point functions can be represented as the following decomposition:
\beq \begin{aligned}
I(\tau_1, \tau_2, \tau_3, \tau_4) = \frac{1}{2} \int_{-\infty}^\infty \frac{d\omega}{2\pi} \Bigg[ &\int_0^\infty \frac{ds}{2\pi} \frac{2h - 1}{\pi \tan(\pi h)} \Psi_{h\omega}(\tau_1, \tau_2) \Psi_{h\omega}^*(\tau_3, \tau_4)\Big|_{h = \frac{1}{2} + is} + \\ + &\sum_{n=1}^\infty \frac{2h-1}{\pi^2} \Psi_{h\omega}(\tau_1, \tau_2) \Psi_{h\omega}^*(\tau_3, \tau_4)\Big|_{h = 2n} \Bigg].
\end{aligned} \eeq
Substituting the ansatz~\eqref{eq:eigen-0} and integrating over the frequencies, we obtain the decomposition which explicitly looks as the conformal four-point function of fields with $\Delta=1$:
\beq \label{eq:set-2}
I(\tau_1, \tau_2, \tau_3, \tau_4) = \frac{1}{2} \frac{\sgn(\tau_{12}) \sgn(\tau_{34})}{|\tau_{12}| |\tau_{34}|} \left[ \int_0^\infty \frac{ds}{2\pi} \frac{2h - 1}{\pi \tan(\pi h)} \Psi_h (\chi) \Big|_{h = \frac{1}{2} + is} + \sum_{n=1}^\infty \frac{2h-1}{\pi^2} \Psi_h (\chi) \Big|_{h=2n} \right], \eeq
where we have denoted $\tau_{12} \equiv \tau_1 - \tau_2$, introduced the $SL(2,\mathbb{R})$-invariant cross-ratio:
\beq \label{eq:cross}
\chi \equiv \frac{\tau_{12} \tau_{34}}{\tau_{13} \tau_{24}}, \eeq
and defined the function $\Psi_h(\chi)$:
\beq \Psi_h(\chi) \equiv \begin{cases}
\frac{\Gamma\left(\frac{h}{2}\right) \Gamma\left(\frac{1-h}{2}\right)}{\sqrt{\pi}} \phantom{.}_2 F_1\left[\frac{h}{2}, \frac{1-h}{2}, \frac{1}{2}, \left(\frac{2-\chi}{\chi}\right)^2\right], \quad &\text{if} \quad \chi > 1, \\
\frac{\cos^2\left(\frac{\pi h}{2}\right)}{\cos(\pi h)} \frac{\Gamma(h)^2}{\Gamma(2h)} \chi^h \phantom{.}_2 F_1\left(h, h, 2h, \chi \right) + (h \rightarrow 1-h), \quad &\text{if} \quad 0 < \chi < 1.
\end{cases} \eeq
Here $\phantom{.}_2 F_1(\cdots)$ is the hypergeometric function. For the details on this calculation see appendix~\ref{sec:product-integral} and papers~\cite{Polchinski,Maldacena-SYK}.

Finally, the decomposition~\eqref{eq:set-2} can be rewritten as the single controur integral:
\beq \label{eq:set-3}
I(\tau_1, \tau_2, \tau_3, \tau_4) = \frac{1}{2} \frac{\sgn(\tau_{12}) \sgn(\tau_{34})}{\tau_{12} \tau_{34}} \int_\mathcal{C} \frac{dh}{2 \pi i} \frac{h - \frac{1}{2}}{\pi \tan\left(\frac{\pi h}{2}\right)} \Psi_h(\chi), \eeq
where the countor $\mathcal{C}$ is defined in the following way:
\beq \label{eq:set-4}
\int_\mathcal{C} \frac{dh}{2 \pi i} \equiv \int_{\frac{1}{2}-i\infty}^{\frac{1}{2}+i\infty} \frac{dh}{2\pi i} + \sum_{n=1}^\infty \text{Res}_{h=2n}. \eeq
In order to rewrite the integral over $ds$ we used the symmetry of the integrand under the change $h \rightarrow 1-h$ and the following indentity:
\beq \frac{2}{\tan(\pi h)} = \frac{1}{\tan\frac{\pi h}{2}} - \frac{1}{\tan\frac{\pi(1-h)}{2}}. \eeq
Of course, the decomposition for the identity operator can also be deduced from the representation theory of $SL(2,\mathbb{R})$ group. One can find more details on this method in~\cite{Kitaev-reps}.

\subsubsection{Four-point function and OPE}
\label{sec:OPE}

To find the conformal contribution to the four-point function, we substitute the eigenvalues and the decomposition of the identity operator~\eqref{eq:set-3} into eq.~\eqref{eq:4p-CFT-1}:
\beq \label{eq:4p-fin-1}
\mathcal{F}_{CFT}(\tau_1, \tau_2, \tau_3, \tau_4) = \frac{\sqrt{4\pi}}{3 N} \frac{\sgn(\tau_{12})}{|J\tau_{12}|^{2\Delta}} \frac{\sgn(\tau_{34})}{|J\tau_{34}|^{2\Delta}} \mathcal{F}_{CFT}(\chi), \eeq
where $\Delta = \frac{1}{4}$ and we have introduced the $SL(2,\mathbb{R})$-invariant function $\mathcal{F}_{CFT}(\chi)$:
\beq \label{eq:4p-fin-2}
\mathcal{F}_{CFT}(\chi) = \int_\mathcal{C} \frac{dh}{2 \pi i} \frac{k(h)}{1 - k(h)} \frac{h - \frac{1}{2}}{\pi \tan\left(\frac{\pi h}{2}\right)} \Psi_h(\chi) \Big|_{h\ne2}. \eeq
In the finite temperature case the expression~\eqref{eq:4p-fin-1} transforms to:
\beq \label{eq:4p-fin-3}
\mathcal{F}_{CFT}(\tau_1, \tau_2, \tau_3, \tau_4) = \frac{\sqrt{4\pi}}{3 N} \frac{1}{\beta J} \frac{\sgn\left(\sin\frac{\pi \tau_{12}}{\beta}\right)}{\big|\sin\frac{\pi \tau_{12}}{\beta}\big|^{2\Delta}} \frac{\sgn\left(\sin\frac{\pi \tau_{34}}{\beta}\right)}{\big|\sin\frac{\pi \tau_{34}}{\beta}\big|^{2\Delta}} \mathcal{F}_{CFT}(\tilde{\chi}), \quad \tilde{\chi} = \frac{\sin\frac{\pi \tau_{12}}{\beta} \sin\frac{\pi \tau_{34}}{\beta}}{\sin\frac{\pi \tau_{13}}{\beta} \sin\frac{\pi \tau_{24}}{\beta}}. \eeq
In~\eqref{eq:set-4} we have to exclude the value $h=2$ because it corresponds to the zero mode of the operator $K_c$, i.e. to the soft mode discussed in the subsection~\ref{sec:4p-S}. However, $h=2$ is not the only solution to the equation $k(h) = 1$ with $k(h)$ from~\eqref{eq:eigen-4}. In fact, this equation has infinitely many real solutions of the form $h_m = 2\Delta + 2 m + 1 + \epsilon_m$, where $\epsilon_m$ is going to zero for large integer $m$ as:
\beq \epsilon_m \approx \frac{3}{2 \pi m}, \quad \text{for} \quad m \gg 1. \eeq
These solutions do not belong to the spectrum of the operator $K_c$, but they correspond to the simple poles of the function $\frac{k(h)}{1 - k(h)} \frac{h-\frac{1}{2}}{\pi \tan\left(\frac{\pi h}{2}\right)} \Psi_h(\chi)$. Hence, we can push the contour $\mathcal{C}$ to the right (Fig.~\ref{fig:pole-1}) and obtain the different decomposition for the function $\mathcal{F}_{CFT}$:
\begin{figure}[t]
\center{\includegraphics[scale=0.3]{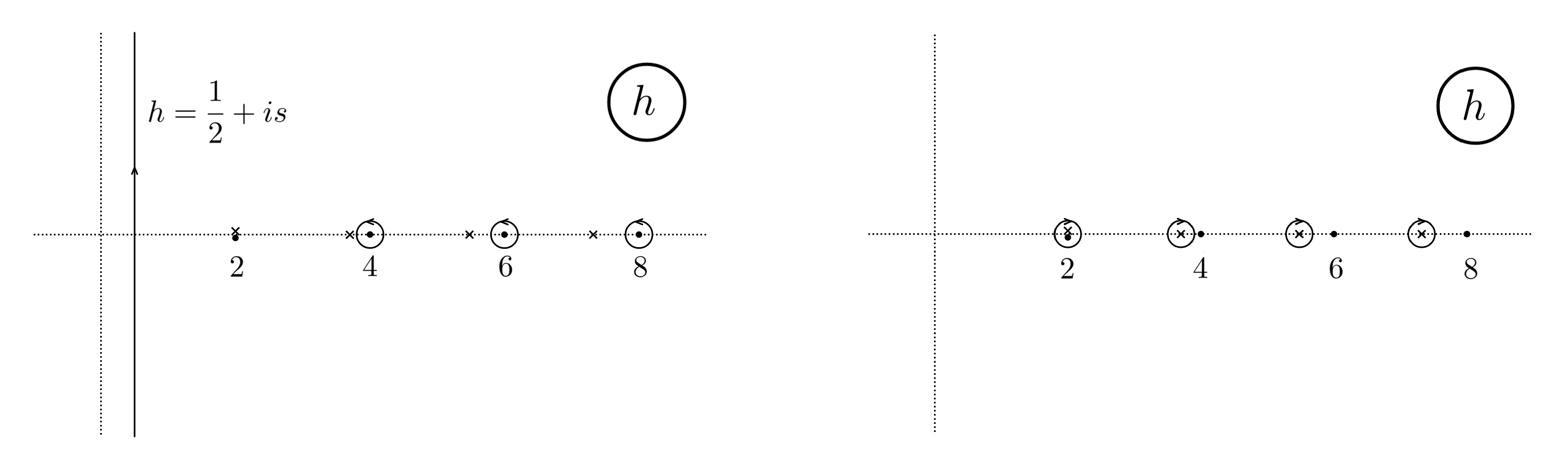}}\caption{To the left: contour $\mathcal{C}$ from the sum~\eqref{eq:4p-fin-2}. To the right: contour from the sum~\eqref{eq:4p-fin-OPE}. Dots denote poles that correspond to the solutions of $\tan\frac{\pi h}{2} = 0$, crosses denote poles that correspond to the solutions of $k(h) = 1$. Note the double pole at $h=2$.} \label{fig:pole-1}
\end{figure}
\beq \label{eq:4p-fin-OPE}
\mathcal{F}_{CFT}(\chi) = \sum_{m=0}^\infty \text{Res}_{h=h_m} \left[ \frac{k(h)}{1 - k(h)} \frac{h-\frac{1}{2}}{\pi \tan\left(\frac{\pi h}{2}\right)} \Psi_h(\chi) \right], \eeq
where $h_0 = 2$ and $h_m$ for $m>0$ have the form mentioned above. However, the contribution from the $h_0 = 2$ pole cancels if one moves away from the IR limit and considers the corrections to the $k(h, \omega)$ near the $h_0 = 2$ (we will not discuss how this happens, for the details see~\cite{Maldacena-SYK}). Thus, for $\chi < 1$ this expansion reproduces the four-point function OPE~\cite{Maldacena-SYK,Sarosi}:
\beq \mathcal{F}_{CFT}(\chi) = \sum_{m=1}^\infty c_m^2 \chi^{h_m} \phantom{.}_2 F_1\left(h_m, h_m, 2h_m, \chi \right) = \vcenter{\hbox{\includegraphics[scale=0.15]{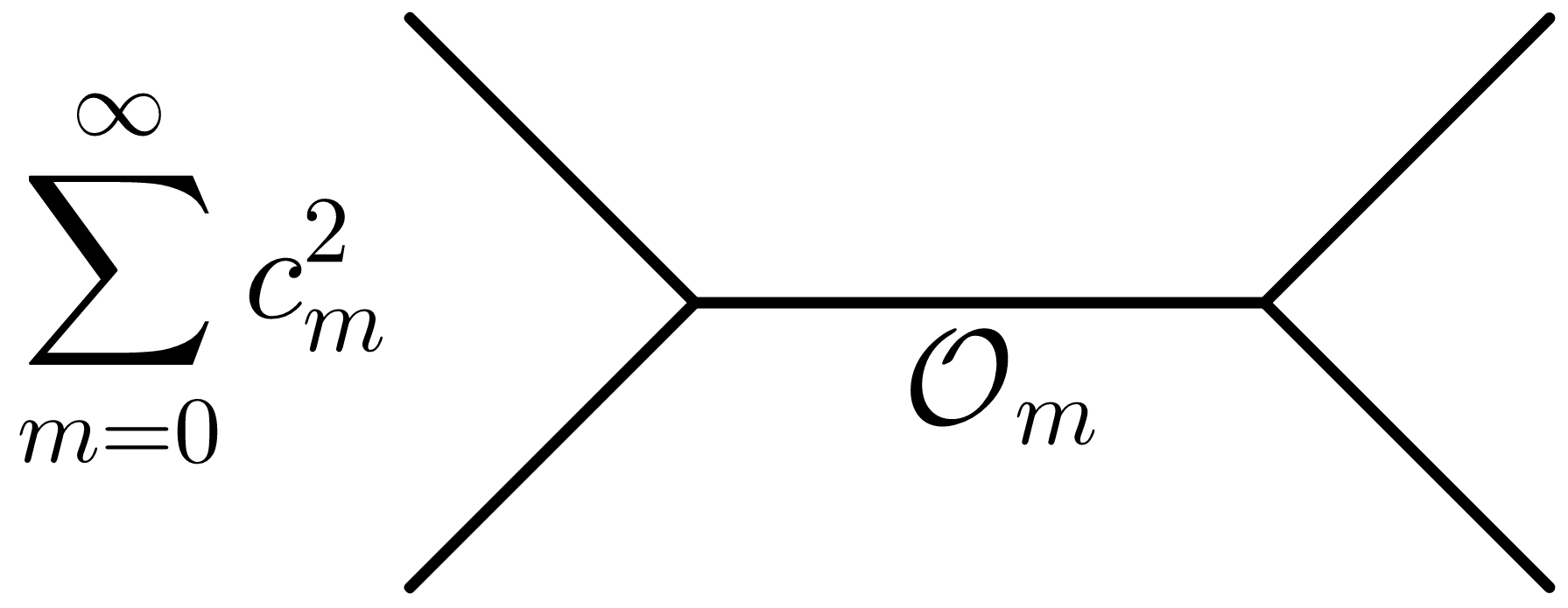}} }, \eeq
where $h_m$ are conformal weights of the corresponding intermediate operators and the coefficients $c_m$ are found from the decomposition of~\eqref{eq:set-3} around $\chi = 0$. The asymptotic behavior of the conformal weights shows that the operators of the OPE are built from two fermion fields, $2m+1$ derivatives and anomalous part that corresponds to the interactions:
\beq \mathcal{O}_m = \sum_{i=1}^N \sum_{k=0}^{2m+1} d_{mk} \pd_\tau^k \chi_i \pd_\tau^{2m+1-k} \chi_i, \eeq
where $d_{mk}$ are some numerical coefficients. The explicit form of the operators $\mathcal{O}_m$ can be found in~\cite{Gross}.

\subsubsection{OTOC and TOC}

In this subsubsection we estimate the conformal contributions to OTOC~\eqref{eq:OTOC}, which corresponds to the function $\mathcal{F}\left(\frac{\beta}{4}+it, -\frac{\beta}{4}+it,0,-\frac{\beta}{2}\right)$, and TOC~\eqref{eq:TOC}, which corresponds to the function $\mathcal{F}\left(\frac{\beta}{2}+it, it,0,-\frac{\beta}{2}\right)$. On the tree level both of these correlators behave as:
\beq \OTOC(t) = \TOC(t) = \tilde{G}\left(\frac{\beta}{2}\right) \tilde{G}\left(\frac{\beta}{2}\right) \approx \frac{\sqrt{\pi}}{2 \beta J} + \mathcal{O}\left(\frac{1}{N}\right), \eeq
in the limit $t \rightarrow \infty$. In subsection~\ref{sec:4p-S} we estimated the leading $\mathcal{O}\left(\frac{1}{N}\right)$ corrections to these correlators which are ensured by the so-called soft mode. Here we find the subleading corrections that have the same order in $\frac{1}{N}$ but are suppressed by the small factor $\frac{1}{\beta J}$. We denote such corrections as $\delta \OTOC(t)$ and $\delta \TOC(t)$.

In the limit $t \rightarrow \infty$ choises of times for both OTOC and TOC give small cross-ratios~\eqref{eq:cross}, $\chi \rightarrow 0$. However, in the limit $t \rightarrow 0$ times of the OTOC correspond to the cross-ratio $\chi \rightarrow 2 - \frac{4 \pi i t}{\beta}$, whereas times of the TOC correspond to $\chi \rightarrow 1 - \frac{\pi^2 t^2}{\beta^2}$. Hence, for the OTOC we need to analytically continue the $\chi > 1$ version of the expression~\eqref{eq:4p-fin-2} to small imaginary cross-ratios $\chi \sim -4 i e^{-\frac{2 \pi t}{\beta}}$:
\small \beq \delta \OTOC(t) = \frac{\sqrt{4\pi}}{3 N} \frac{1}{\beta J}  \int_\mathcal{C} \frac{dh}{2 \pi i} \frac{k(h)}{1 - k(h)} \frac{h - \frac{1}{2}}{\pi \tan\left(\frac{\pi h}{2}\right)} \frac{\Gamma\left(\frac{h}{2}\right) \Gamma\left(\frac{1-h}{2}\right)}{\sqrt{\pi}} \phantom{.}_2 F_1\left[\frac{h}{2}, \frac{1-h}{2}, \frac{1}{2}, \left(\frac{2-\chi}{\chi}\right)^2\right]_{h\ne2}. \eeq \normalsize
For the TOC we just need to take the limit $\chi \sim 4 e^{-\frac{2 \pi t}{\beta}} \rightarrow 0$:
\small \beq \delta \TOC(t) = \frac{\sqrt{4\pi}}{3 N} \frac{1}{\beta J}  \int_\mathcal{C} \frac{dh}{2 \pi i} \frac{k(h)}{1 - k(h)} \frac{h - \frac{1}{2}}{\pi \tan\left(\frac{\pi h}{2}\right)} \left[\frac{\cos^2\left(\frac{\pi h}{2}\right)}{\cos(\pi h)} \frac{\Gamma(h)^2}{\Gamma(2h)} \chi^h \phantom{.}_2 F_1\left(h, h, 2h, \chi \right) + (h \rightarrow 1-h)\right]_{h\ne2}. \eeq \normalsize
To evaluate the integral along the contour $\mathcal{C}$ we use the following trick. First of all, we define the function $k_R(h)$:
\beq k_R(h) \equiv \frac{\cos\left[\frac{\pi (2h-1)}{4}\right]}{\cos\left[\frac{\pi (2h+1)}{4}\right]} k(h), \eeq
which has two useful properties. On the one hand, for any real even $h$ this function coincides with the eigenvalue $k(h)$, so we can substitute $k(h) \rightarrow k_R(h)$ into the descrete sum in~\eqref{eq:4p-fin-2}. On the other hand, $k_R(h) = 1$ in the unique point of the complex plane, $h=2$. Hence, we can freely\footnote{$\Psi_h(\chi)$ has simple poles at the points $h=1,3,5,\cdots$, but these poles are cancelled by zeroes of $\left[\tan\left(\frac{\pi h}{2}\right)\right]^{-1}$.} pull the contour that circles $h=2,4,6,\cdots$ back to the line $h = \frac{1}{2} + is$ (Fig.~\ref{fig:pole-2}). After this operation we get the single integral over the line plus the pole at $h=2$: 
\begin{figure}[t]
\center{\includegraphics[scale=0.3]{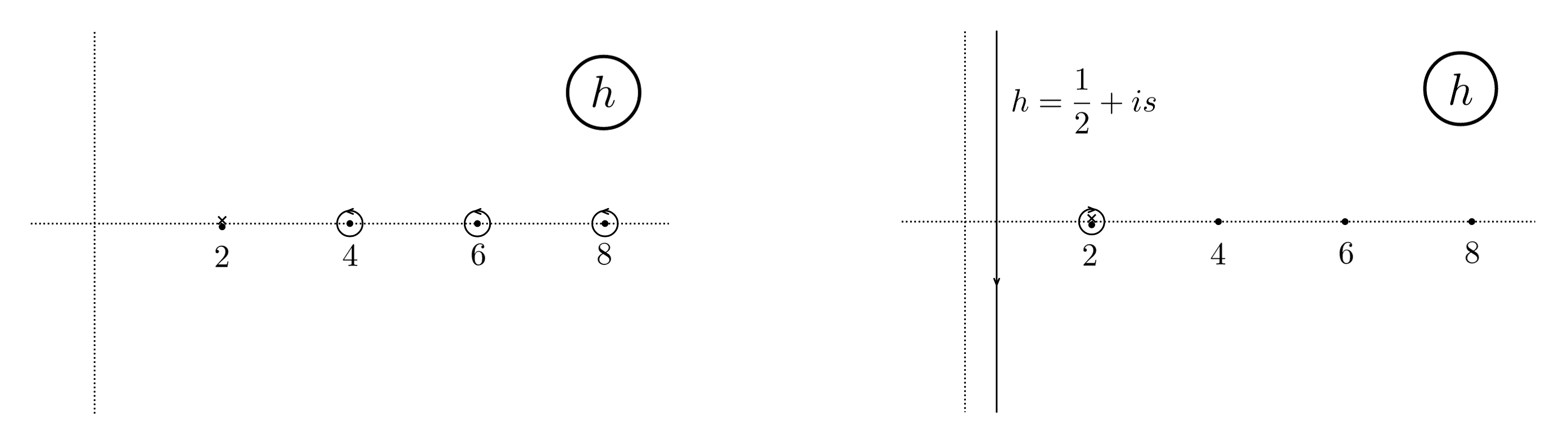}}\caption{To the left: sum over the poles into~\eqref{eq:4p-fin-2} with $k(h) \rightarrow k_R(h)$. To the right: result of pushing the contour to the left. Note that $k_R(h) = 1$ only for $h =2$, so we do not get contributions like in the right part of Fig.~\ref{fig:pole-1}.} \label{fig:pole-2}
\end{figure}
\beq \label{eq:4p-fin-4} \begin{aligned}
\mathcal{F}_{CFT}(\chi) &= \int_{-\infty}^\infty \frac{ds}{2\pi} \left[ \frac{k(h)}{1-k(h)} - \frac{k_R(h)}{1-k_R(h)}\right] \frac{h-\frac{1}{2}}{\pi \tan\left(\frac{\pi h}{2}\right)} \Psi_h(\chi) - \\ &- \text{Res}_{h=2} \left[ \frac{k_R(h)}{1-k_R(h)} \frac{h-\frac{1}{2}}{\pi \tan\left(\frac{\pi h}{2}\right)} \Psi_h(\chi) \right].
\end{aligned} \eeq
The integral in the first line rapidly converges and does not grow in the limit $\chi \rightarrow 0$, because in this limit function $\Psi_{\frac{1}{2} + s}(\chi) \sim \chi^\frac{1}{2}$ (this is true for both OTOC and TOC cases). Therefore, for our purposes this integral can be neglected.

At the same time, the pole in the second line does give a growing contribution to OTOC. Moreover, this is a double pole, which gives the combination of the function $\Psi_h$ and its derivative $\pd_h \Psi_h$. This combination grows faster than exponentially:
\beq \label{eq:OTOC-CFT}
\delta \OTOC(t) \approx \frac{C_1}{\beta J N} e^{\frac{2 \pi t}{\beta}} + \frac{C_2}{\beta J N} \frac{2 \pi t}{\beta} e^{\frac{2 \pi t}{\beta}}, \eeq 
where $C_1$ and $C_2$ are some positive numerical constants. However, this does not mean that bound~\cite{MSS} is violated in SYK model. Indeed, the contribution~\eqref{eq:OTOC-CFT} is multiplied by the small factor $\frac{1}{\beta J}$. Hence, the contribution of the conformal part is smaller than the contribution of the soft mode, at least untill the growth of the OTOC saturates. This means that the expression~\eqref{eq:OTOC-CFT} can be interpreted in terms of the leading correction to the Lyapunov exponent~\cite{Maldacena-SYK,Kitaev,Sarosi}:
\beq \label{eq:4p-fin-5}
\OTOC(t) \rightarrow \OTOC(t) + \delta \OTOC(t) \approx \frac{\sqrt{\pi}}{2 \beta J} \left[ 1 - \text{const} \frac{\beta J}{N} e^{\kappa t} \right], \quad \text{for} \quad \beta \ll t \ll \beta \log\frac{N}{\beta J}, \eeq
where ``const'' is a positive $\mathcal{O}(1)$ numerical factor and $\kappa$ is the corrected Lyapunov exponent; both factor and exponent are equal to the corresponding values from~\eqref{eq:OTOC-S} with $\mathcal{O}\left(\frac{1}{\beta J}\right)$ corrections:
\beq \label{eq:4p-fin-6}
\kappa \approx \frac{2 \pi}{\beta} \left( 1 - \frac{6.05}{\beta J} + \cdots \right). \eeq
One can also check that the pole in the second line in~\eqref{eq:4p-fin-4} does not give any growing with time contributions into the TOC. This contribution is of the ordrer $\mathcal{O}\left(\frac{1}{N}\right)$, i.e. the approximate indentity for the whole TOC is the same as in~\eqref{eq:TOC}.

We emphasize that OTOC rapidly decays only well before the scrambling time $t_* \sim \beta \log \frac{N}{\beta J}$. At greater times our approximations do not work, and other types of diagrams (e.g. multiple parallel ladders) also generate significant corrections to~\eqref{eq:OTOC-S}. So one expects that the rate of the decay slows down before OTOC is eventually saturated~\cite{Sarosi,MSS,Maldacena-SYK,Kitaev}. This conjecture was confirmed in~\cite{Bagrets-1702}, where SYK OTOCs were evaluated for arbitrary times. Namely, there was established a new time scale $t_M = \frac{N \log N}{64 \sqrt{\pi} J}$, after which exponential decay of OTOCs is replaced by a power law: $\OTOC(t) \sim (t/t_M')^{-6}$, where $t_M' = t_M$ if $\beta \ll t_M$ and $t_M' = \beta^{-1}$ in the opposite case.

Let us also emphasize again that identities~\eqref{eq:4p-fin-5} and~\eqref{eq:4p-fin-6} were obtained in the limit $1 \ll \tau J < \beta J \ll N$, i.e. only for small but non-zero temperature. However, recently it was argued that in the large $q$ limit, where $q$ is the number of fermions in the interaction vertex, similar identities hold for arbitrary temperatures and couplings~\cite{Streicher,Choi}.

\section{2D dilaton gravity}
\label{sec:JT}

The other remarkable theory which exibits a chaotic behavior is two-dimensional dilaton gravity coupled to matter. Correlation functions of the boundary operators corresponding to bulk matter fields in this model behave similarly to the correlation functions of the fermion fields in SYK model. However, we emphasize that the behavior of these models coincides only in the low energy limit. Two-dimensional dilaton gravity has been extensively studied in~\cite{Almheiri, Maldacena-JT, Jensen, Engelsoy}. In this section we review it.

\subsection{Dilaton gravity as the near-horizon limit of extremal black hole}
\label{sec:extremal}

First of all, let us show that in the near-horizon limit space-time of the 4D extremal black hole factorizes into the product of 2D anti-de Sitter space and 2D sphere. The metric and the electromagnetic field of the charged Reissner--Nordstr{\"o}m black hole are as follows:
\beq \label{eq:RN} \begin{aligned}
ds^2 &= -\frac{(r-r_+)(r-r_-)}{r^2} dt^2 + \frac{r^2}{(r-r_+)(r-r_-)} dr^2 + r^2 d\Omega^2, \\
r_\pm &= Q l_p + E l_p^2 \pm \sqrt{2 Q E l_p^3 + E^2 l_p^4}, \\
F_{rt} &= \frac{Q}{r^2}.
\end{aligned} \eeq
Here $M$ is the mass and $Q$ is the electrical charge of the black hole, $d\Omega^2$ is the metric on the two-sphere with unit radius. Also $l_p = \sqrt{G}$ is the Planck length ($G$ is the usual 4D Newton constant), and the excitation energy above extremality is $E = M - \frac{Q}{l_p}$. Obviously, for $E=0$ horizons $r_+$ and $r_-$ coincide and the black hole becomes extremal. Note that in this case $M$ and $Q$ are not independent, so the Planck length is the only dimensionful parameter of the extremal black hole.

In order to take the near-horizon limit of the extremal black hole, we introduce the variable $z$:
\beq z \equiv \frac{Q^2 l_p^2}{r - r_+}, \eeq
and take the limit $r \rightarrow r_+$, $l_p \rightarrow 0$ while keeping $z = \text{const}$. This is the simplest combination of $r-r_+$ and $l_p$ with the dimensionality of length which does not vanish in the limit $r \rightarrow r_+$ (we introduce the factor $Q^2$ for the convenience). It is straightforward to see that the metric~\eqref{eq:RN} factorizes into the sum of $AdS_2$ and $S_2$ in the limit in question:
\beq \label{eq:extremal-0}
ds^2 \approx Q^2 l_p^2 \left( \frac{-dt^2 + dz^2}{z^2} + d\Omega^2 \right). \eeq
Now let us show that some type of the excitations above the near horizon extremal black hole~\eqref{eq:extremal-0} are described by the two-dimensional dilaton gravity~\cite{Sarosi, Nayak, Grumiller-1, Thomi}. Namely, we consider static, spherically symmetric ansatz for the metric:
\beq ds^2 = h_{ij}(x^0,x^1) dx^i dx^j + \Phi^2(x^0,x^1) d\Omega^2, \eeq
where $i,j=0,1$, $x^0=t$, $x^1=r$, $h_{ij}$ and $\Phi$ are some functions to be determined. The determinant of the metric ($g = \det g_{\mu\nu}$), Ricci-scalar ($R_g$) and square of the electromagnetic tensor ($F_{\mu\nu}^2$) are as follows:
\beq \begin{aligned}
\sqrt{-g} &= \sqrt{-h} \cdot \Phi^2 \sin\theta, \\
R_g &= R_h + \frac{2}{\Phi^2} - 4 \nabla^2 \log\Phi - 6 h^{mn} \nabla_m \log\Phi \nabla_n\log\Phi, \\
F_{\mu\nu}^2 &= \frac{2 Q^2}{\Phi^4},
\end{aligned} \eeq
where $\nabla_k$ denotes the covariant derivative wrt the metric $h_{ij}$. In the second line we used that the unit sphere has constant curvature $R_{(\theta,\phi)} = 2$. Subsituting these formulae into the Einstein--Hilbert action:
\beq I = -\frac{1}{16 \pi l_p^2} \int d^4x \sqrt{-g} \left[ R_h - \frac{l_p^2}{4} F_{\mu\nu}^2 \right], \eeq
using Stokes' theorem (we assume that corresponding boundary terms at flat space-time infinity vanish) and integrating over the angular degrees of freedom, we obtain the following two-dimensional theory\footnote{Of course, one can also consider other theories of 2D dilaton gravity, e.g. theories with a different type of the potential. A comprehensive review of such theories can be found in~\cite{Grumiller-1}.}:
\beq I = -\frac{1}{4 l_p^2} \int d^2x\sqrt{-h} \left[ \Phi^2 R_h + 2 (\nabla \Phi)^2 + 2 - \frac{2 Q^2 l_p^2}{\Phi^2} \right]. \eeq
The field $\Phi$ is usually referred to as the dilaton field. Note that the Weyl transformation shifts the potential and the coefficient in front of the kinetic term:
\beq h_{ij} \rightarrow h_{ij} \Phi^{-\frac{\lambda}{2}} \quad \text{leads to} \quad 2 \rightarrow 2 - \lambda, \quad 2 - \frac{2 Q^2 l_p^2}{\Phi^2} \rightarrow \Phi^{-\frac{\lambda}{2}} \left( 2 - \frac{2 Q^2 l_p^2}{\Phi^2} \right), \eeq
so we can get rid of the kinetic term for the field $\Phi$:
\beq \label{eq:extremal-1}
I = -\frac{1}{4 l_p^2} \int d^2x\sqrt{-h} \left[ \Phi^2 R_h + 2 - \frac{2 Q^2 l_p^2}{\Phi^2} \right]. \eeq
Since now the dilaton is non-dynamical, the extremum of this action is achieved at some constant value $\Phi_0$ which determines the curvature of the spacetime. Moreover, the curvature is always negative, i.e. the extremum corresponds to the $AdS_2$ space:
\beq \delta_\Phi I = 0 \quad \text{implies} \quad R_h = -\frac{2 Q^2 l_p^2}{\Phi_0^4} = - \frac{2}{L^2}, \eeq
where we have defined the radius of the $AdS_2$ as $L = \frac{\Phi_0^2}{|Q| l_p}$. Substituting $L^2 \approx Q^2 l_p^2$ from~\eqref{eq:extremal-0}, one can estimate the critical value of the dilaton field: $\Phi_0 \approx |Q| l_p$. As we have expected, in the leading order this theory reproduces the near-horizon limit of the extremal black hole with the gravitational radius $r_\pm \approx \Phi_0$. Let us consider excitations above the extremality, which in this picture correspond to small deformations of the dilaton field:
\beq \label{eq:extremal-3}
\Phi^2 = \Phi_0^2 + \phi(x,t), \quad \phi(x,t) \ll \Phi_0^2, \eeq
and expand the action~\eqref{eq:extremal-1} up to the second order in $\frac{\phi}{\Phi_0^2}$:
\beq \label{eq:extremal-2} \begin{aligned}
I \approx &-\frac{1}{2 l_p^2} \int d^2 x \sqrt{-h} - \frac{ \Phi_0^2}{4 l_p^2} \left[ \int d^2x \sqrt{-h} \left( R_h + \frac{2}{L^2} \right) + 2 \int_{bdy} \mathcal{K} \right] - \\ &- \frac{1}{4 l_p^2} \left[ \int d^2x \sqrt{-h} \, \phi \left( R_h + \frac{2}{L^2} \right) + 2 \int_{bdy} \phi_b \mathcal{K} \right].
\end{aligned} \eeq
Here we have restored the appropriate boundary terms at the $AdS_2$ boundary\footnote{Note that this is not the same as flat space-time boundary of the 4D theory.} to make the minimal action finite (we will check this below) and introduced the trace of the extrinsic curvature:
\beq \label{eq:extremal-4}
\mathcal{K} = -\frac{h_{ab} T^a T^c \nabla_c n^b}{h_{ab} T^a T^b}, \eeq
where $T^a$ and $n^a$ are tangent and unit normal vectors to the boundary curve\footnote{In higher dimensional case boundary surface has two tangent vectors $T_1^a$ and $T_2^a$, so this expression must be modified to:
\beqs \mathcal{K} = -\frac{h_{ab} T_1^a T_2^c \nabla_c n^b}{h_{ab} T_1^a T_2^b}. \eeqs} of $AdS_2$. Also we have denoted for short $\phi \big|_{bdy} = \phi_b$.

The first term in~\eqref{eq:extremal-2} is proportional to the volume of the $AdS_2$ space which is infinite but constant. The second term is the ordinary 2D Einstein gravity. This expression is purely topological, i.e. it just gives the Euler characteristic of the manifold due to the Gauss-Bonnet theorem. Hence, both of the terms under discussion do not affect the equations of motion.

At the same time, the last term in the sum~\eqref{eq:extremal-2} does describe a non-trivial dynamics of the remaining fields. The corresponding action:
\beq \label{eq:JT}
I_{JT} = -\frac{1}{16 \pi G} \left[ \int d^2x \sqrt{-h} \, \phi \left(R_h + \frac{2}{L^2} \right) + 2 \int_{bdy} \phi_b \mathcal{K} \right], \eeq
is usually referred to as Jackiw--Teitelboim 2D gravity theory~\cite{Jackiw, Teitelboim}. Note that we have rescaled the Newton constant. Also note that $\phi$ and $G^{-1}$ always come together and form a dimensionless combination, so it is convenient to define dimensionless dilaton and Newton constant. In the following sections we will study the dynamical implications of the action~\eqref{eq:JT} more thoroughly.

A more detailed derivation of the theory~\eqref{eq:JT} from the near-horizon limit of extremal black hole can be found e.g. in~\cite{Thomi, Nayak}. Also note that this theory can be obtained by a reduction of some other higher-dimensional models~\cite{Kolekar, Grumiller-1}.

\subsection{Pure $AdS_2$ and its symmetries}
\label{sec:AdS-basics}

Before discussing the Jackiw--Teitelbom theory, let us first consider pure $AdS_2$ space to set up the notations and reveal some useful properties of the space.

First of all, it is convenient to set the radius of the space to unity, $L=1$, because it can be easily restored on dimensional grounds. Below we will consider such space if it is not stated otherwise.

Second, we will work in Euclidean signature. On the one hand, it is natural from the holographic point of view, because eventually we are interested in correlation functions of operators in the dual boundary theory (see subsection~\ref{sec:AdS-4p}): similarly to SYK case (section~\ref{sec:treatment}), we evalute some type of correlation functions in the Euclidean signature and then analytically continue them to Lorentzian times (see~\cite{Akhmedov-1802} for the discussion of the analytical continuation of $AdS$ correlation functions).

On the other hand, in Euclidean signature $AdS_2$ is just the hyperbolic disk (Lobachevsky space\footnote{We do not distinguish between the upper half-plane and unit disk because they can be mapped into each other by the M{\"o}bius transformation: $w \rightarrow \frac{iw+1}{w+i}$, where $w = t + iz$. The metrics on the plane and the disk are related by the same transformation. In particular, curves of constant $t$ and $z$ on the Fig.~\ref{fig:AdS-1} should be interpreted as the mappings of the corresponding curves on the hyperbolic plane.}), which is fully covered by the Poincar{\'e} and Rindler coordinates (see figure~\ref{fig:AdS-1}):
\begin{figure}[t]
\begin{center}
\begin{minipage}[t]{0.45\linewidth}
\includegraphics[width=1\linewidth]{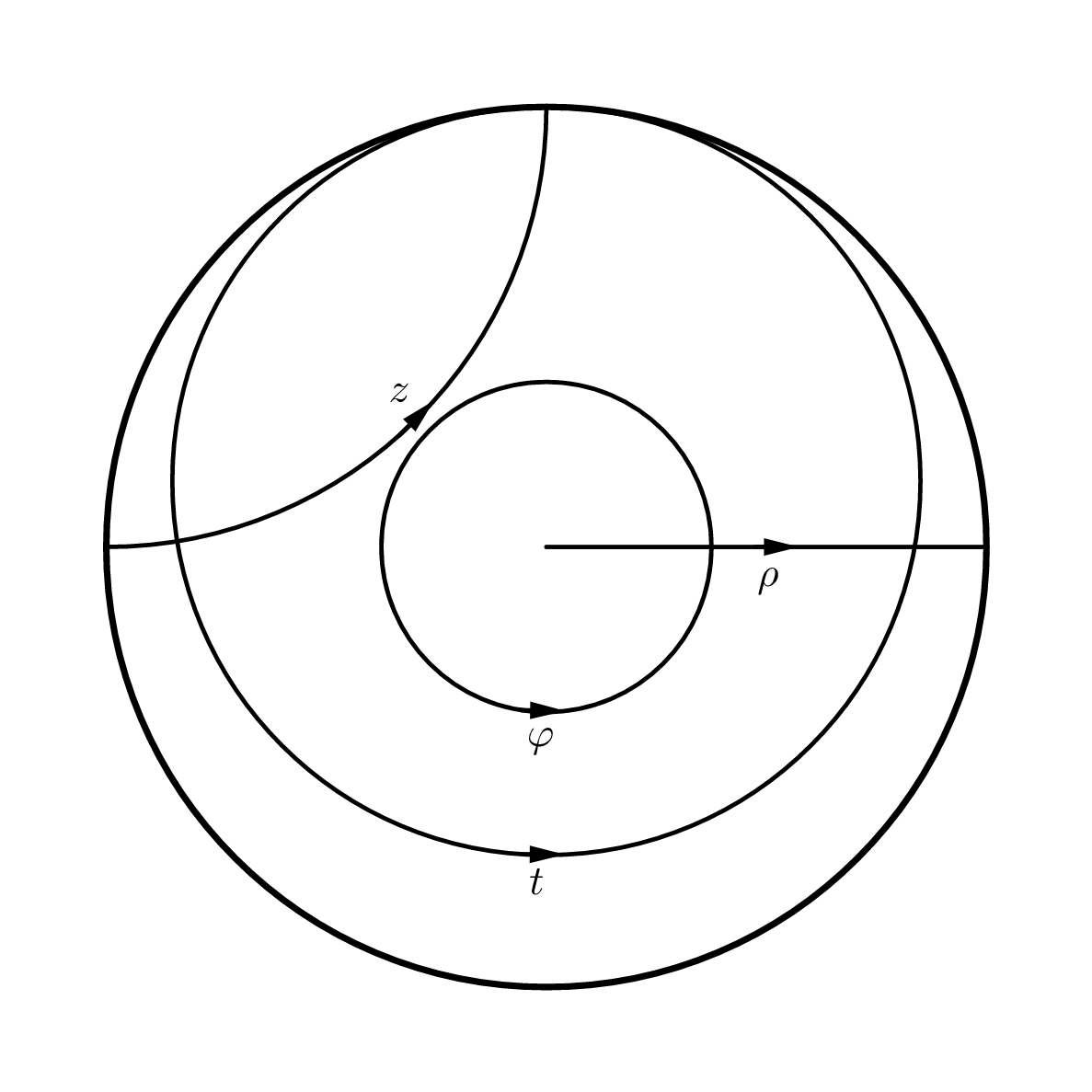}
\caption{Curves of constant $t$, $z$, $\varphi$ and $\rho$. Arrows show the direction in which the complementary coordinate increases} 
\label{fig:AdS-1}
\end{minipage}
\hfill 
\begin{minipage}[t]{0.45\linewidth}
\includegraphics[width=1\linewidth]{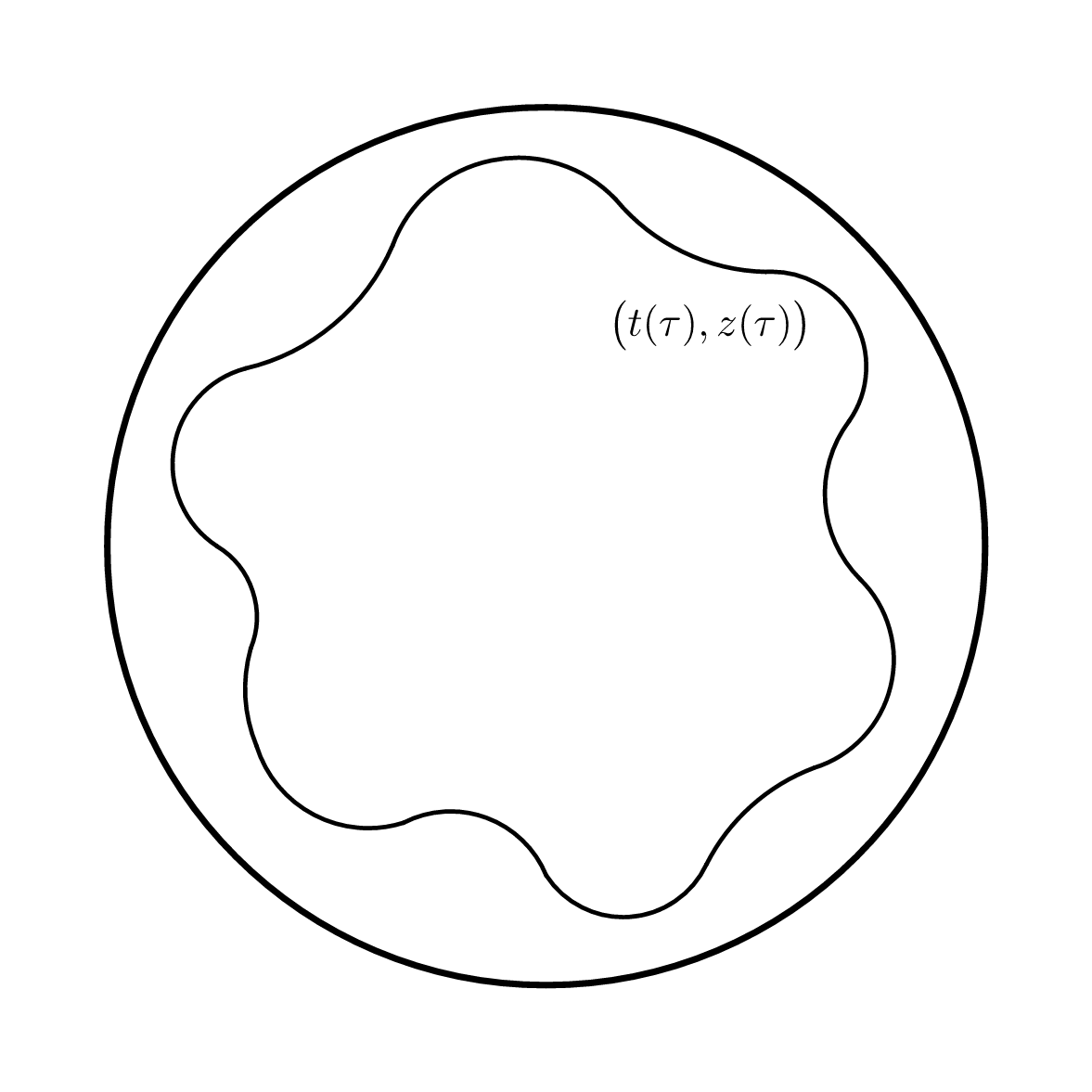}
\caption{Cutoff of the $AdS_2$ space}
\label{fig:AdS-2}
\end{minipage}
\end{center}
\end{figure}
\beq \begin{aligned}
ds^2 &= \frac{dt^2 + dz^2}{z^2}, \quad &\text{(Poincar{\'e}}) \\
ds^2 &= d\rho^2 + \sinh^2\rho d\varphi^2. \quad &\text{(Rindler)}
\end{aligned} \eeq
One can change between these coordinates using the following identities:
\beq \label{eq:AdS-2}
\tanh \frac{\rho}{2} \cos \varphi = - \frac{2t}{t^2 + (z+1)^2}, \quad \tanh \frac{\rho}{2} \sin \varphi = \frac{t^2 + z^2 - 1}{t^2 + (z+1)^2}. \eeq
Note that $t$ runs from $-\infty$ to $\infty$ and $\varphi$ runs from $-\pi$ to $\pi$ (in fact, this coordinate is periodic: $\varphi \sim \varphi + 2\pi$). Also note that in Lorentzian signature Poincar{\'e} coordinates ($ds^2 = \frac{-d\hat{t}^2 + dz^2}{z^2}$) cover only half of the spacetime and Rindler coordinates ($ds^2 = d\rho^2 - \sinh^2\rho d\hat{\varphi}^2$) cover even smaller region (e.g. see~\cite{Maldacena-JT,Spradlin}).

Finally, in practice one should cut off $AdS_2$ space at some curve that is close to the boundary (Fig.~\ref{fig:AdS-2}). Otherwise, the volume of the space and the length of the boundary-boundary geodesics are infinite. This cutoff corresponds to the UV cutoff in putative dual boundary theory. To implement such a cutoff we fix the boundary value of the metric:
\beq ds \big|_{bdy} = \sqrt{\frac{ds^2}{d\tau^2}} d\tau = \sqrt{\frac{(t')^2 + (z')^2}{z^2}} d\tau = \frac{d\tau}{\epsilon}, \eeq
which implies that the boundary curve has large proper length:
\beq S = \int ds = \int_0^\beta \frac{d\tau}{\epsilon} = \frac{\beta}{\epsilon} \rightarrow \infty, \eeq
where the time on the boundary theory runs in the interval $\tau \in [0, \beta)$ and prime denotes the derivative over $\tau$. The limit $S \rightarrow \infty$ corresponds to $\epsilon \rightarrow 0$. Note that in this limit coordinates of the curve are not independent:
\beq \label{eq:AdS-3}
\frac{(t')^2 + (z')^2}{z^2} = \frac{1}{\epsilon^2}, \quad \text{hence}, \quad z(\tau) = \epsilon t'(\tau) + \mathcal{O}(\epsilon^3). \eeq
Thus, the function $t(\tau)$ unambiguously determines the boundary curve.

As soon as the interior of the space is the same for all boundary curves, the geometry of the clipped space is determined by the shape of the boundary curve, i.e. by the single function $t(\tau)$. However, we remind that Euclidean $AdS_2$ space is invariant under the transformations from the isometry group $SO(2,1) \simeq SL(2,\mathbb{R})/\mathbb{Z}_2$, i.e. under translations and rotations. Hence, the functions $t(\tau)$ and $\tilde{t}(\tau)$, which are related by such a transformation:
\beq \label{eq:isometry}
t(\tau) \rightarrow \tilde{t}(\tau) = \frac{a t(\tau) + b}{c t(\tau) + d}, \quad \text{where} \quad ad - bc = 1 \quad \text{and} \quad a,b,c,d \in \mathbb{R}, \eeq
describe the same geometry. This statement is obvious if we rewrite the Poincar{\'e} metric in terms of complex coordinates, $w = t + i z$. The transformations that map the upper half plane into itself are as follows: 
\beq w \rightarrow \frac{aw + b}{cw + d}, \quad \text{where} \quad ad - bc = 1 \quad \text{and} \quad a,b,c,d \in \mathbb{R}, \eeq
which gives~\eqref{eq:isometry} in the limit $\epsilon \rightarrow 0$.

\subsection{Schwarzian theory}
\label{sec:AdS-Sch}

In this section we consider the Jackiw--Teitelboim theory~\eqref{eq:JT} on the clipped Poincar{\'e} disk and show that it effectively reduces to the one-dimensional theory with Schwarzian action. First, let us consider the bulk part of the action~\eqref{eq:JT}:
\beq I_{bulk} = -\frac{1}{16 \pi G} \int d^2x \sqrt{h} \, \phi \left(R_h + 2 \right). \eeq
The equation of motion for the dilaton establishes the constraint $R_h + 2 = 0$, i.e. simply tells that the metric is that of $AdS_2$. This is true even if we add matter fields, because they are not directly coupled to the dilaton. The equations of motion for the metric are as follows:
\beq T_{ij}^\phi \equiv \frac{1}{8 \pi G} \left( \nabla_i \nabla_j \phi - h_{ij} \nabla^2 \phi + h_{ij} \phi \right) = 0, \eeq
which determines the behavior of the dilaton field:
\beq \phi = \frac{a + b t + c (t^2 + z^2)}{z}, \eeq
where $a,b,c$ are integration constants. Note that near the boundary dilaton blows up:
\beq \label{eq:Dilaton}
\phi \big|_{bdy} \approx \frac{1}{\epsilon} \frac{a + b t(\tau) + c t^2(\tau)}{t'(\tau)} \equiv \frac{\phi_r(\tau)}{\epsilon}, \eeq
where we have used~\eqref{eq:AdS-3} and for convenience defined the ``renormalized'' boundary field $\phi_r(\tau)$. However, we assume that this large field is still smaller than the extremal value, $\frac{\phi_r}{\epsilon} \ll \Phi_0^2 \approx Q^2 l_p^2$ due to~\eqref{eq:extremal-3}. 

Now let us evaluate the boundary term. The tangent and normal vectors to the curve $\left( t(\tau), z(\tau) \right)$ in the Poincar{\'e} metric are $T^a = \bem t' \\ z' \eem$ and $n^a = \frac{z}{\sqrt{(t')^2 + (z')^2}} \bem -z' \\ t' \eem$ correspondingly. Hence, using~\eqref{eq:extremal-4} and~\eqref{eq:AdS-3} one readily obtains the trace of the extrinsic curvature:
\beq \label{eq:K}
\mathcal{K} = \frac{t'\left(t'^2+z'^2+z z''\right)-zz't''}{\left(t'^2 + z'^2\right)^{3/2}} = 1+ \epsilon^2 \Sch\left[t(\tau),\tau\right] + \mathcal{O}(\epsilon^4). \eeq
Substituting~\eqref{eq:Dilaton} and~\eqref{eq:K} into the boundary part of the action~\eqref{eq:JT} and changing to the integration over the time on the boundary, we obtain the following action:
\beq I_{JT}^{min} = 0 + I_{bdy} = -\frac{1}{8 \pi G} \int_{bdy} ds \frac{\phi_r(\tau)}{\epsilon} \mathcal{K} = - \frac{1}{8 \pi G} \int_0^\beta \frac{d\tau}{\epsilon} \frac{\phi_r(\tau)}{\epsilon} \Big[ 1 + \epsilon^2 \Sch\left[ t(\tau), \tau \right] + \mathcal{O} (\epsilon^4) \Big]. \eeq 
The divergent term (of the order of $\mathcal{O}(\epsilon^{-2})$) contributes to the ground state energy of the theory and should be treated using the holographic renormalizations~\cite{Skenderis, deBoer, Akhmedov-RG}. This method as applied to 2D dilaton gravity was extensively studied in~\cite{Grumiller-2, Grumiller-3, Cvetic, Jensen}. Here we just assume that the divergent term can be omitted\footnote{We emphasize that the only safe way to get the correct action and observables is honest holographic renormalization, because the mentioned crude method is sometimes misleading~\cite{Grumiller-2, Davis}. However, for the theory~\eqref{eq:JT} this crude method gives the correct result. A thorough discussion of boundary conditions, boundary counterterms and derivation of the Schwarzian action in 2D dilaton gravity can be found in~\cite{Grumiller-2, Grumiller-3, Gonzalez, Grumiller-4, Grumiller-5}.}. Thus, in the leading order in $\epsilon$ we obtain the following action:
\beq I_{JT}^{min} \approx -\frac{1}{8 \pi G} \int_0^\beta d\tau \phi_r(\tau) \Sch\left[ t(\tau), \tau \right]. \eeq
It is straightforward to check that the variation of this action over $t(\tau)$ reproduces the relation~\eqref{eq:Dilaton}.

Moreover, the time dependence of the $\phi_r(\tau)$ can be removed by the rescaling the time on the boundary theory. In order to do this we define a new coordinate $\tilde{\tau}$ such that $d\tilde{\tau} = \frac{\bar{\phi}_r d\tau}{\phi_r^2(\tau)}$, where $\bar{\phi}_r$ is some positive dimensionless constant (we remind that we consider dimensionless dilaton and Newton constant), and use the composition law for the Schwarzian\footnote{$\Sch\left[f\left(g(\tau)\right), \tau\right] = (g')^2 \Sch\left[f(g), g\right] + \Sch[g,\tau]$.}:
\beq \label{eq:Sch-ac}
I_{bdy} \approx -\frac{\bar{\phi}_r}{8 \pi G} \int_0^{\tilde{\beta}} d\tilde{\tau} \Sch\left[ t(\tilde{\tau}), \tilde{\tau}\right]. \eeq
The integral of the second term, $\phi_r \Sch\left[\tilde{\tau},\tau\right] = -2 \phi_r''$, is zero due to the periodicity $\phi_r'(\tau+\beta) = \phi_r'(\tau)$ (the boundary curve is smooth and closed). So in what follows we consider constant boundary values of the dilaton without loss of generality.

It is also convenient to change to the Rindler coordinates using the map $t(\tau) = \tan\frac{\varphi(\tau)}{2}$, which follows from the near-boundary limit ($z \rightarrow 0$) of the identities~\eqref{eq:AdS-2}:
\beq \Sch\left[t, \tau\right] = \Sch\left[\varphi, \tau\right] + \frac{(\varphi')^2}{2}. \eeq
Varying the corresponding action wrt $\varphi$ we obtain the following equation of motion:
\beq \label{eq:Sch-eom}
 \frac{\Sch\left[\varphi,\tau\right]'}{\varphi'} - \varphi'' = 0, \eeq
which has a linear in time solution:
\beq \label{eq:Sch-sln}
\varphi(\tau) = \frac{2 \pi \tau}{\beta}. \eeq
We choose the coefficient of the linear dependence in such a way that the Rindler time is periodic with the period $2 \pi$, $\varphi \sim \varphi + 2\pi$. This solution can be associated to the boundary theory at the temperature $\beta$. In what follows we will consider excitations over this solution. For convenience we set $\beta = 2 \pi$.

Note that the equation~\eqref{eq:Sch-eom} is a fourth-order non-linear differential equation that potentially has many sophisticated solutions. We do not know all of them. As a consequence, we cannot explicitly check whether the solution~\eqref{eq:Sch-sln} is the true minimum of the action~\eqref{eq:Sch-ac} or not. However, we expect the latter to be true on physical grounds.

Finally, let us consider fluctuations of the boundary curve near the minimal solution~\eqref{eq:Sch-sln}:
\beq \label{eq:Sch-flu}
\varphi(\tau) \approx \tau + \delta\varphi(\tau). \eeq
As in SYK model (see subsection~\ref{sec:4p-S}) we find the effective action for such fluctuations:
\beq \label{eq:Sch-app}
I_S = -\frac{\bar{\phi}_r}{16 \pi G} \int_0^{2\pi} d\tau \left[ (\delta\varphi')^2 - (\delta\varphi'')^2 \right] + \mathcal{O}\left(\delta\varphi^3\right), \eeq
and estimate their correlation function (compare with~\eqref{eq:4p-S-6}):
\beq \label{eq:Sch-G}
\langle \delta\varphi(\tau) \delta\varphi(0)\rangle_S \approx \frac{4 G}{\bar{\phi}_r} \sum_{m\ne-1,0,1} \frac{e^{i m \tau}}{m^2 (m^2 - 1)} = \frac{4 G}{\bar{\phi}_r} \left[ - \frac{\left(|\tau|-\pi\right)^2}{2} + \left(|\tau|-\pi\right)\sin|\tau| + 1 + \frac{\pi^2}{6} + \frac{5}{2} \cos|\tau| \right]. \eeq
Note that we excluded the modes that correspond to translations and rotations from $SL(2,\mathbb{R})$, because they are not dynamical. We will need this expression to evaluate the corrections to the correlators in the boundary theory (subsection~\ref{sec:AdS-4p}).

\subsection{Matter fields}
\label{sec:AdS-matter}

Let us add matter fields to the theory~\eqref{eq:JT}. The simplest action would be:
\beq I_{m} = \frac{1}{2} \int d^2x \sqrt{h} \left[ h^{ab} \pd_a \xi \pd_b \xi + m^2 \xi^2 \right]. \eeq
The solution to the corresponding equation of motion which is finite in the bulk but divergent in the limit $z \rightarrow 0$ is as follows:
\beq \label{eq:matter-1}
\xi(t,z) = z^{1-\Delta} \xi_r(t) + \cdots, \quad \text{where} \quad \Delta = \frac{1}{2} + \sqrt{\frac{1}{4} + m^2}, \eeq
and $\xi_r(t)$ is the boundary value of the field $\xi(t,z)$; the function $\xi_r(t)$ unambiguously determines the field $\xi(t,z)$ if it is finite in the bulk. We have denoted the subleading contribution in the limit $z \rightarrow 0$ as ``$\cdots$''. According to the $AdS/CFT$ prescription~\cite{Maldacena-AdS,Gubser,Aharony,Witten-AdS}, the function $\xi_r(t)$ can be interpreted as the source for the operator with the conformal dimension $\Delta$. Hence, the effective theory for matter fields which propagate in $AdS_2$\footnote{We remind that matter fields do not affect the constraint $R_h + 2 =0$, see the beginning of the subsection~\ref{sec:AdS-Sch}.} and satisfy boundary conditions~\eqref{eq:matter-1} is as follows (for the derivation see e.g.~\cite{Freedman}):
\beq \label{eq:matter-2}
I_{m-bdy} = - D \int dt dt' \frac{\xi_r(t) \xi_r(t')}{|t - t'|^{2\Delta}}, \quad \text{where} \quad D = \frac{\left(\Delta - \frac{1}{2}\right) \Gamma(\Delta)}{\sqrt{\pi} \Gamma\left(\Delta - \frac{1}{2} \right)}. \eeq
This action implicitly depends on the form of the boundary curve. In order to reveal this dependence we use~\eqref{eq:AdS-3} and rewrite the boundary condition in terms of the time on the boundary:
\beq \xi_r(t,z) \approx z^{1-\Delta} \xi_r(t) = \epsilon^{1-\Delta} \left[t'(\tau)\right]^{1-\Delta} \xi_r\left[t(\tau)\right] = \epsilon^{1-\Delta} \xi_r(\tau). \eeq
where we have introduced the ``renormalized'' field $\xi_r(\tau) \equiv \left[t'(\tau)\right]^{1-\Delta} \xi_r\left[t(\tau)\right]$. Substituting this definition into the action~\eqref{eq:matter-2} we obtain:
\beq I_{m-bdy} = - D \int d\tau d\tau' \left[ \frac{t'(\tau) t'(\tau')}{\left( t(\tau) - t(\tau') \right)^2} \right]^{\Delta} \xi_r(\tau) \xi_r(\tau'). \eeq
Thus, in the quasiclassical limit $G \rightarrow 0$ the boundary partition function with the source $\xi_r(\tau)$ looks as follows:
\beq \label{eq:matter-3}
Z\left[\xi_r(\tau)\right] = e^{ - I_0 - I_{Sch} - I_{m-bdy}}, \eeq
where $I_0$ denotes the ground state free energy. This term is naively divergent (in particular, it includes the divergent term which we have obtained in subsection~\ref{sec:AdS-Sch}), so it should be renormalized~\cite{Sarosi,Cvetic,Jensen}. However, it does not depend on the shape of the boundary and we just omit it in what follows.

Moreover, in the limit $G \rightarrow 0$ the contribution of the matter term is negligible (at least if $\Delta$ grows slower than $G^{-2/3}$, see~\cite{Sarosi,Maldacena-JT}), so the partition function~\eqref{eq:matter-3} is saturated at the extremum of the Schwarzian action. This limit correponds to the large $N$ limit in the dual boundary CFT. Hence, the two-point correlation function of the operators in the dual theory in the leading order is as follows:
\beq \label{eq:AdS-2p}
\langle V(\tau) V(\tau') \rangle = \frac{1}{Z\left[\xi_r\right]} \frac{\pd^2 Z\left[ \xi_r \right]}{\pd \xi_r(\tau) \pd \xi_r(\tau')} \Bigg|_{\xi_r = 0} = \left[ \frac{t'(\tau) t'(\tau')}{\left( t(\tau) - t(\tau') \right)^2} \right]^{\Delta} = \frac{1}{\left(2 \sin\frac{\tau - \tau'}{2}\right)^{2\Delta}}, \eeq
where we substituted the saddle point solution~\eqref{eq:Sch-sln} and set $\beta = 2\pi$. Here operator $V(\tau)$ is the conjugate to $\xi_r(\tau)$ according to the $AdS/CFT$ dictionary. Of course, this argumentation also holds for many-point correlation function.

There are two possible types of corrections to this expression. The first one is the corrections due to interactions in the bulk, including interaction between matter fields and backreaction to the shape of the boundary. The second one is ``quantum gravity'' loop corrections due to the fluctuations of $t(\tau)$ and $\xi(t,z)$ near the classical values (we remind that for finite $G$ the right hand side of~\eqref{eq:matter-3} is the functional integral over the bulk fields). In the limit $G \rightarrow 0$ the leading corrections come from the fluctuations of the boundary shape~\eqref{eq:Sch-flu}. In the next subsection we evaluate the contribution of such fluctuations into four-point correlation functions.

\subsection{Four-point correlation function, TOC and OTOC}
\label{sec:AdS-4p}

Following~\cite{Maldacena-JT,Jensen} in this subsection we evaluate the first ``quantum gravity'' correction to the four-point function in the ``nearly $AdS_2$'' theory. In this subsection the calculations are very similar to those that we have already discussed for SYK model in section~\ref{sec:treatment}. As in SYK model, it is convenient to define the connected part of the four-point function:
\beq \label{eq:AdS-4p-0}
\mathcal{F}(\tau_1, \tau_2, \tau_3, \tau_4) \equiv \left\langle V(\tau_1) V(\tau_2) W(\tau_3) W(\tau_4) \right\rangle - \left\langle V(\tau_1) V(\tau_2) \right\rangle \left\langle W(\tau_3) W(\tau_4) \right\rangle. \eeq
For simplicity we consider operators $V$ and $W$ which have the same conformal dimension~$\Delta$ and dual to different free fields in the bulk. First, thus we avoid the cross-channels. Second, two-point correlation functions of such operators rapidly decay under the evolution in the Lorentzian time: $\langle V(\tau_1 + it) W(\tau_2) \rangle \sim e^{- \frac{2 \pi \Delta}{\beta} t} \approx 0$ for $ t \gg \beta$ (here we restored $\beta$ in~\eqref{eq:AdS-2p}). We will need this property to evaluate OTOC and TOC.

Let us find the first order in $G$ correction to the function $\mathcal{F}$. To do this, we consider small fluctuations\footnote{Due to the action~\eqref{eq:Sch-app} such fluctuations are of the order $\delta \varphi \sim \sqrt{G/\bar{\phi}_r}$.} on top of the ``classical'' boundary curve:
\beq t(\tau) = \tan \frac{\varphi(\tau)}{2} = \tan\frac{\tau + \delta \varphi(\tau)}{2}, \eeq
and expand the two-point function~\eqref{eq:AdS-2p} to the third order in $\delta \varphi$:
\beq \left[ \frac{t'(\tau) t'(\tau')}{\left( t(\tau) - t(\tau') \right)^2} \right]^{\Delta} = \frac{1}{\left(2 \sin\frac{\tau - \tau'}{2}\right)^{2\Delta}} \left[ 1 + \mathcal{B}(\tau, \tau') + \mathcal{C}(\tau, \tau') + \mathcal{O}\left(\delta \varphi^3\right) \right], \eeq
where
\beq \begin{aligned}
\mathcal{B}(\tau_1, \tau_2) &= \Delta \left( \delta \varphi'(\tau_1) + \delta \varphi'(\tau_2) - \frac{\delta \varphi(\tau_1) - \delta \varphi(\tau_2)}{\tan \frac{\tau_{12}}{2}} \right), \\
\mathcal{C}(\tau_1, \tau_2) &= \frac{\Delta}{\left( 2 \sin \frac{\tau_{12}}{2} \right)^2} \Big[ \left(1 + \Delta + \Delta \cos \tau_{12} \right) \left( \delta \varphi(\tau_1) - \delta \varphi(\tau_2) \right)^2 + \\ &\phantom{=\frac{\Delta}{\left( 2 \sin \frac{\tau_{12}}{2} \right)^2} \Big[}+ 2 \Delta \sin \tau_{12} \left( \delta \varphi(\tau_1) - \delta \varphi(\tau_2) \right) \left( \delta \varphi'(\tau_1) + \delta \varphi'(\tau_2) \right) - \\ &\phantom{=\frac{\Delta}{\left( 2 \sin \frac{\tau_{12}}{2} \right)^2} \Big[}- \left( \cos \tau_{12} - 1 \right) \left( \Delta \left( \delta \varphi'(\tau_1) + \delta \varphi'(\tau_2) \right)^2 - \delta \varphi'(\tau_1)^2 - \delta \varphi'(\tau_2)^2 \right) \Big].
\end{aligned} \eeq
Here we denoted $\tau_{12} = \tau_1 - \tau_2$. Using this expansion we average the generating functional~\eqref{eq:matter-3} over the fluctuations of the boundary shape and find the effective action:
\beq \begin{aligned}
-I_{eff}[\xi_V, \xi_W] &= \log\left\langle e^{-I_{m-bdy}[\xi_V] - I_{m-bdy}[\xi_W]}\right\rangle_S = \\ &= D \int d\tau_1 d\tau_2 \Big[ 1 + \left\langle \mathcal{C}(\tau_1, \tau_2) \right\rangle_S \Big] \frac{\xi_V(\tau_1) \xi_V(\tau_2) + (\xi_V \leftrightarrow \xi_W)}{\left(2 \sin\frac{\tau_1 - \tau_2}{2}\right)^{2\Delta}} + \\ &+ \frac{D^2}{2} \int d\tau_1 d\tau_2 d\tau_3 d\tau_4 \left\langle \mathcal{B}(\tau_1, \tau_2) \mathcal{B}(\tau_3, \tau_4) \right\rangle_S \frac{\xi_V(\tau_1) \xi_V(\tau_2) \xi_V(\tau_3) \xi_V(\tau_4)  + (\xi_V \leftrightarrow \xi_W)}{\left(2 \sin\frac{\tau_1 - \tau_2}{2}\right)^{2\Delta} \left(2 \sin\frac{\tau_3 - \tau_4}{2}\right)^{2\Delta}} + \\ &+ D^2 \int d\tau_1 d\tau_2 d\tau_3 d\tau_4 \left\langle \mathcal{B}(\tau_1, \tau_2) \mathcal{B}(\tau_3, \tau_4) \right\rangle_S \frac{\xi_V(\tau_1) \xi_V(\tau_2) \xi_W(\tau_3) \xi_W(\tau_4)}{\left(2 \sin\frac{\tau_1 - \tau_2}{2}\right)^{2\Delta} \left(2 \sin\frac{\tau_3 - \tau_4}{2}\right)^{2\Delta}} + \mathcal{O}(G^2),
\end{aligned} \eeq
where the sources $\xi_V$, $\xi_W$ are dual to the operators $V$, $W$ correspondingly and $\langle \cdots \rangle_S$ denotes the averaging over the linearized Schwarzian action~\eqref{eq:Sch-app}:
\beq \langle O \rangle_S \equiv \frac{\int \mathcal{D} \delta \varphi \, O e^{-I_{Sch}[\delta \varphi]}}{\int \mathcal{D} \delta \varphi \, e^{-I_{Sch}[\delta \varphi]}}. \eeq
Note that $\langle \mathcal{B}(\tau_1, \tau_2)\rangle_S = 0$, because $\mathcal{B}$ is linear in $\delta \varphi$. Differentiating the effective generating functional, we find the connected four-point function:
\beq \mathcal{F}(\tau_1, \tau_2, \tau_3, \tau_4) = \frac{\pd^4 e^{-I_{eff}[\xi_V, \xi_W]}}{\pd \xi_V(\tau_1) \pd \xi_V(\tau_2) \pd \xi_W(\tau_3) \pd \xi_W(\tau_4)} \Big|_{\xi_V = 0, \; \xi_W = 0} = \frac{\langle \mathcal{B}(\tau_1, \tau_2) \mathcal{B}(\tau_3, \tau_4) \rangle_S}{\left(2 \sin\frac{\tau_1 - \tau_2}{2}\right)^{2\Delta} \left(2 \sin\frac{\tau_3 - \tau_4}{2}\right)^{2\Delta}}. \eeq
Thus, we need to calculate the expectation value of the product of two $\mathcal{B}$s. Using the propagator~\eqref{eq:Sch-G} and taking into account that
\beq \langle \delta \varphi'(\tau_1) \delta \varphi(\tau_2) \rangle_S = \sgn \left( \tau_1 - \tau_2 \right) \langle \delta \varphi(\tau_1) \delta \varphi(\tau_2) \rangle'_S, \quad \langle \delta \varphi'(\tau_1) \delta \varphi'(\tau_2) \rangle_S = \langle \delta \varphi(\tau_1) \delta \varphi(\tau_2) \rangle''_S, \eeq
we find that this average significantly depends on the order of the Euclidean times due to sign factors. As in SYK model, there are two essentially different orderings (expressions for other orderings follow from the symmetries of $\mathcal{F}$ discussed in the section~\ref{sec:treatment}):
\beq \begin{aligned}
\text{OPE:} \quad 2 \pi > \tau_1 > \tau_2 > \tau_3 > \tau_4 > 0, \\
\text{OTO:} \quad 2 \pi > \tau_1 > \tau_3 > \tau_2 > \tau_4 > 0.
\end{aligned} \eeq
For the first type of ordering the connected four-point function is as follows:
\beq \label{eq:AdS-4p-1}
\frac{\mathcal{F}(\tau_1, \tau_2, \tau_3, \tau_4)}{G(\tau_1, \tau_2) G(\tau_3, \tau_4)} = \frac{16 G \Delta^2}{\bar{\phi}_r} \left( \frac{\tau_{12}}{2 \tan \frac{\tau_{12}}{2}} - 1 \right) \left( \frac{\tau_{34}}{2 \tan \frac{\tau_{34}}{2}} - 1 \right) + \mathcal{O}(G^2). \eeq
Here $G(\tau_1, \tau_2)$ denotes the tree-level two-point functions~\eqref{eq:AdS-2p} of operators $V$ and $W$. For the second type of ordering the expression for the connected four-point function is more cumbersome:
\beq \label{eq:AdS-4p-2}
\begin{aligned}
\frac{\mathcal{F}(\tau_1, \tau_2, \tau_3, \tau_4)}{G(\tau_1, \tau_2) G(\tau_3, \tau_4)} &= \frac{16 G \Delta^2}{\bar{\phi}_r} \left( \frac{\tau_{12}}{2 \tan \frac{\tau_{12}}{2}} - 1 \right) \left( \frac{\tau_{34}}{2 \tan \frac{\tau_{34}}{2}} - 1 \right) + \\ &+ \frac{8 \pi G \Delta^2}{\bar{\phi}_r} \left( \frac{\sin\frac{\tau_{12} + \tau_{34}}{2} - \sin\frac{\tau_{13}+\tau_{24}}{2}}{\sin\frac{\tau_{12}}{2} \sin\frac{\tau_{34}}{2}} + \frac{\tau_{23}}{\tan\frac{\tau_{12}}{2} \tan\frac{\tau_{34}}{2}} \right) + \mathcal{O}(G^2).
\end{aligned} \eeq
In a similar way we also find the $\mathcal{O}(G)$ correction to the two-point functions:
\beq \label{eq:AdS-4p-3}
\begin{aligned}
\frac{\langle V(\tau_1) V(\tau_2) \rangle}{G(\tau_1, \tau_2)} = 1 &+ \frac{G \Delta}{\bar{\phi}_r} \frac{1}{\left(\sin \frac{\tau_{12}}{2}\right)^2} \Big[ 2 + 4 \Delta + \tau_{12} (\tau_{12} - 2 \pi) (\Delta + 1) + \\ &+ \left( \Delta \tau_{12} (\tau_{12} - 2 \pi) - 4 \Delta - 2 \right) \cos \tau_{12} + 2 ( \pi - \tau_{12}) (2 \Delta + 1) \sin \tau_{12} \Big] + \mathcal{O}(G^2).
\end{aligned} \eeq
The correction to the $\langle W W \rangle$ correlator is the same.

Finally, we restore $\beta$, substitute appropriate Euclidean times into the correlator~\eqref{eq:AdS-4p-0} and analytically continue~\eqref{eq:AdS-4p-1} and~\eqref{eq:AdS-4p-2} to non-zero Lorentzian times to obtain TOC and OTOC. For OTOC we consider the following set of complex times:
\beq \tau_1 = \frac{\beta}{4} + it, \quad \tau_2 = -\frac{\beta}{4} + it, \quad \tau_3 = 0, \quad \tau_4 = -\frac{\beta}{2}, \eeq
In the pure imaginary case ($t = 0$) this choice corresponds to the OTO ordering, so we need to analytically continue~\eqref{eq:AdS-4p-2}:
\beq \begin{aligned}
\OTOC(t) &\equiv \tr\left[\rho^{\frac{1}{4}} V(t) \rho^{\frac{1}{4}} W(0) \rho^{\frac{1}{4}} V(t) \rho^{\frac{1}{4}} W(0) \right] = \\ &= \mathcal{F}\left(\frac{\beta}{4} + it, -\frac{\beta}{4} + it, 0, -\frac{\beta}{2} \right) + \left\langle V\left(\frac{\beta}{2}\right) V\left(0 \right) \right\rangle \left\langle W\left(\frac{\beta}{2} \right) W\left(0 \right) \right\rangle \approx \\ &\approx \left( \frac{\pi}{\beta} \right)^{4 \Delta} \left[ 1 - 2 \Delta^2 \frac{\beta G}{\bar{\phi}_r} e^{\frac{2 \pi t}{\beta}} \right], \quad \text{for} \quad \beta \ll t \ll \beta \log \frac{\bar{\phi}_r}{\beta G}. 
\end{aligned} \eeq
Here $\rho$ denotes the density matrix in the corresponding boundary CFT. Note that we neglect the $\mathcal{O}(G)$ contributions from~\eqref{eq:AdS-4p-2} and~\eqref{eq:AdS-4p-3} which do not grow with $t$. We demand $t \gg \beta$ to exclude possible contribution from ``mixed'' correlators of the form $\langle V W \rangle$, which decay at such times. Also note that gaussian approximation that we used to obtain this result works well only for relatively small times, i.e. until the decay of the OTOC is saturated. For larger times this correlator should be calculated more carefully.

For TOC we consider the different set of times
\beq \tau_1 = \frac{\beta}{2} + it, \quad \tau_2 = it, \quad \tau_3 = 0, \quad \tau_4 = -\frac{\beta}{2}, \eeq
which corresponds to the OPE ordering at the beginning of the Lorentzian time evolution, $t = 0$. Thus we analytically continue the correlator~\eqref{eq:AdS-4p-1} and obtain the following expression:
\beq \TOC(t) \equiv \tr\left[ V(t) \rho^{\frac{1}{2}} V(t) W(0) \rho^{\frac{1}{2}} W(0) \right] \approx \left( \frac{\pi}{\beta} \right)^{4\Delta} \left[ 1 + \text{const} \frac{G}{\bar{\phi}_r} \right], \eeq
which weakly deviates from the tree-level value even for large evolution times.

\section{Examples of chaotic behavior (instead of conclusion)}
\label{sec:examples}

Instead of conclusion let us briefly review the most notable examples of chaotic systems, i.e. models with exponentially growing $C(t)$ and rapidly decaying OTOC. All these models are considered in the quasiclassical limit (large $N$ or small $G$ limit) and somehow model all-to-all couplings; furthermore, only small fluctuations above the equilibrium state are considered. So the calculation of the correlation functions are similar in all cases. In particular, in these models the leading contribution to OTOC is ensured by ladder diagrams.

\subsection{SYK model / 2D dilaton gravity}

First of all, let us briefly recall the main properties of SYK model. This is a quantum mechanical model of $N \gg 1$ Majorana fermions with all-to-all couplings $J_{ijkl}$, which are distributed randomly and independently, i.e. accordingly to the gaussian distribution with an average square $\overline{J_{ijkl}^2} = \frac{3! J^2}{N^3}$ (no sum assumed). Such a choice of couplings allows one to introduce a kind of $\frac{1}{N}$ expansion for the disorder averaged correlation functions. In particular, disorder averaged corrections to two-point and four-point functions are determined by the so-called ``melonic'' (Fig.~\ref{fig:melonic-1}) and ``ladder'' (Fig.~\ref{fig:ladder}) diagrams.

Using such a diagrammatics, one finds that in the limit $1 \ll \beta J \ll N$, which corresponds to small but non-zero temperature ($T = 1/\beta$), the exact two-point correlation function exponentially decays in Lorentzian time:
\beq G_c^\beta(t) \approx \frac{\pi^{\frac{1}{4}}}{\sqrt{2 \beta J}} \frac{\sgn(t)}{\big| \sinh\frac{\pi t}{\beta} \big|^{\frac{1}{2}}} \sim e^{-t/t_d}, \quad \text{for} \quad t \gg t_d = \frac{2 \beta}{\pi}, \eeq
time-ordered correlator is approximately equal to the product of two-point functions:
\beq \TOC(t) \approx G_c^\beta\left(-\frac{i \beta}{2}\right) G_c^\beta\left(-\frac{i \beta}{2}\right) \approx \frac{\sqrt{\pi}}{2 \beta J}, \quad \text{for} \quad t \gg t_d, \eeq
and out-of-time-ordered correlator rapidly saturates:
\beq \OTOC(t) \approx  \frac{\sqrt{\pi}}{2 \beta J}\left[1  - \frac{\Delta^2}{2 C} \frac{\beta J}{N} e^{\kappa t} \right], \quad \text{for} \quad t_d \ll t \ll t_* = \beta \log\frac{N}{\beta J}, \eeq
where $C$ is some positive numerical constant, $\Delta = \frac{1}{4}$ is effective conformal dimension of fermions and $\kappa \approx \frac{2 \pi}{\beta} \left(1 - \frac{6.05}{\beta J} + \cdots \right)$ is Lyapunov exponent. Thus, the expectation value of the square of commutator grows exponentially:
\beq \begin{aligned}
C(t) &= 2 \times \TOC(t) - \OTOC\left(t - \frac{i \beta}{4} \right) - \OTOC\left(t + \frac{i \beta}{4} \right) \approx \\ &\approx \frac{\text{const}}{N} 2 \cos\left(\frac{\beta \kappa}{4}\right) e^{\kappa t} \approx \frac{\text{const}}{N} \frac{6 \pi}{\beta J} e^{\kappa t}.
\end{aligned} \eeq
Note that the prefactor of the growing exponent is non-zero because $\kappa$ is not exactly equal to the maximal value $\frac{2 \pi}{\beta}$.

One can find the detailed derivation of these identities in the sections~\ref{sec:basics},~\ref{sec:treatment} of the present paper, papers~\cite{Kitaev, Polchinski, Maldacena-SYK, Sarosi, Jevicki-1, Jevicki-2, Rosenhaus-1807} and talks~\cite{Kitaev-talks}.

It is worth stressing that a pure boson analog of SYK model:
\beq I = \int d\tau \left[ \frac{1}{2} \sum_{i=1}^N \left(\frac{d\phi^i}{d\tau}\right)^2 + \sum_{i,j,k,l=1}^N J_{ijkl} \phi^i \phi^j \phi^k \phi^l \right], \eeq
is not self-consistent, in particular it has no reasonable exact solution~\cite{SUSY-SYK}. At the same time, supersymmetric analogs of SYK model are well defined~\cite{SUSY-SYK,Fu}.

SYK model is also closely related to Jackiw--Teitelboim (JT) gravity, i.e. two-dimensional ``near-$AdS_2$'' gravity with dilaton~\cite{Almheiri,Maldacena-JT,Jensen}. It can be shown that this theory is effectively one-dimensional, since its dynamics is determined by the shape of the boundary curve. Furthermore, in the IR limit the effective action of this theory exactly coincides with the effective action of SYK model. In both cases this  action appears due to the symmetry wrt $SL(2,\mathbb{R})$ trasformations. Therefore it is not surprising that in the semiclassical limit the behavior of correlation functions in JT gravity is similar to the one of corresponding quantities in SYK model:
\begin{align}
G(t) &\approx \left(\frac{\pi}{\beta \sinh\frac{\pi t}{\beta}}\right)^{2 \Delta} \sim e^{-t/t_d}, \quad &&\text{for} \quad t \gg t_d = \frac{\beta}{2 \pi \Delta}, \\
\TOC(t) &\approx \left( \frac{\pi}{\beta} \right)^{4 \Delta}, \quad &&\text{for} \quad t \gg t_d , \\
\OTOC(t) &\approx \left( \frac{\pi}{\beta} \right)^{4 \Delta} \left[ 1 - 2 \Delta^2 \frac{\beta G}{\bar{\phi}_r} e^{\kappa t} \right], \quad &&\text{for} \quad t_d \ll t \ll t_* = \beta \log \frac{\bar{\phi}_r}{\beta G},
\end{align}
where $\Delta$ is the conformal dimension of the operators dual to free matter fields in the bulk, $G$ is 2D Newton constant, $\bar{\phi}_r$ is the boundary value of the dilaton and $\kappa \approx \frac{2 \pi}{\beta}$ is the Lyapunov exponent.

The details on the derivation of the correlation functions and other properties of 2D dilaton gravity can be found in section~\ref{sec:JT}, papers~\cite{Maldacena-JT,Jensen,Almheiri,Engelsoy,Sarosi} and talks~\cite{Kitaev-talks}.

Note that JT gravity can be derived as a near-horizon limit of an extremal black hole~\cite{Nayak,Kolekar}, and $AdS_2$ space exibits the same causal properties as higher-dimensional $AdS$ black holes. This opens a way to use JT gravity and SYK model as toy models of many complex black hole phenomena, e.g. as toy models of traversable wormhole~\cite{Maldacena-1704, Maldacena-1804, Maldacena-1807, Maldacena-1912}.

However, it is worth stressing that JT gravity incorporates only the lowest-energy features of SYK model (which are described by the Schwarzian action) and hence cannot be considered as a complete gravity dual of this model. In fact, at the present moment such a dual is far from being known. The main problem is that the complete gravity dual should reproduce the non-local action~\eqref{eq:effective-3} that describes the dynamics of the bilinear fields $G$ and $\Sigma$. This requires to couple the theory to an infinite number of massive bulk fields (each with $\mathcal{O}(1)$ mass), but it is not known how to do this. A more detailed discussion of the putative SYK gravity dual can be found in~\cite{Sarosi,Gross,Gross-1710}.

\subsection{Generalizations of SYK model}
\label{sec:tensor}

All the remarkable properties of SYK model, including solvability in the large $N$ limit, emergence of conformal symmetry in IR and saturation of ``bound on chaos'', are based on the averaging of correlation functions over the quenched disorder, i.e. over random implementations of coupling constants. This means that SYK model is not really a quantum mechanical model; in particular, one cannot find a unitary operator that generates time evolution in this model. Thus, generalizations of SYK model, which mimick it in the large $N$ limit without the quench disorder, are of great interest. Here we present three examples of such models.

The first example is Gurau--Witten model proposed in~\cite{Witten-1610,Gurau-1611}:
\beq \label{eq:GW}
I_{GW} = \int_0^\beta d\tau \left[\frac{1}{2} \sum_{c=0}^3 \left(\sum_{\mathbf{a}^{c}} \chi_{\mathbf{a}^{c}}^{c} \frac{d}{d\tau} \chi_{\mathbf{a}^{c}}^{c}\right) + \frac{J}{N^{3/2}} \sum_{\mathbf{a}^{0} \mathbf{a}^{1} \mathbf{a}^{2} \mathbf{a}^{3}} \chi_{\mathbf{a}^{0}}^{0} \chi_{\mathbf{a}^{1}}^{1} \chi_{\mathbf{a}^{2}}^{2} \chi_{\mathbf{a}^{3}}^{3} \prod_{c_{1}<c_{2}} \delta_{a^{c_1 c_2} a^{c_2 c_1}}\right], \eeq
where $\chi^c$ are real fermionic fields and $\tau$ is Euclidean time. For every color $c$, the field $\chi^c$ lives in a vector representation of $O(N)^3$, i.e. it is a rank three tensor with indexes $\textbf{a}^c =\left \{ a^{c d}, d \ne c \right\}$, each of which runs in the range $1 \ldots N$. The full symmetry group of the model is $O(N)^6$.  For simplicity we present only the model with four-fermion vertex, general expressions can be found in~\cite{Witten-1610,Gurau-1611}. 

The second example is uncolored fermionic tensor model, or Klebanov--Tarnoposky model~\cite{Klebanov-1,Klebanov-2,Klebanov-3}:
\beq \label{eq:KT}
I_{KT} = \int_0^\beta d\tau \left[ \frac{i}{2} \sum_{abc} \chi^{abc} \frac{d}{d\tau} \chi^{abc} - \frac{g}{4} \sum_{a_1 a_2 b_1 b_2 c_1 c_2} \chi^{a_1 b_1 c_1} \chi^{a_1 b_2 c_2} \chi^{a_2 b_1 c_2} \chi^{a_2 b_2 c_1} \right],  \eeq
where $\chi^{abc}$ is rank-three fermionic tensor, indexes $a,b,c$ are indistinguishable and run in the range $1 \ldots N$. The full symmetry of the model is $O(N)^3$.

The third example mimicks SYK model by replacing random couplings $J_{ijkl}$ with a light boson tensor field~\cite{Nishinaka}:
\beq \label{eq:NT}
I_{NT} = \int_0^\beta d\tau \sum_{i<j<k<l} \frac{1}{2 \epsilon} \left[  \left( \frac{d \phi_{ijkl}}{d\tau}\right)^2 + m^2 \left(\phi_{ijkl}\right)^2 \right] + I_{SYK},  \eeq
where $\epsilon = \frac{3!}{\pi} \frac{m J^2}{N^3}$, $m \beta \ll 1$ and $I_{SYK}$ is the standard SYK action~\eqref{eq:SYK-action} with $J_{ijkl} = \phi_{ijkl}$.

We will not review models~\eqref{eq:GW},~\eqref{eq:KT} and~\eqref{eq:NT} in details, the only important point for us is that they reproduce SYK diagrammatics in the large $N$ limit. The derivation of this and other remarkable properties of SYK-like tensor models can be found in~\cite{Witten-1610, Gurau-1611, Klebanov-1, Klebanov-2, Klebanov-3, Nishinaka, Klebanov-4, Pakrouski, Popov, Gaitan, Choudhury, Bulycheva, Krishnan, Bonzom-tensor, Giombi, Ferrari-1, Ferrari-2, Azeyanagi-1, Azeyanagi-2}. Therefore, one can expect that these models are described by the same effective action and have the same properties as SYK model.

The other notable extension of SYK model is the complex SYK model~\cite{Sachdev, Gu-1910, Bulycheva-1706}:
\beq I_{CSYK} = \int_0^\beta d\tau \left[ \sum_{i=1}^N \chi_i^\dagger(\tau) \dot{\chi}_i(\tau) - \sum_{j_1 < j_2, k_1 < k_2} J_{j_1 j_2, k_1 k_2} \mathcal{A} \left\{ \chi_{j_1}^{\dagger} \chi_{j_2}^{\dagger} \chi_{k_1} \chi_{k_2} \right\} \right], \eeq
where $\mathcal{A}\{\cdots\}$ denotes the antisymmetrized product of operators and randomly distributed couplings $J_{j_1 j_2, k_1 k_2}$ have zero mean and variance $\overline{\left| J_{j_1 j_2, k_1 k_2}\right|^2} = \frac{2 J^2}{N^3}$. This theory has both $SL(2, \mathbb{R})$ and $U(1)$ symmetry. Similarly to its real predecessor, in the IR limit complex SYK is described by the Schwarzian action with an additional term corresponding to the $U(1)$ mode. A thorough discussion of this model and its applications can be found in~\cite{Sachdev, Gu-1910, Bulycheva-1706, Davison}.

\subsection{$CFT_2$ with large central charge / shock waves in $AdS_3$}

BTZ black hole and 2D $CFT$ with large central charge were among the first systems where OTOCs were calculated~\cite{Shenker-1306, Shenker-1312, Roberts-1409, Shenker-1412, Roberts-1412}. Let us briefly review the main ideas of this calculation.

First of all, in the subsection~\ref{sec:scramblers} we noticed that OTOC of local operators $V$ and $W$ can be represented as a two-sided correlation function in a perturbed thermofield double state, see formulae~\eqref{eq:scramblers-1} and~\eqref{eq:scramblers-2}. If the left and right systems are $CFT$s with $AdS$ duals then the pure state~\eqref{eq:TFD} is dual to an eternal $AdS$ Schwarzschild black hole with inverse temperature $\beta$~\cite{Maldacena-TFD}. In particular, if both systems are 2D $CFT$, $| TFD \rangle$ describes a BTZ black hole. In this picture operator $V_L(t)$ acting on  the pure $| TFD \rangle$ is dual to a particle injected near the left boundary at the moment $t$ in the past. According to holographic dictionarty~\cite{Maldacena-AdS,Gubser,Aharony,Witten-AdS}, the mass of the particle is $m_V = \frac{\Delta_V}{2 L}$, where $L$ is the radius of $AdS$ space and $\Delta_V$ is the conformal dimension of $V$ (we assume that $\Delta_V \gg 1$). In general, such a perturbation distorts the geometry of the space. Hence, one needs to estimate this distortion in order to evaluate the two-sided correlator and OTOC.

Without going into details, one obtains that the distorted geometry is described by a so-called shock wave~\cite{Cornalba,Shenker-1306,Shenker-1312}. In a nutshell, this solution is obtained by gluing the metrics of the initial black hole (of mass $M$) and the black hole that swallowed the injected particle (of mass $M + m_V$) in such way that the time at the boundary flows continuously and the radius of unit circle is continuous across the gluing surface. For small masses of the injected particle, $m_V \ll M$, the metric of the shock wave is as follows:
\beq ds^2 = -\frac{4 L^2}{(1 + UV)^2} dU dV + R^2 \left( \frac{1 - UV}{1 + UV} \right)^2 d\phi^2 + \frac{4 L^2}{(1 + UV)^2} \frac{m_V}{4 M} e^{\frac{R t}{L^2}} \delta(U) dU^2, \eeq
where $U = u$, $V = v + \frac{m_V}{4 M} e^{R t/L^2} \theta(u)$, $u$ and $v$ are standard Kruskal coordinates and $R$ is the radius of the black hole. In this metric the geodesic distance between two points close to the left and right boundaries is:
\beq \frac{d}{L} \approx 2 \log \frac{2 r}{R} + 2 \log \left[ \cosh \frac{R (t_R - t_L)}{2 L^2} + \frac{m_V}{8 M} e^{\frac{R t}{L^2} - \frac{R (t_R + t_L)}{2 L^2}} \right], \eeq
where $t_L$, $t_R$ are time coordinates and $r$ is radial coordinate of the left and right end points of the geodesic. For simplicity we assume that the anglular coordinates of the end points coincide. Subtracting the divergent contribution and setting $t_L = t_R = 0$, one ontains the following two-sided correlation function in the semiclassical limit ($G \rightarrow 0$):
\beq \label{eq:shock-1} \begin{aligned}
\OTOC(t) &\approx \Big\langle TFD \Big| V_L^\dagger(t) W_L(0) W_R(0) V_L(t) \Big| TFD \Big\rangle \sim e^{-m_W d} \sim \\ &\sim \left[ 1 + \frac{m_V}{8 M} e^{\frac{R t}{L^2}} \right]^{-2 L m_W} \sim \left[1 + C_1 \frac{m_V L}{S} e^{\frac{2 \pi t}{\beta}} \right]^{-2 L m_W}, \quad \text{for} \quad t \ll t_* = \frac{\beta}{2 \pi} \log S,
\end{aligned} \eeq
where $m_W = \frac{\Delta_W}{2 L}$, $\Delta_W \gg 1$ is the conformal dimension of $W$ and $C_1$ is a positive numerical constant. Here we have used identities for the temperature $\beta = \frac{2 \pi L^2}{R}$, mass $M = \frac{R^2}{8 G L^2}$ and entropy $S = \frac{\pi R}{2 G}$ of BTZ black hole. Also we assumed that the black hole is large, $R \sim L$, so that $S \sim \frac{R^2}{G L}$ and $C_1 = \mathcal{O}(1)$. A detailed derivation of~\eqref{eq:shock-1} and the related discussion can be found in~\cite{Shenker-1306,Roberts-1409,Shenker-1412}.

Finally, under these assumptions one can obtain the correlation function in the boundary $CFT$ with large central charge $c = \frac{3 L}{2 G}$:
\beq \label{eq:shock-2}
\OTOC(t) \sim \left[ 1 + C_2 \frac{\Delta_V}{c} e^{\frac{2 \pi t}{\beta}} \right]^{-\Delta_W}, \quad \text{for} \quad t \ll t_* \sim \frac{\beta}{2 \pi} \log c, \eeq
where $C_2$ is another positive $\mathcal{O}(1)$ numerical constant. One can also obtain this answer without holography, considering different analytical continuations of the Euclidean four-point function and using Virasoro conformal block of the identity operator~\cite{Roberts-1412,Fitzpatrick,Turiaci}.

Note that both black hole entropy and central charge measure the number of degrees of freedom of the corresponding systems, hence, for both~\eqref{eq:shock-1} and~\eqref{eq:shock-2} scrambling time $t_* \sim \beta \log N$. This saturates the bound of the fast scrambling conjecture. The Lyapunov exponent $\kappa = \frac{2 \pi}{\beta}$ also saturates the corresponding bound. However, we remind that~\eqref{eq:shock-1} reproduces only the leading contribution in the limit $G \rightarrow 0$, while the complete answer captures quantum corrections too. As was shown in~\cite{Shenker-1412}, such corrections increase the scrambling time and reduce the growth rate of OTOCs.

\subsection{Large $N$ Hermitian matrix $\Phi^4$ model}

A remarkable example of chaotic, but not maximally chaotic, model is the large $N$ matrix scalar quantum field theory with quartic self-interaction, which was considered in~\cite{Stanford-1512}:
\beq \label{eq:weak-0}
I = \int d^4 x \frac{1}{2} \tr \Big[ \left( \pd_ \mu \Phi \right)^2 - m^2 \Phi^2 - g^2 \Phi^4 \Big], \eeq
where $\Phi$ is Hermitian $N \times N$ matrix. The `t Hooft coupling is $\lambda = g^2 N \ll 1$. Summing the leading contributions in the limit $N \rightarrow \infty$, $g \rightarrow 0$, $\lambda = \text{const}$ and taking the integral over the spatial coordinates one obtains an integro-differential equation for the averaged square of commutator:
\beq \label{eq:weak-1}
\frac{d}{dt} C(t) = M \circ C(t), \eeq
where $M$ is some integral operator specified in~\cite{Stanford-1512} and
\beq \label{eq:weak-2}
C(t) = \frac{1}{N^4} \sum_{abcd} \int d^3\textbf{x} \, \tr\left( \rho^{\frac{1}{2}} \left[\Phi_{ab}(x), \Phi_{cd}(0)\right] \rho^{\frac{1}{2}} \left[\Phi_{ab}(x), \Phi_{cd}(0)\right] \right). \eeq
As in the conformal part of SYK four-point function (subsection~\ref{sec:4p-CFT}), the leading contribution to~\eqref{eq:weak-2} is provided by ladder diagrams, with operator $M$ adding an extra rung to the ladder. The largest eigenvalue of the equation~\eqref{eq:weak-1} is nothing but Lyapunov exponent $\kappa$ that determines the growth rate of $C(t) \sim e^{\kappa t}$. Numerically diagonalizing~\eqref{eq:weak-1} one can show that for small inverse temperatures, $m \beta \ll 1$, the exponent is as follows:
\beq \label{eq:weak-3}
\kappa \approx 0.025 \frac{\lambda^2}{\beta^2 m}. \eeq
In the case of zero bare mass, $m = 0$, one should substitute into~\eqref{eq:weak-3} the thermal mass $m_{th}^2 = \frac{2 \lambda}{3 \beta^2}$ generated by one-loop corrections to two-point functions:
\beq \kappa \approx 0.025 \frac{\lambda^2}{\beta^2 m_{th}} \approx 0.031 \frac{\lambda^{3/2}}{\beta}. \eeq
There is also another way to find the Lyapunov exponent~\eqref{eq:weak-3} which relies on an analogy between epidemic growth and scrambling. Let us consider the theory~\eqref{eq:weak-0} as a gas of $N^2$ interacting particles. The one-particle distribution function $f(t,\mathbf{p})$ of this gas satisfies (in the leading order) the linearized Boltzmann equation:
\beq \label{eq:weak-4}
\frac{\pd}{\pd t} f(t,\mathbf{p}) = \int \frac{d^3\mathbf{q}}{(2\pi)^3} \frac{1}{2 E_\mathbf{q}} \Big[ R^\wedge(\mathbf{p},\mathbf{q}) - R^\vee(\mathbf{p},\mathbf{q}) \Big] f(t,\mathbf{q}), \eeq
where $E_\mathbf{p} = \sqrt{m^2 + \mathbf{p}^2}$, $\mathbf{p}$ is three-dimensional momentum, functions $R^\wedge(\mathbf{p}, \mathbf{q})$ and $R^\vee(\mathbf{p}, \mathbf{q})$ measure increase and decrease of the particle density in the phase space cell $\mathbf{p}$ associated with the phase space cell $\mathbf{q}$. Note that the loss of particles is caused by two distinct processes: annihilation and outflow of particles to other cells. These processes are described by functions $2 \Gamma_\mathbf{p} \delta(\mathbf{p} - \mathbf{q})$ and $R^\vee(\mathbf{p}, \mathbf{q}) - 2 \Gamma_\mathbf{p} \delta(\mathbf{p} - \mathbf{q})$ correspondingly. The gain is only due to the inflow from other cells. For simplicity we assume that the system is spatially homogeneous.

Now let us use this qualitative model to estimate how quickly a local perturbation spreads throughout the system (i.e. estimate how quickly the system scrambles). Imagine that we injected into the system a contagious particle which infects other particles when they collide. In the early stages of the epidemic the rate of its growth is determined by the gross flow passing through the phase space cell, i.e. by the sum of inflow and outflow:
\beq \label{eq:weak-5}
\frac{\pd}{\pd t} f_{OTOC}(t,\mathbf{p}) = \int \frac{d^3\mathbf{q}}{(2\pi)^3} \frac{1}{2 E_\mathbf{q}} \frac{\sinh\frac{\beta E_\mathbf{q}}{2}}{\sinh\frac{\beta E_\mathbf{p}}{2}} \Big[ R^\wedge(\mathbf{p},\mathbf{q}) + R^\vee(\mathbf{p},\mathbf{q}) - 4 \Gamma_\mathbf{p} \delta(\mathbf{p} - \mathbf{q}) \Big] f_{OTOC}(t,\mathbf{q}). \eeq 
To obtain this equation we changed the sign of the outflow term in~\eqref{eq:weak-4} and divided the function $f(t,\mathbf{p})$ by $\sinh\frac{\beta E_\mathbf{p}}{2}$. The function $f_{OTOC}(t,\mathbf{p})$ measures the infected particle density. If this qualitative picture is applicable to the system~\eqref{eq:weak-0} and infected particles are analogs of particles affected by a perturbation, then the epidemic growth is equivalent to scrambling. Hence, one expects that the growth rate of $f_{OTOC}(t,\mathbf{p})$ coincides with the growth rate of $C(t)$.

Indeed, it was shown in~\cite{Aleiner,Grozdanov} that equation~\eqref{eq:weak-5} can be deduced from the IR limit of Bethe-Salpeter equation for OTOC (in this limit Bethe-Salpeter equations decouple). Therefore, one can evaluate the Lyapunov exponent by diagonalizing~\eqref{eq:weak-5} instead of~\eqref{eq:weak-1}. In particular, this method reproduces the result~\eqref{eq:weak-3} in the limit $N \gg 1$, $m \beta \ll 1$. Note that this approach also can be applied to other weakly coupled systems.

\section*{Acknowledgements}

Authors would like to thank F.~K.~Popov, A.~Milekhin, A.~Yu.~Morozov, V.~A.~Rubakov, P.~I.~Arseev, V.~V.~Losyakov, U.~Moschella, A.~S.~Gorsky, A.~Dymarsky, D.~Grumiller, D.~A.~Galante, L.~A.~Akopyan, E.~N.~Lanina, R.~O.~Sharipov and E.~S.~Trunina for useful comments and discussions. Especially we would like to thank E.~T.~Akhmedov for sharing of his ideas and support throughout the work. Also we would like to thank Hermann Nicolai and Stefan Theisen for the hospitality at the Albert Einstein Institute, Golm, where the work on this project was partly done. This work was supported by the grant from the Foundation for the Advancement of Theoretical Physics and Mathematics ``BASIS''.

\appendix

\section{One-dimensional Majorana fermions}
\label{sec:majorana}
\setcounter{equation}{0}

In this section we consider representations of the one-dimensional Majorana fermions~\cite{Sarosi,Kitaev-talks,Kitaev-wires,Kourkoulou}:
\beq \label{eq:majorana-3} \{ \chi_i, \chi_j \} = \delta_{ij}, \quad i,j = 1 \ldots N, \eeq
where $\chi_i = \chi_i^\dagger$. For convenience we restrict ourselves to the even number\footnote{As was shown in~\cite{Cotler, Garcia-Garcia}, the spectrum of SYK model also depends on the value of $N\;\text{mod}\;8$, but we will not discuss this point here.} $N = 2K$. In this case we can combine even and odd operators into non-Hermitian ones:
\beq \label{eq:majorana-1}
c_i = \frac{1}{\sqrt{2}} \left(\chi_{2i} - i \chi_{2 i+1}\right), \quad c_i^\dagger =  \frac{1}{\sqrt{2}} \left(\chi_{2i} + i \chi_{2i+1}\right), \quad i = 1 \ldots K, \eeq
which obey the standard anticommutation relations as a corollary of~\eqref{eq:majorana-3}:
\beq \label{eq:majorana-0}
\{ c_i, c_j \} = \{ c_i^\dagger, c_j^\dagger \} = 0, \quad \{ c_i, c_j^\dagger \} = \delta_{ij}. \eeq
They are the creation and annihilation operators of the fermion field. Hence, we can build the standard representation for the fermionic modes using these operators. Namely, we define the vacuum state as the state that is annihilated by all annihilation operators: $c_i | 0 \rangle = 0$, and build $k$-particle states using creation operators: $\left(c_1^\dagger\right)^{n_1} \cdots \left(c_K^\dagger\right)^{n_K} | 0 \rangle$, where $n_i = 0,1$ and $n_1 + \cdots + n_K = k$. There are $2^K$ such states. One can also build an explicit representation for these operators using $2^K \times 2^K$ matrices~\cite{Sarosi}, but we do not need it in this paper.

Let us calculate finite-temperature two-point correlation functions using this representation. We remind that in the free theory~\eqref{eq:SYK-action}, $J_{ijkl} = 0$, Hamiltonian is identically zero, so we rewrite the thermal average as follows:
\beq \label{eq:majorana-2}
\langle \mathcal{T} \chi_i(\tau) \chi_j(0) \rangle_\beta = \frac{\tr \left[ \mathcal{T} e^{-\beta H_0} \chi_i(\tau) \chi_j(0) \right]}{\tr \left[ e^{-\beta H_0} \right]} = \frac{\tr \left[ \mathcal{T} \chi_i(\tau) \chi_j(0) \right]}{\tr \left[ 1 \right]} = \theta(\tau) \frac{\tr \left[ \chi_i \chi_j \right]}{\tr \left[ 1 \right]} + \theta(-\tau) \frac{\tr \left[\chi_j \chi_i \right]}{\tr \left[ 1 \right]}, \eeq
where the trace denotes summation over all possible states. The last identity is valid for $\tau \in \left[-\frac{\beta}{2}, \frac{\beta}{2}\right)$, for other times we use the antiperiodycity of the propagator under the change $\tau \rightarrow \tau + \beta$ to restore the correct answer. Obviously, the average~\eqref{eq:majorana-2} equals zero if $|i - j| > 1$, because in this case it contains averages of different creation and annihilation operators which anticommute with each other (e.g. $\langle c_1 c_2 \rangle_\beta$ or $\langle c_1 c_2^\dagger \rangle_\beta$). The case $|i - j| \le 1$ is more subtle. Let us separately consider four averages that correspond to this case:
\begin{align}
\tr\left[\chi_{2i} \chi_{2i} \right] &= \frac{1}{2} \tr \left[ c_i c_i^\dagger + c_i^\dagger c_i \right] = \frac{1}{2} 2^{K-1} \left[ \langle 0 | c_i c_i^\dagger + c_i^\dagger c_i | 0 \rangle + \langle 1_i | c_i c_i^\dagger + c_i^\dagger c_i | 1_i \rangle \right] = 2^{K-1}, \\
\tr\left[\chi_{2i+1} \chi_{2i+1} \right] &= \frac{1}{2} \tr \left[ c_i c_i^\dagger + c_i^\dagger c_i \right] = \frac{1}{2} 2^{K-1} \left[ \langle 0 | c_i c_i^\dagger + c_i^\dagger c_i | 0 \rangle + \langle 1_i | c_i c_i^\dagger + c_i^\dagger c_i | 1_i \rangle \right] = 2^{K-1}, \\
\tr\left[\chi_{2i} \chi_{2i+1} \right] &= \frac{i}{2} \tr \left[ c_i c_i^\dagger - c_i^\dagger c_i \right] = \frac{i}{2} 2^{K-1} \left[ \langle 0 | c_i c_i^\dagger - c_i^\dagger c_i | 0 \rangle + \langle 1_i | c_i c_i^\dagger - c_i^\dagger c_i | 1_i \rangle \right] = 0, \\
\tr\left[\chi_{2i+1} \chi_{2i} \right] &= \frac{i}{2} \tr \left[ c_i^\dagger c_i - c_i c_i^\dagger \right] = \frac{i}{2} 2^{K-1} \left[ \langle 0 | c_i^\dagger c_i - c_i c_i^\dagger| 0 \rangle + \langle 1_i | c_i^\dagger c_i - c_i c_i^\dagger| 1_i \rangle \right] = 0,
\end{align}
where we denoted $|1_i\rangle \equiv c_i^\dagger |0\rangle$ and used the relations~\eqref{eq:majorana-1}. Now it is easy to find that the free Wightman function is:
\beq \langle \chi_i \chi_j \rangle_\beta = \frac{\tr\left[\chi_i \chi_j\right]}{\tr[1]} = \frac{1}{2} \delta_{ij}. \eeq
Taking formal limit $\beta \rightarrow \infty$ we find the free zero-temperature Feynman propagator:
\beq  \langle \mathcal{T} \chi_i(\tau) \chi_j(0) \rangle = \frac{1}{2} \sgn\tau \delta_{ij}. \eeq
This expression is also valid for the finite-temperature Feynman propagator for $\tau \in \left[-\frac{\beta}{2}, \frac{\beta}{2}\right)$. At the same time, the propagator is antiperiodic under the change $\tau \rightarrow \tau + \beta$. Using this property we restore the finite-temperature correlation function for all $\tau$:
\beq \langle \mathcal{T} \chi_i(\tau) \chi_j(0) \rangle = \frac{1}{2} \sgn\left(\sin\frac{\pi \tau}{\beta} \right) \delta_{ij}. \eeq
Finally, one can prove Wick's theorem for $n$-point correlation functions using representation~\eqref{eq:majorana-1} and standard textbook argumentation~\cite{Peskin,Evans}. However, note that for a non-equilibrium initial state this theorem does not work (e.g. see~\cite{Trunin}). This is a pecularity of the $(0+1)$-dimensional quantum field theory.

\section{Functional integral over Majorana fermions}
\label{sec:majorana-integral}
\setcounter{equation}{0}

Different one-dimensional Majorana fermions anticommute, but the square of the single fermion is not zero (see the relations~\eqref{eq:anticommutator}). Hence, these fermions cannot be described by neither normal nor Grassmann numbers. This means that a naive definition of the functional integral over Majorana fermions (e.g.~\eqref{eq:effective-0}) is unclear: how can one integrate over the variables $\chi_i$ if one does not even know what algebra they obey?

However, in the appendix~\ref{sec:majorana} we have shown that the set of $N=2K$ one-dimensional Majorana fermions can be rewritten in terms of the ordinary Dirac fermions:
\beq \chi_{2i} = \frac{1}{\sqrt{2}} \left( \psi_i + \bar{\psi}_i \right), \quad \chi_{2i+1} = \frac{i}{\sqrt{2}} \left( \psi_i - \bar{\psi}_i \right), \quad i = 1 \ldots K, \eeq
which become grassmanian upon quantization~\eqref{eq:majorana-0}. So we define the measure of the integration as follows:
\beq \mathcal{D} \chi_{2i} \mathcal{D} \chi_{2i+1} = \left| \frac{\pd(\chi_{2i}, \chi_{2i+1})}{\pd(\psi_i, \bar{\psi}_i)} \right| \mathcal{D} \psi_i \mathcal{D} \bar{\psi}_i = \mathcal{D} \psi_i \mathcal{D} \bar{\psi}_i. \eeq 
Using these definitions we calculate the gaussian integral from the subsection~\ref{sec:effective}. Now it is an ordinary integral over Grassmann variables:
\beq \begin{aligned}
\int \left( \prod_{i=1}^{2K} \mathcal{D} \chi_i \right) &\exp\left[-\frac{1}{2} \sum_{i=1}^{2K} \int d\tau d\tau' \, \chi_i(\tau) A(\tau, \tau') \chi_i(\tau')\right] = \\ = & \int \left( \prod_{i=1}^K \mathcal{D} \psi_i \mathcal{D} \bar{\psi}_i \right) \exp\left[- \sum_{i=1}^{K} \int d\tau d\tau' \, \bar{\psi}_i(\tau) A(\tau, \tau') \psi_i(\tau')\right] = \tr \log  \sqrt{A(\tau, \tau')}.
\end{aligned} \eeq
Here we used that the function $A(\tau, \tau') = - \delta(\tau-\tau') \pd_\tau - \Sigma(\tau,\tau')$ is antisymmetric under the change $\tau \leftrightarrow \tau'$. Note that the factor $\frac{1}{2}$ is cancelled, as it was expected.

\section{Correlator of energy fluctuations in SYK model}
\label{sec:energy-fluctuations}
\setcounter{equation}{0}

Following~\cite{Maldacena-SYK}, in this appendix we show that the expression~\eqref{eq:4p-S-8} is nothing but the correlator of the energy fluctuations. In order to to do this, we need the following thermodynamic relation:
\beq \label{eq:C-1}
-\log Z = \beta F = N \left[ -S_0 - \frac{2 \pi^2 C}{\beta J} + \mathcal{O}\left(\frac{1}{(\beta J)^2}\right) \right] + \beta E_0 + \frac{3}{2} \log(\beta J) + \text{const} + \mathcal{O}\left(\frac{1}{N}\right), \eeq
where $E_0$ is the ground state energy, $S_0$ is the zero-temperature entropy and coefficient $C$ is the same coefficient as in the Schwarzian action~\eqref{eq:Sch-4}. The derivation of this identity can be found in~\cite{Maldacena-SYK,Jevicki-2,Cotler}.

Varying~\eqref{eq:C-1}, in the leading order in $N$ and $\beta J$ we find that:
\beq \label{eq:C-2}
\delta E = 4 \pi^2 C \frac{N \delta \beta}{\beta^3 J}. \eeq
Now let us consider a small variation of the temperature in the propagator~\eqref{eq:DS-7}:
\beq \frac{G_c^{\beta + \delta \beta}(\tau)}{G_c^\beta(\tau)} = 1 - \frac{2 \Delta}{\beta} \left( 1 - \frac{\pi \tau}{\beta \tan \frac{\pi \tau}{\beta}} \right) \delta \beta. \eeq
Substituting~\eqref{eq:C-2} and averaging over the quantum fluctuations, we obtain the connected four-point function:
\beq \frac{\langle G_c^{\beta + \delta \beta}(\tau_1, \tau_2) G_c^{\beta + \delta \beta}(\tau_3, \tau_4) \rangle}{G_c^\beta(\tau_1, \tau_2) G_c^\beta(\tau_3, \tau_4)} - 1 = \frac{\Delta^2}{4 \pi^4 C^2} \left( 1 - \frac{\pi \tau_{12}}{\beta \tan\frac{\pi \tau_{12}}{\beta}} \right) \left( 1 - \frac{\pi \tau_{34}}{\beta \tan\frac{\pi \tau_{34}}{\beta}} \right) \frac{\beta^4 J^2}{N^2} \left\langle (\delta E)^2 \right\rangle. \eeq
The average square of the energy fluctuation can be found from~\eqref{eq:C-1}:
\beq \left\langle (\delta E)^2 \right\rangle = \pd_\beta^2 \log Z = \frac{4 \pi^2 C}{\beta^3 J}, \eeq
so the correlation function is as follows:
\beq \frac{\langle G_c^{\beta + \delta \beta}(\tau_1, \tau_2) G_c^{\beta + \delta \beta}(\tau_3, \tau_4) \rangle}{G_c^\beta(\tau_1, \tau_2) G_c^\beta(\tau_3, \tau_4)} - 1 = \frac{\Delta^2}{\pi^2 C} \frac{\beta J}{N} \left( 1 - \frac{\pi \tau_{12}}{\beta \tan\frac{\pi \tau_{12}}{\beta}} \right) \left( 1 - \frac{\pi \tau_{34}}{\beta \tan\frac{\pi \tau_{34}}{\beta}} \right). \eeq
This expression coincides with~\eqref{eq:4p-S-8} for $\Delta = \frac{1}{4}$.

\section{Integral over the product of two eigenfunctions}
\label{sec:product-integral}
\setcounter{equation}{0}

To deduce the explicitly $SL(2,\mathbb{R})$-invariant decompositions for the identity operator and four-point function, we need to calculate the following integral:
\beq A(h) = \int_{-\infty}^\infty \frac{d\omega}{2\pi} \Psi_{h\omega}(\tau_1,\tau_2) \Psi_{h\omega}(\tau_3,\tau_4)^*. \eeq
Using the symmetry of the integral under the changes $h \rightarrow 1-h$, $\omega \rightarrow -\omega$ and substituting the eigenfunctions~\eqref{eq:eigen-2} we obtain the following expression:
\beq \label{eq:app-B-1}
A (h)= \frac{2}{\pi} \int_0^\infty d\omega \cos\left(\omega(\tau_{13} + \tau_{24})\right) \begin{cases} \frac{\cos^2\left(\frac{\pi h}{2}\right)}{\cos^2(\pi h)} J_{h-\frac{1}{2}}(\omega |\tau_{12}|) J_{h-\frac{1}{2}}(\omega |\tau_{34}|) + \left(h \rightarrow 1-h \right), & h = 2n, \\ \frac{\sin^2\left(\frac{i \pi s}{2}\right)}{\sin(i \pi s)} J_{is}(\omega |\tau_{12}|) J_{is}(\omega |\tau_{34}|) + \left(s \rightarrow -s \right), & h = \frac{1}{2} + is. \end{cases} \eeq
Then we use the identity established in appendix~D of~\cite{Polchinski} as a generalization of Eq.~6.612 from~\cite{Gradshteyn}:
\beq \int_0^\infty dx \cos(a x) J_\nu(b x) J_\nu(cx) = \frac{1}{\pi \sqrt{bc}} \begin{cases} Q_{\nu-\frac{1}{2}}(z), & z > 1, \\ \tilde{Q}_{\nu-\frac{1}{2}}(z), & |z| < 1, \\ -\sin(\pi\nu) Q_{\nu-\frac{1}{2}}(z), & z < -1, \end{cases} \eeq
where we have introduced the variable $z$:
\beq z = \frac{-a^2 + b^2 + c^2}{2 b c}, \eeq
and defined the function $\tilde{Q}_\nu(z)$, which is analytic on the real interval $z\in(-1,1)$:
\beq \tilde{Q}_\nu(z) \equiv \frac{1}{2} \left[ Q_\nu(z+i0) + Q_\nu(z-i0) \right]. \eeq
Here $Q_\nu(z)$ is the usual Legendre function of the second kind, while $\tilde{Q}_\nu(z)$ is reffered to as the Legendre function on the cut. Applying this identity to the integral~\eqref{eq:app-B-1} and using Eqs.~8.335,~8.820,~9.134 from~\cite{Gradshteyn} we obtain the required formula:
\beq \label{eq:app-B-2}
A(h) = \frac{1}{\sqrt{|\tau_{12} \tau_{34}|}} \begin{cases}
\frac{\Gamma\left(\frac{h}{2}\right) \Gamma\left(\frac{1-h}{2}\right)}{\sqrt{\pi}} \phantom{.}_2 F_1\left[\frac{h}{2}, \frac{1-h}{2}, \frac{1}{2}, \left(\frac{2-\chi}{\chi}\right)^2\right], &\quad \text{if} \quad \chi > 1, \\
\frac{\cos^2\left(\frac{\pi h}{2}\right)}{\cos(\pi h)} \frac{\Gamma(h)^2}{\Gamma(2h)} \chi^h \phantom{.}_2 F_1\left(h, h, 2h, \chi \right) + (h \rightarrow 1-h), &\quad \text{if} \quad 0 < \chi < 1,
\end{cases} \eeq
where $\chi$ is the $SL(2,\mathbb{R})$-invariant cross-ratio. Note that the function~\eqref{eq:app-B-2} is invariant wrt the transformation $\chi \rightarrow \frac{\chi}{\chi - 1}$, which allows one to calculate $A(h)$ for negative cross-ratios.

\end{document}